\documentclass[fleqn,usenatbib]{mnras}

\usepackage{newtxtext,newtxmath}
\usepackage[T1]{fontenc}
\usepackage{ae,aecompl}

\usepackage{graphicx}	% Including figure files
\usepackage{amsmath}	% Advanced maths commands
\usepackage{bbold}
\usepackage{scalerel}

\def\bTheta{\boldsymbol{\Theta}}
\def\btheta{\boldsymbol{\theta}}

\def\bxi{\boldsymbol{\xi}}
\def\boeta{\boldsymbol{\eta}}
\def\bdelta{\boldsymbol{\delta}}
\def\bchi{\boldsymbol{\chi}}
\def\bDelta{\boldsymbol{\Delta}}
\def\bphi{\boldsymbol{\phi}}
\def\bchi{\boldsymbol{\chi}}
\def\bmu{\boldsymbol{\mu}}
\def\bsigma{\boldsymbol{\sigma}}
\def\bsigmatilde{\boldsymbol{\tilde \sigma}}
\def\blambda{\boldsymbol{\lambda}}
\def\bSigma{\boldsymbol{\Sigma}}

\def\bflam{\boldsymbol{f}}
\def\bflamtilde{\boldsymbol{\tilde f}}

\def\bx{\boldsymbol{x}}
\def\bt{\boldsymbol{t}}
\def\bthat{\boldsymbol{\hat t}}

\def\bs{\boldsymbol{s}}
\def\bstilde{\boldsymbol{\tilde s}}
\def\bsDR{\boldsymbol{s}_{\scaleto{\rm DR}{4pt}}}
\def\bsDRmed{\boldsymbol{s}_{\scaleto{\rm DR, med}{4pt}}}
\def \xhiz {\langle x_{\rm HI}(z)\rangle}

\def\bSDR{\boldsymbol{S}_{\scaleto{\rm DR}{4pt}}}
\def\bA{\boldsymbol{A}}
\def\bC{\boldsymbol{C}}
\def\bT{\boldsymbol{T}}
\def\bS{\boldsymbol{S}}

\def\kms{{\rm km\,s^{-1}}}
\def\bSigma{\boldsymbol{\Sigma}}
\def\logtq{\log_{10}(t_{\rm Q}\slash {\rm yr})}
\newcommand{\ccirc}{\kern0.05ex\vcenter{\hbox{$\scriptstyle\circ$}}\kern0.05ex}
\newcommand{\diag}[1]{{\rm diag}(#1)}

\title[IGM Damping Wings Towards Quasars]{Precisely Measuring the Cosmic Reionization History from IGM Damping Wings Towards Quasars}

\newcommand{\beq}{\begin{equation}}
\newcommand{\eeq}{\end{equation}}
\def\be{\begin{equation}}
\def\ee{\end{equation}}
\def\bea{\begin{eqnarray}}
\def\eea{\end{eqnarray}}



\def \logTd6 {\hbox{log$( T/6 \kev)$} }

\def\myputfigure#1#2#3#4#5%
{\vskip#5pt\makebox[0pt]{\hskip#2in
\includegraphics[width=#3\textwidth]{#1}}\vskip#4pt\hfill}

\def \kms            {~{\rm km~s}^{-1}}

\def \kev       {{\rm\ keV}}

\author[Hennawi et al.]{
Joseph F. Hennawi,$^{1,2}$\thanks{E-mail: joe@physics.ucsb.edu (JFH)}
Timo Kist,$^{1}$
Frederick B. Davies,$^{3}$
and John Tamanas$^{4}$
\\
$^{1}$Leiden Observatory, Leiden University, P.O. Box 9513, 2300 RA Leiden,
The Netherlands\\
$^{2}$Department of Physics, University of California, Santa Barbara, CA 93106, USA\\
$^{3}$Max-Planck-Institut für Astronomie, Königstuhl 17, 69117 Heidelberg, Germany\\
$^{4}$Department of Physics, University of California, Santa Cruz, CA 95064
}

\date{Accepted XXX. Received YYY; in original form ZZZ}

\pubyear{2024}

\begin{document}
\label{firstpage}
\pagerange{\pageref{firstpage}--\pageref{lastpage}}
\maketitle

\begin{abstract}
We introduce a new approach for analyzing the IGM damping wings
imprinted on the proximity zones of quasars in the epoch of
reionization (EoR).  Whereas past work has typically forgone the additional
constraining power afforded by the blue side continuum
($\lambda \lesssim 1280\,$\AA) and/or opted not to model the large correlated IGM transmission 
fluctuations in the proximity zone ($\lambda \lesssim 1216\,$\AA), we construct a
generative probabilistic model for the entire spectrum accounting for
all sources of error -- the stochasticity induced by patchy
reionization, the impact of the quasar's ionizing radiation on the
IGM, the unknown intrinsic spectrum of the quasar, and spectral
noise. This principled Bayesian method allows us to marginalize out
nuisance parameters associated with the quasar's radiation and its unknown intrinsic
spectrum to precisely measure
the IGM neutral fraction, $\langle x_{\rm HI}\rangle$.  A key element
of our analysis is the use of dimensionality reduction (DR) to
describe the intrinsic quasar spectrum via a small
number of nuisance parameters. Using a large sample of $15,559$
SDSS/BOSS quasars at $z \gtrsim 2.15$ we trained and quantified the
performance of six distinct DR methods, and find that a six parameter
PCA model (five coefficients plus a normalization)
performs best, with complex machine learning approaches providing no
advantage.  By conducting statistical inference on 100 realistic mock
EoR quasar spectra, we demonstrate the 
reliability of the
credibility contours that we obtain on $\langle x_{\rm HI}\rangle$ and the quasar
lifetime, $t_{\rm Q}$. 
The new method introduced
here will transform IGM damping wings into a precision probe of
reionization, on the same solid methodological and statistical footing
as other precision cosmological measurements.
\end{abstract}

%%%%%%%%%%%%%%%%%%%%%%%%%%%%%%%%%%%%%%%%%%%%%%%%%%

%%%%%%%%%%%%%%%%% BODY OF PAPER %%%%%%%%%%%%%%%%%%

\section{Introduction}

About 380,000\,yr after the Big Bang,
primordial plasma recombined to form the first atoms, releasing the
CMB and initiating the cosmic `dark ages', which prevailed
until radiation from stars and
black holes in primeval galaxies 
reionized the Universe.  Understanding how this epoch of reionization (EoR) emerged
and the nature of the early sources that drove it are
among the most important open questions in cosmology and key science drivers for numerous major observatories
(Planck, LOFAR, SKA, HERA, Keck, VLT, HST, Euclid, JWST). 

Our understanding  of the evolution of the average intergalactic medium (IGM) neutral fraction with cosmic time, $\langle x_{\rm HI}(z)\rangle$, currently rests upon two pillars. The first is the CMB electron scattering optical depth $\tau_e$, which provides  an integral constraint on $\langle x_{\rm HI}(z)\rangle$,  but leaves its shape poorly determined. The second is the Gunn-Peterson (GP) Ly$\alpha$ opacity measured towards distant $z\gtrsim 6$ quasars, which only robustly constrains the end of  reionization \citep[e.g][]{McGreer15,Jin23}, because the overly sensitive Ly$\alpha$
transition saturates for neutral fractions of $\langle x_{\rm HI}\rangle \gtrsim 10^{-4}$. 
The final Planck CMB constraints on reionization history,  which incorporate the GP Ly$\alpha$ opacity constraint
on the end of the reionization, indicate that the IGM was 50\% neutral at a redshift in the range
$z_{\rm reion}\simeq 5.9-8.0$ \citep[$2\sigma$;][]{Planck20},
considerably lower than the $z_{\rm reion} \sim 17$ initially inferred
by WMAP \citep{Spergel03}, and pulling reionization 
into the realm of the highest-$z$ quasars known.

Low frequency radio observations of the 21cm line have been touted as  the premier probe of reionization and are steadily increasing in sensitivity \citep[e.g.][]{HERA22}. They aim to detect the minuscule cosmic 21cm background beneath foregrounds and instrumental systematics that are $10^5$ times larger \citep{Cheng18}, but have yet to provide quantitative constraints  
on $\langle x_{\rm HI}(z)\rangle$. Similarly,  the so-called kinetic Sunyaev-Zeldovich (kSZ) effect, which must be disentangled from other small-scale  secondary CMB anistropies  as well as post-reionization kSZ contributions  \citep[e.g.][]{Dore04,Ferraro18}, also holds promise, but current  $\xhiz$ constraints are weak \citep{Nikolic23} and model-dependent \citep{Zahn12,George15}.

In a neutral IGM the GP optical depth is so large, $\tau_{\rm Ly\alpha}\sim 10^5$, that the red Lorentzian wing of the Ly$\alpha$ absorption 
cross-section can imprint an observable IGM damping wing on the spectrum of a background astronomical source \citep{Miralda98}. Multiple methods have been proposed to leverage this unique signature to obtain quantitative constraints on $\langle x_{\rm HI}(z)\rangle$ using either high-$z$ galaxies or quasars as the background sources. For quasars, one can either analyze the damped Ly$\alpha$ absorption signature arising from the IGM in the immediate vicinity of the quasar, or in the foreground (closer to the observer) where it could be imprinted upon the lower-$z$ fluctuating Ly$\alpha$ forest transmission \citep{Malloy15}.\footnote{In practice the demarcation between these two regimes depends on the extent of the region that is overionized by the quasar, i.e. the size of its proximity zone.}  Whereas the former case is the subject of this paper, several recent studies investigate the latter in the vicinity of individual Gunn-Peterson troughs embedded in the $5.5 \lesssim z \lesssim 6$ Ly$\alpha$ forest \citep{Becker24}, or stacks thereof \citep{Spina24,Zhu24} with the hope of
constraining $\xhiz$. In addition, IGM damping wings strongly suppress Ly$\alpha$ emission lines from EoR galaxies, and it has been argued that constraints on  $\langle x_{\rm HI}(z)\rangle$ can be obtained from the statistics of Ly$\alpha$ line strengths \citep{Dijkstra11,Mesinger15},  motivating a large body of work exploring this technique \citep[e.g.][]{Mason18,Mason19,Hoag19,Jung20}. Another approach recently enabled by the exquisite spectra of EoR galaxies provided by JWST is to try to measure the IGM damping wing signature from the continua of individual or stacked galaxy spectra \citep{Umeda23,Keating23a}.

However, \citet{Heintz24b} \citep[see also][]{Heintz23,Umeda23,Deugenio23,Chen24} recently demonstrated that a significant fraction $\sim 60-70\%$ of EoR galaxies with $5.5 \lesssim  z \lesssim 13$ exhibit strong {\it intrinsic} damped Ly$\alpha$ (DLA) absorption with $N_{\rm HI} \gtrsim 10^{21}\,{\rm cm}^{-2}$, 
far stronger than the corresponding intrinsic DLAs seen in star-forming galaxies at lower-$z$ ($z \lesssim 4$), and importantly, than the expected damping wings
from a neutral IGM (typical $N_{\rm HI} \sim 10^{21}\,{\rm cm}^{-2}$).  It is perhaps not surprising 
that this intrinsic absorption, which arises from the ISM or circumgalactic medium (CGM) of galaxies, should rapidly evolve as one approaches the EoR.
Estimates of the ultraviolet background (UVB) from  the statistics of Ly$\alpha$ forest transmission indicate  rapid evolution towards $z\gtrsim 5.5$ \citep{Davies18ABC,Gaikwad23,Davies24}, which is further supported by the similarly rapidly evolving 
mean-free path of ionizing photons \citep[$\lambda_{\rm mfp}$;][]{Becker21,Zhu23}, since the UVB is proportional to the mean-free path. CGM absorbers in ionization equilibrium with this evolving UVB are thus expected to be stronger and more abundant  in the EoR \citep{Bolton13}, which is empirically supported by both the strong increase in the abundance of low-ionization metal absorption lines at $z\gtrsim 6$ \citep{Becker19,Christensen23} as well as the strong redshift evolution in the occurrence of DLAs in star-forming galaxies observed by  \citet{Heintz24b}. Regardless of the physical explanation, \emph{the increased prevalence of strong intrinsic DLA absorption in EoR galaxies calls into question the entire enterprise of using galaxies, whether via the statistics of their Ly$\alpha$ lines or via damping wing absorption imprinted on their continua, as background sources to probe reionization.} No study to date has demonstrated that the intrinsic ISM/CGM absorption in galaxies can be disentangled from the damping wing absorption induced by the neutral IGM. Further obstacles arise from the poorly understood intrinsic nebular continuum shape of galaxy spectra near the Ly$\alpha$ line \citep{Raiter10,Byler17}, and 
from the challenge of modeling their Ly$\alpha$ emission lines, which is complicated by resonant scattering effects \citep{Sadoun17}. 

In contrast, quasars provide several advantages over galaxies as background sources for IGM damping wing measurements. First, quasars are far brighter, allowing one to obtain high signal-to-noise ratio ($\gtrsim 10$) and  high-resolution ($R\sim 3000-10000$) spectra with far less telescope time than required for galaxies. Second, analogous to the way O-stars transform their nearby ISM, the quasar's own
ionizing radiation sources a giant Mpc-scale HII region
\citep{CenHaiman00} known as a \emph{proximity zone}, manifest as
enhanced Ly$\alpha$ transmission near the quasar itself \citep[e.g.][]{Fan06,Eilers17a}  Although this reduces the strength of the IGM damping wing and adds a nuisance physical process to the modeling, it provides the
great advantage that neutral gas in the ISM/CGM of the quasar's host galaxy is completely ionized away, 
eliminating systematics associated with poorly understood ISM/CGM absorption. This simplifies the modeling dramatically, as one only needs to treat 1D radiative transfer of the quasar radiation through the diffuse IGM, which is a well posed problem that can be solved \emph{ab initio} --- IGM density fluctuations can be predicted from first principles and the IGM damping wing
strength (being an integral constraint) is insensitive to the exact details of the reionization topology at fixed $\langle x_{\rm HI}\rangle$ \citep{Davies18b,Chen23a,Keating23a}. 
Furthermore, this approach is also largely insensitive
\footnote{For a quasar shining into a completely neutral IGM, the size of its surrounding ionized region is primarily set by the total number of ionizing photons emitted over the age of the Universe, and is largely insensitive to its light curve due to the long recombination timescale relative to the Hubble time. In contrast, for a quasar in a highly ionized IGM, the light curve can more strongly affect IGM transmission in the proximity zone. This sensitivity arises if the quasar's radiation varies on timescales comparable to the \textit{equilibration time}, $t_{\rm eq} \equiv \Gamma_{\rm HI}^{-1} \sim 3 \times 10^4~{\rm yr}$, the timescale on which the IGM responds to quasar radiation or recombines to its baseline neutral fraction in the absence of the quasar \citep{Davies20}. Observations of quasar proximity zones are broadly consistent with a simple \textit{light-bulb} light curve \citep{Eilers17a,Davies20}, with lognormally distributed lifetimes $t_{\rm Q}$ centered around $\sim 10^6~{\rm yr}$ and a standard deviation of $\sim 1$ dex \citep{Eilers20a,Eilers21,Khrykin21,Morey21,Satyavolu23a,Dominika24}. Thus, even in an ionized IGM, current evidence suggests light-curve variability has limited impact on proximity zone transmission, though this remains an important topic for future study.} to the details of the quasar's radiative history \citep{DaviesBH19,Davies20}, which can be encapsulated by a single nuisance parameter, $t_{\rm Q}$, the quasar lifetime (assuming a `light-bulb' light curve).  The third advantage is that modeling the intrinsic quasar spectrum near Ly$\alpha$ is straightforward. Because there is very little evidence that quasar spectra evolve from $z\sim 2-7$ \citep{Shen19}, large training sets of thousands of $z\sim 2-3$ quasar spectra from surveys like SDSS/DESI  can be used to train empirical models  for their intrinsic spectra \citep[e.g.][]{Greig17a,Davies18a}.  

The intrinsic quasar spectrum constitutes a nuisance stochastic process that must be marginalized out to obtain 
constraints on the astrophysical parameters $\langle x_{\rm HI}\rangle$ and $t_{\rm Q}$. Indeed, a critical aspect of measuring the IGM damping wing signature in EoR quasar spectra
is obtaining a reliable estimate  of and uncertainty for the intrinsic unabsorbed quasar spectrum in the region near the Ly$\alpha$ line. 
To date, the approach adopted in most past work is to use the spectrum redward of the absorbed Ly$\alpha$ region (e.g. $\lambda > 1280$\,\AA) to predict the blue part of the unabsorbed quasar spectrum in the vicinity of the Ly$\alpha$ line (e.g. $\lambda < 1280$\,\AA),
an idea first suggested by \citet{Suzuki05} in the context of studies of the $z\lesssim 4$ Ly$\alpha$ forest \citep{Paris11,Lee12,Lee15,Eilers17a}.
\citet{Mortlock11} were the first to apply this red-blue prediction approach to search for an IGM damping wing imprinted on a $z=7.1$ quasar spectrum. This approach was elaborated upon by \citet{Greig17a}, who adopted a parametric model for the unabsorbed spectrum -- a sum of a power law plus multiple Gaussian emission line components --  and used the statistical correlations between these parameters, inferred from a large training dataset, to generate the red-blue predictions. 
Building on past Ly$\alpha$ forest work using principal component analysis \citep[PCA;][]{Paris11,Lee12,Lee15,Eilers17a}, 
\citet{Davies18b} developed a PCA-based red-blue continuum prediction algorithm \citep[see also][]{Eilers17b}, trained on a set of $>10,000$ quasar spectra at $z \gtrsim 2$ 
from the SDSS/BOSS DR12 quasar catalog \citep{Paris17}. They quantified the covariant uncertainties in the predicted continua, and found them to have a typical relative error of $\sim 6-12\%$ at the rest-frame wavelengths most relevant for IGM damping wing absorption. These studies motivated the development of  a plethora of red-blue continuum prediction methods 
for studying quasar proximity zones and IGM damping wings \citep{Dominika20,Fathi20,Reiman20,Chen22} as well 
as the Ly$\alpha$ forest \citep{Bosman21,Liu21}. \citet{Greig24b} recently conducted a detailed comparison 
of the  $\sim 10$ red-blue
quasar continuum prediction pipelines in existence \citep[see also][]{Bosman21}, and found that they all yield roughly comparable precision of $\sim 10-20\%$ at the relevant blue-side rest-frame wavelengths near the Ly$\alpha$ line.  

The discovery of the first $z > 7$ quasar ULASJ1120$+$0641 \citep{Mortlock11} led to  significant interest 
in using such sources to obtain quantitative constraints on reionization \citep{Mortlock11,Bolton11,Keating15,Greig17b}. \citet{Greig17b} combined the \citet{Greig17a} continuum prediction algorithm with semi-numerical simulations of the reionization topology \citep{Mesinger16} to model the distribution of IGM damping wing strengths as a function of $\langle x_{\rm HI}(z)\rangle$, but their analysis pipeline only fits rest-frame  wavelengths redward of the Ly$\alpha$ line. However it is well known that the size of the proximity zone and the strength of the damping wing are sensitive to the quasar lifetime $t_{\rm Q}$ \citep{Bolton07b,Bolton11,Keating15,Eilers17a,Davies18a}, which can vary from $10^4-10^8$~{\rm yr} \citep{Eilers17a,Eilers21,Khrykin21}, and thus must be treated as a nuisance parameter. \citet{Davies18b} presented the 
first complete model of the proximity zone and damping wing region of quasar spectra by combining the \citet{Davies18a} estimator for the intrinsic quasar continuum and its associated uncertainty, with a model for the small-scale density fluctuations in the IGM, a description of the reionization topology surrounding the massive dark matter halos hosting quasars \citep{Davies22}, and time-dependent ionizing photon radiative transfer \citep{Davies16a}. To date, the \citet{Davies18b} modeling approach has been applied to four $7 \lesssim  z \lesssim 7.5$ quasars \citep{Banados18,Davies18a,Wang20a,Yang20b} yielding robust constraints on $\xhiz$ that are competitive with the CMB.
Furthermore, the two independent modeling pipelines of \citet{Greig17b} and \citet{Davies18a}, which adopt distinct approaches for treating the impact of the quasar radiation, reionization topology, and the intrinsic quasar continuum yield results in very good agreement \citet{Greig22}. Progress on analyzing larger samples of quasars, and specifically $z < 7$ quasars for which $\langle x_{\rm HI}(z)\rangle$ should be lower, has recently been made by 
\citet{Dominika24} who analyzed stacks of a sample of 18 quasars at $6 \lesssim z \lesssim 7$, and by \citet{Greig24a}
who applied the \citet{Greig17b} pipeline to a sample of 42 quasar spectra at $5.8 \lesssim z \lesssim 6.6$. 

These IGM damping wing analyses show that, after marginalizing out  nuisance parameters describing the quasar's intrinsic continuum and lifetime, each quasar measures  $\langle x_{\rm HI}(z_{\rm QSO})\rangle$ to  $\sim 10-25\%$ precision at the quasar redshift, $z_{\rm QSO}$
\citep[see also][]{Kist25a}. Higher precision constraints on $\langle x_{\rm HI}(z)\rangle$ thus require averaging independent measurements over large statistical samples of quasars. The recently launched ESA/Euclid satellite 
is poised to discover over 100 quasars with $7.0 \lesssim z \lesssim 7.5$, and $\sim$ 25 quasars beyond the current record of $z = 7.6$,
including $\sim$ 8 beyond $z = 8.0$ \citep{Warren19}.  The \emph{James Webb Space Telescope} (JWST) will obtain exquisite spectra of these Euclid EoR quasars, many of which will be too faint to be observed with ground-based telescopes in a reasonable observing allocation.
Indeed, five existing $z > 6.8$ quasars have already been observed by JWST \citep[e.g.][]{Christensen23}.  The combination of large Euclid quasar samples and sensitive  JWST spectra have the potential to revolutionize the study 
of IGM damping wings towards quasars and  constrain the cosmic reionization history to unprecedented precision.  

But exploiting the tremendous potential of these facilities to yield precise measurements of $\langle x_{\rm HI}(z)\rangle$ requires 
the quantitative study of reionization using 
quasar damping wings to be
on the same solid methodological and statistical footing as other precision cosmological measurements. It is well known that the CMB electron scattering optical depth, $\tau_e$, is degenerate with
other cosmological parameters such as the Hubble constant, $H_0$, the amplitude of matter fluctuations, $\sigma_8$, or the 
sum of neutrino masses $\Sigma m_\nu$. Growing tensions between CMB determinations of these parameters and their values measured from late-time probes such as baryon acoustic oscillations \citep[e.g.][]{DESI24BAO}, weak-lensing \citep[e.g.][]{HSCWL19,KIDSWL22,DESWL22} the cosmic distance ladder \citep[e.g.][]{Riess22,SHOES23}, and laboratory beta-decay experiments 
\citep[e.g.][]{DiValentino22,Kreish20}  has spawned an industry of research on these putative anomalies. 
It is telling that this enormous body of work only uses \citet{Planck20} reionization constraints  because measurements from the spectra of distant astronomical objects are, apparently, not yet considered credible.

The goal of this paper is to introduce a framework that will elevate the study of IGM damping wings towards quasars to be a precision cosmological probe of reionization.  The most significant limitation of past IGM damping wing studies is that they are fundamentally sub-optimal for two reasons. First,  the commonly adopted red-blue prediction approach forgoes the additional continuum constraining power afforded by spectral pixels blueward of the typically chosen dividing line of $1280$\,\AA.  But these blue spectral pixels  ($\lambda < 1280$\,\AA) contain an abundance of information about the intrinsic
quasar continuum and the astrophysical parameters $\langle x_{\rm HI}\rangle$ and $t_{\rm Q}$.  The most intuitive way to see this is that red-blue continuum prediction  provides no mechanism to explicitly prevent the blue-side continuum from lying several spectral noise standard deviations, $\sigma_\lambda$, below the observed spectrum, $f_\lambda$, for large swaths of pixels, although such continua are clearly unphysical.  Furthermore, jointly fitting the spectral range $\lambda < 1280$\,\AA, where the smooth damping wing absorption is imprinted, for both the continuum and the astrophysical parameters, $\langle x_{\rm HI}\rangle$ and $t_{\rm Q}$,  will surely help break the degeneracy between damping wing strength and intrinsic continuum shape.  Second, nearly all past work has not modeled the highly absorbed proximity zone region blueward of the Ly$\alpha$ line ($\lambda \lesssim 1216$\,\AA), owing to the dual challenge of modeling both the impact of the quasar radiation and the large correlated IGM transmission fluctuations 
at these wavelengths. Exceptions are \citet{Davies18b} who performed ionizing radiative transfer \citep[see also][]{Bolton11} and used  simulation based inference to approximate the intractable likelihood for the correlated transmission fluctuations, and \citet{Dominika24}, who similarly performed radiative transfer and modeled the correlated transmission fluctuations with a Gaussian likelihood  --- accurate only because they analyzed stacked and hence effectively Gaussianized spectra (although stacking is suboptimal and significantly degrades the precision on $\langle x_{\rm HI}(z)\rangle$\footnote{The constraints from the \citet{Dominika24} stacked spectra are scarcely more precise than those from individual quasars obtained by other workers \citep{Greig17b,Davies18a,Greig22,Greig24a}.}). 
But an accurate likelihood that allows one to fit these absorbed proximity zone pixels would
clearly provide information about both the
astrophysical parameters and the underlying continuum, since the former determines the distribution of IGM transmission 
fluctuations, which in turn constrains the latter in a statistical sense (and vice versa).  Finally, to be taken seriously as a precision 
cosmological probe, one must use mock observations to establish that the measurements and the parameter uncertainties quoted are reliable, which has to date never been undertaken.  

Building upon the modeling approach presented in \citet{Davies18a,Davies18b} we present an improved technique for constraining reionization with EoR quasar spectra and establish the statistical robustness of the inferred astrophysical parameter constraints.  Whereas past  work failed to exploit the full constraining power of the blue side spectral region, our key innovation is the construction of a single Bayesian likelihood for the entire spectrum, allowing us to fit the continuum and the IGM damping wing signature \emph{simultaneously}.     We use dimensionality reduction (DR) to describe the intrinsic quasar continuum with a parametric model, but in contrast with most previous IGM damping wing work there is no red-blue prediction, but rather a single \emph{latent variable model} for the entire quasar spectrum. Using a large training set of $15,559$ SDSS/BOSS quasars at $z \gtrsim 2.15$ we trained and quantified the performance of six distinct dimensionality reduction methods, including machine learning approaches, and find that a six parameter PCA model (five PCA coefficients plus a normalization) performs best, with complex machine learning providing no improvements in performance.  All sources of error -- the stochasticity induced by the ionization topology, the unknown QSO lifetime $t_{\rm Q}$, continuum reconstruction errors, and spectral noise -- are accounted for in a principled manner, allowing us to marginalize out all  continuum nuisance parameters. Finally, by conducting statistical inference on 100 realistic mock EoR quasar spectra,
we show that our posterior distributions pass a \emph{coverage test}, 
establishing the reliability of the credibility contours that we obtain on $\xhiz$ and $t_{\rm Q}$ from this new method.

In our companion paper \citet{Kist25a} we quantify the precision with which IGM damping wings analyzed with this new inference
approach can measure  the astrophysical parameters, $\langle x_{\rm HI}\rangle$ and $\logtq$, and the dependence of this precision on the  dimensionality of the DR latent variable model, as well as on the spectral resolution, signal-to-noise ratio, and spectral coverage of the quasar spectra that are analyzed. 

The structure of this paper is as follows. In  \S~\ref{sec:formalism} we derive the expression for the likelihood of the quasar spectrum that is at the heart of the technique. An exploration of the six dimensionality reduction methods and a description of the training data and procedure are the subject of \S~\ref{sec:DR}. In \S~\ref{sec:sims} we summarize the \citet{Davies18b} approach for simulating quasar transmission spectra with IGM damping wing absorption and quasar lifetime effects, describe how these simulations are used to determine 
the ingredients required for the quasar spectrum likelihood, and explain our procedure for creating mock quasar spectra. In \S~\ref{sec:inference} we describe our Hamiltonian Monte Carlo (HMC) based statistical inference, 
show examples of the inference on mock spectra, present the procedure and results from the coverage testing, and compare the accuracy of our continuum reconstructions to previous work based on red-blue prediction. Finally, we summarize and conclude in \S~\ref{sec:summary}. Appendix~\ref{appendix:coverage} introduces the formalism behind our coverage tests as well as a novel approach to reweight the HMC parameter samples to guarantee that we pass a coverage test even if our original posterior distributions are overconfident. 

In this work we assume a flat $\Lambda$CDM cosmology with $h=0.685$, $\Omega_b = 0.047$, $\Omega_m = 0.3$, $\Omega_\Lambda = 0.7$,  and $\sigma_8 = 0.8$.

\section{Formalism}
\label{sec:formalism}

Our goal is to derive an expression for the likelihood of a quasar spectrum $\bflam$ with noise vector $\bsigma$ (with elements $f_\lambda$ and $\sigma_\lambda$,  respectively) to be observed in a possibly neutral IGM.
If we unrealistically assume perfect knowledge of the IGM transmission field, $\bt$, and the underlying
unabsorbed quasar spectrum, $\bs$, then because the spectral noise is Gaussian distributed,  the probability of measuring $\bflam$ is simply
\begin{equation}
  P(\bflam | \bsigma, \bt, \bs) = \mathcal{N}(\bflam;\bt \ccirc \bs, \bSigma)\label{eqn:gaussF}, 
\end{equation}
where $\mathcal{N}(\bflam;\boldsymbol{\mu}, \boldsymbol{K})$ is the standard normal distribution with mean $\bmu$ and covariance matrix $\boldsymbol{K}$,
$\ccirc$ represents an element wise (Hadamard) product of vectors, and
$\bSigma\equiv \diag{\bsigma}$
is the diagonal matrix formed from
the measured noise vector $\bsigma$ (throughout we denote vectors with bold lowercase letters/symbols
and matrices with bold capital letters/symbols).

The intrinsic quasar spectrum $\bs$ and the IGM transmission $\bt$
are \emph{latent stochastic processes}, which is to say that they are random variables that are not directly
observable. Instead, they are related to observables, but we  must marginalize over their probability distributions in order to measure the astrophysical parameters of interest. We
will adopt this approach in what follows. 

First consider the intrinsic
quasar spectrum, $\bs$ ---  our knowledge is clearly limited by
spectral noise, and, blueward of rest-frame Ly$\alpha$, by IGM
absorption. Furthermore, quasar spectra cannot be modeled from first
principles. As a result it is common to adopt a data-driven approach and describe $\bs$ with a
DR algorithm, of which principal component
analysis \citep[PCA; e.g.][]{Suzuki05,Suzuki06,Paris11,Davies18a} is the simplest example. This results in
a parametric model, $\bsDR(\boeta)$, where $\boeta$ is a new latent
variable describing the unabsorbed quasar spectrum which lives in a space with
dimensionality lower than the number of spectral pixels
(i.e. wavelengths) in $\bs$. For example, in a PCA decomposition, 
$\boeta$
would be the vector of PCA coefficients plus an overall normalization parameter. 
Since the unabsorbed spectral pixels redward of the Ly$\alpha$ line provide considerable
information about $\boeta$, it is advantageous to \emph{fit} for these parameters, rather than completely marginalize
over the quasar continuum stochastic process $\bs$. 

However, no DR algorithm is perfect, which motivates defining the \emph{relative
reconstruction error}
\be
\bdelta \equiv \frac{\bs - \bsDR(\boeta)}{\bs} \label{eqn:delta},
\ee
where division here is understood to be element wise (i.e. analogous to the Hadamard products of vectors defined in eqn.~(\ref{eqn:gaussF})). 
It then follows that $\bs = \bsDR(\boeta)\ccirc(\boldsymbol{1} + \bdelta)$ to lowest order in $\bdelta$. We assume
that $\bdelta$ is a stochastic process that follows a Gaussian probability distribution function
(PDF) given by
\be
\mathcal{N}(\bdelta; \langle \bdelta\rangle, \bDelta) \label{eqn:pdelta},
\ee
where $\langle \bdelta\rangle$ and $\bDelta$ are the mean
and covariance matrix of the relative reconstruction error, $\bdelta$,
respectively, which can be empirically determined by applying the DR
algorithm to a `test' dataset. We carry out this procedure in \S~\ref{sec:DR}, where it is shown that a Gaussian form
is indeed a very good approximation. Given these assumptions, we can
finally write for the PDF of the latent variable $\bs$:
\be
P(\bs| \boeta)  =\mathcal{N}(\bs ; \langle \bs(\boeta) \rangle, \bC_{\bs}(\boeta))\label{eqn:gaussS},
\ee
where we have defined $\langle \bs(\boeta)\rangle \equiv
\bsDR(\boeta)\ccirc (\mathbf{1} + \langle \bdelta\rangle)$, $\bC_{\bs}(\boeta) =
\bSDR(\boeta)\bDelta \bSDR(\boeta)$, and $\bSDR(\boeta) \equiv
\diag{\bsDR(\boeta)}$.

Next, consider the stochastic process governing the IGM transmission
$\bt$. While noise and the unknown continuum, $\bs$, also limits our knowledge of this
latent variable, the primary source of stochasticity is `cosmic variance',
resulting from the unknown initial conditions of the Universe. To make this more concrete, consider how
realizations of $\bt$ are generated. A common approach is to
post-process cosmological hydrodynamical simulation outputs with
ionizing radiative transfer \citep{Bolton07a,Davies16a,Davies18b,Chen21a,Satyavolu23a,Zhou23}, as we
discuss further in \S~\ref{sec:sims}. As such, the IGM
transmission, $\bt(\bphi, \btheta)$ depends on a vector, $\bphi$, which are the random phases and amplitudes of Gaussian distributed complex latent variables used to initialize the simulation, which represents the unknown initial conditions of the Universe in the vicinity of the quasar, as well as
a vector of astrophysical parameters $\btheta$, which for IGM damping
wing analysis, would be the average IGM neutral fraction $\langle
x_{\rm HI}\rangle$ and the quasar lifetime $t_{\rm Q}$, in the
simplest description \citep[e.g.][]{Davies18b}. Our aim is to measure
$\btheta$, whereas it is clearly computationally intractable to
attempt to fit for the latent variables $\bphi$, unless one had an
incredibly fast way of simulating the IGM and performing the radiative transfer (RT). As it
is unclear whether fitting for $\bphi$ would be advantageous
\citep[but see][]{Horowitz19} and it is clearly computationally intractable, the obvious strategy is to marginalize over the initial conditions,  $\bphi$.

Based on the foregoing considerations, we perform the marginalization
over the relative reconstruction error, $\bdelta$, and the initial conditions, $\bphi$, to arrive at an expression for the likelihood of the data $\bflam$ given observed noise vector $\bsigma$, astrophysical model parameters $\btheta$, and DR latent variables $\boeta$:
\be
  L(\bflam | \bsigma, \btheta, \boeta) = \iint \! P(\bflam | \bsigma, \bt(\bphi, \btheta), \bs) P(\bs | \boeta) P(\bphi)\mathrm{d}\bs \mathrm{d}\bphi. \label{eqn:marg0}
\ee 
Using the definition of the Dirac delta function
\be
1 = \int \delta_{\rm D}[\bt - \bt(\bphi, \btheta)] \, \mathrm{d}\bt,
\ee
we can  introduce an additional integral over the stochastic variable $\bt$ in eqn.~(\ref{eqn:marg0}) 
giving
\be
L(\bflam | \bsigma, \btheta, \boeta)   = \iiint \! P(\bflam | \bsigma, \bt, \bs) P(\bs | \boeta) \delta_{\rm D}[\bt-\bt(\bphi, \btheta)] P(\bphi)  \mathrm{d}\bs \mathrm{d}\bphi \mathrm{d}\bt. 
\ee
The probability distribution of IGM transmission, $P(\bt | \btheta)$, can now be defined as 
\be
P(\bt | \btheta) = \int \delta_{\rm D}[\bt - \bt(\bphi, \btheta)] P(\bphi) \mathrm{d}\bphi, 
\ee
which is the {\it pushforward distribution} of the prior over initial conditions, $P(\bphi)$, through the deterministic mapping $\bt(\bphi, \btheta)$. Intuitively, $P(\bt | \btheta)$ describes the distribution of IGM transmission fields that results from drawing random initial conditions $\bphi$ according to $P(\bphi)$ and evolving them forward under the simulation specified by the astrophysical parameters $\btheta$. Finally, we can write the likelihood as
\be
L(\bflam | \bsigma, \btheta, \boeta) = \iint \! P(\bflam | \bsigma, \bt, \bs) P(\bs | \boeta) P(\bt | \btheta) \mathrm{d}\bt\mathrm{d}\bs. \label{eqn:marg} 
\ee

Note that two of the PDFs under the integral in eqn.~(\ref{eqn:marg}),
$P(\bflam | \bsigma, \bt, \bs)$ and $P(\bs | \boeta)$ have a Gaussian form, but $P(\bt | \btheta)$ poses a challenge since,
whether at random locations in the Universe \citep[e.g.][]{Lee15,Davies18ABC} or in quasar proximity zones \citep{Davies18b}, it is well
known that the PDF of the IGM transmission is non-Gaussian. While generating samples from $P(\bt | \btheta)$ is straightforward -- 
simply randomly select
IGM transmission skewers generated from a simulation with parameters $\btheta$ --- there exists no tractable analytical
expression for $P(\bt | \btheta)$.  It thus follows that
it is impossible to derive an exact analytical
expression for the desired likelihood both because $P(\bt | \btheta)$ is intractable, and because even if an expression
for it existed,  it would be extremely challenging to perform the high dimensional marginalization integrals in
eqn.~(\ref{eqn:marg}).

Our approach 
going forward is to approximate $P(\bt | \btheta)$ with a Gaussian
form, 
\be
P(\bt| \btheta) = \mathcal{N}(\bt ; \langle \bt(\btheta)\rangle, \bC_{\bt}(\btheta))\label{eqn:gaussT},
\ee
where $\langle \bt (\btheta)\rangle$ and $\bC_{\bt}(\btheta)$ are the mean IGM transmission and its covariance, which are easily
measured from realizations using forward simulations of quasar proximity zones.
With this approximation, we can now obtain an approximate analytic expression for the likelihood. 
Substituting the Gaussian PDFs for $\bflam$, $\bs$, and $\bt$ from
eqns.~(\ref{eqn:gaussF}), (\ref{eqn:gaussS}), and (\ref{eqn:gaussT})
into the marginalization integral in eqn.~(\ref{eqn:marg}) gives
\be
\begin{split}
  L(\bflam | \bsigma, \btheta, \boeta) = \iint &\mathcal{N}(\bflam;\bt \ccirc \bs, \bSigma) \mathcal{N}(\bs ; \langle \bs(\boeta) \rangle, \bC_{\bs}(\boeta))\times \\
  &\!\!\mathcal{N}(\bt ; \langle \bt(\btheta)\rangle, \bC_{\bt}(\btheta)) \mathrm{d}\bs  \mathrm{d}\bt.\\
\end{split}
\ee
Rearranging to express the integral over $\bt$ in terms of the variable $\bt \ccirc \bs$ gives
\be
\begin{split}
  L(\bflam | \bsigma, \btheta, \boeta) = \int &\left[\int \mathcal{N}(\bflam;\bt \ccirc \bs, \bSigma)
    \mathcal{N}(\bt \ccirc \bs; \langle \bt\rangle \ccirc \bs, \bS\bC_{\bt}\bS) \mathrm{d}(\bt \ccirc \bs) \right] \times \\
  &\mathcal{N}(\bs ; \langle \bs \rangle, \bC_{\bs}) \mathrm{d}\bs,  \\
\end{split}
\ee
where $\bS \equiv \diag{\bs}$ and we suppress explicit dependencies of the means and covariances on
$\btheta$ and $\boeta$ for notational brevity.  Exploiting the fact that Gaussians
are closed under convolution, the integral in brackets can be
analytically evaluated giving
\be
  L(\bflam | \bsigma, \btheta, \boeta) = \int \mathcal{N}(\bflam; \langle \bt \rangle \ccirc \bs, \bSigma + \bS \bC_{\bt}\bS)\mathcal{N}(\bs ; \langle \bs \rangle, \bC_{\bs}) \mathrm{d}\bs. \label{eqn:sinteg} 
\ee
Analogous to above, we can rearrange eqn.~(\ref{eqn:sinteg}) in terms of the variable  $\langle \bt\rangle \ccirc \bs$ giving
\be
\begin{split}
  L(\bflam | \bsigma, \btheta, \boeta) = \int &\mathcal{N}(\bflam; \langle \bt \rangle \ccirc \bs, \bSigma + \bS \bC_{\bt}\bS)~\times\\
  &\!\mathcal{N}(\langle \bt \rangle \ccirc \bs ; \langle \bt \rangle \ccirc \langle \bs \rangle, \langle \bT \rangle \bC_{\bs} \langle \bT\rangle) \mathrm{d}(\langle \bt \rangle \ccirc \bs). \label{eqn:stinteg}\\
\end{split}
\ee
where $\langle \bT \rangle \equiv \diag{\langle \bt \rangle}$.  While the form of the eqn.~(\ref{eqn:stinteg})
suggests one use the closure of Gaussians under convolutions again to  evaluate the integral over  $\langle \bt\rangle \ccirc \bs$, note that $\bs$ also now appears in the covariance of the first normal distribution
via the term $\bS \bC_{\bt}\bS$ (recall that $\bS \equiv \diag{\bs}$), which instead renders this integral intractable.
To make progress we approximate
\be
\bs  = \bsDR(\boeta)\ccirc (\mathbf{1} + \bdelta) \approx  \bsDR(\boeta)\ccirc (\mathbf{1} + \langle \bdelta\rangle) \equiv \langle \bs (\boeta) \rangle \label{eqn:sbar}
\ee
in the problematic covariance term such that
\be
\bS \bC_{\bt}\bS \approx \langle \bS\rangle \bC_{\bt} \langle \bS \rangle
\ee
where $\langle \bS \rangle \equiv \diag{\langle \bs (\boeta)\rangle} = {\rm diag}[\bsDR(\boeta)\ccirc(\mathbf{1} + \langle \bdelta\rangle)]$.
This removes the $\bs$ dependence from the covariance\footnote{This approximation suppresses the modulation of
the covariance by the fluctuations due to the relative reconstruction error $\bdelta$. While these fluctuations are
small, ignoring them is not obviously mathematically justifiable in terms of an expansion in powers of $\bdelta$. But
given the already crude approximation of a Gaussian
transmission PDF (i.e. eqn.~(\ref{eqn:gaussT})), this
inconsistency is tolerable as it yields a closed form analytical expression} so that the likelihood becomes
\be
\begin{split}
  L(\bflam | \bsigma, \btheta, \boeta) =& \int \mathcal{N}(\bflam; \langle \bt \rangle \ccirc \bs, \bSigma + \langle \bS \rangle \bC_{\bt}\langle \bS\rangle)~\times\\
  &~~~~~\mathcal{N}(\langle \bt \rangle \ccirc \bs ; \langle \bt \rangle \ccirc \langle \bs \rangle, \langle \bT \rangle \bC_{\bs} \langle \bT\rangle) \mathrm{d}(\langle \bt \rangle \ccirc \bs). \label{eqn:lhood_final}\\
  =&~~\mathcal{N}(\bflam; \langle \bt \rangle \ccirc \langle \bs\rangle, \bSigma + \langle \bS \rangle \bC_{\bt} \langle \bS \rangle +  \langle \bT \rangle \bC_{\bs} \langle \bT\rangle),\\
\end{split}
\ee
where the last equality again follows from the closure of Gaussians under convolution. 

The primary virtue of the likelihood in eqn.~(\ref{eqn:lhood_final})
is that it operates on the entire quasar spectrum, and constitutes a
significant departure from the now standard approach in quasar IGM damping wing analysis of using
the spectrum redward of the Ly$\alpha$ line to predict the intrinsic
spectrum bluward of Ly$\alpha$, and then performing inference on the resulting 
normalized spectrum
\citep[e.g.][]{Davies18b,Dominika20,Fathi20,Reiman20,Chen22}. 
Clearly pixels blueward of the typical red-blue dividing line of $\simeq 1280$\AA\, contain an abundance of information
about the intrinsic spectrum and the astrophysical paramters $\btheta$.  
As it operates on the entire spectrum, the likelihood in eqn.~(\ref{eqn:lhood_final}) fully incorporates information from
pixels with $\lambda < 1280$\AA, where the smooth damping wing absorption is imprinted. Fitting this region helps break the degeneracy between damping wing
strength and intrinsic spectrum shape. Finally, even partly absorbed pixels in the
proximity zone constrain $\bsDR(\boeta)$, since each model
$\btheta$ predicts the distribution of $\bt$, which in turn
constrains $\bsDR(\boeta)$ in a statistical sense.
Because $t_\lambda = \exp(-\tau_\lambda)$ and $s_\lambda = f_\lambda\slash t_\lambda$, the
uncertainty in the continuum arising from stochastic IGM fluctuations
$|\delta s_{\lambda}\slash s_\lambda| = |\delta t_\lambda\slash
t_\lambda| = |\delta \tau_\lambda|$ will be the smallest at low
optical depth, and hence the highest transmission, $\bt$, 
inner proximity zone pixels (i.e. closest to the Ly$\alpha$ wavelength $1215.67$\AA in the rest-frame)
arising from regions  illuminated by the quasars intense radiation will contain the most
information about the intrinsic spectrum $\bs$.  
This also suggests that observations
with resolution sufficient to spectrally resolve transmission
spikes in the proximity zone could afford additional constraining power \citep[but see][]{Kist25a}.

The main disadvantage of the likelihood in eqn.~(\ref{eqn:lhood_final}) is that it is approximate, with
the most significant errors incurred from assuming a Gaussian form for $P(\bt | \btheta)$ in eqn.~(\ref{eqn:gaussT}), 
which we investigate in detail in \S~\ref{sec:poor_coverage}. In \S~\ref{sec:inference}, we show that this approximation yields biased and overconfident parameter inference, 
but a strategy for mitigating these shortcomings is introduced. In our companion paper we conduct more detailed tests of the precision and
fidelity of the statistical inference delivered by eqn.~(\ref{eqn:lhood_final}) and better understand the conditions under which
the Gaussian approximation for the PDF of the proximity zone transmission is valid \citep{Kist25a}. 

Finally, we note that our approach bares some resemblance to the likelihoods derived by \citet{Garnett17} and \citet{Sun23} in the context of the lower-$z$ Ly$\alpha$ forest.  However, our likelihood is more accurate since both \citet{Garnett17} and \citet{Sun23} incorporate IGM absorption via an additive noise term.  Although it simplifies the math, this approximation surely breaks down for the highly opaque IGM of interest to us here, whereas our analysis treats IGM absorption as multiplicative and is
both more accurate and applicable to low-$z$ and high-$z$ IGM absorption alike. Furthermore, \citet{Garnett17}  assume that the intrinsic quasar spectrum, $\bs$ in our notation, follows a Gaussian distribution, whereas \citet{Sun23} assumes that the latent variables describing the intrinsic spectrum, $\boeta$ in our notation, 
are Gaussian distributed and that the covariance of the continuum reconstruction errors, $\bC_{\bs}(\boeta)$ in our notation, is a diagonal matrix. In contrast, 
we only assume Gaussianity for the relative reconstruction error, $\bdelta$, which is a far weaker assumption, and treat fully covariant continuum reconstruction errors (see Fig.~\ref{fig:resid_corr}). 
The main advantage of their approaches relative to ours is that they present a method to determine the quasar continuum dimensionality reduction model, $\bsDR(\boeta)$ directly from a low-$z$ Ly$\alpha$ forest dataset in addition to the astrophysical parameters, whereas in our approach, we 
derive the latent variable model from an external training set, which as we discuss in the next section, comes from SDSS/BOSS $2.15 \lesssim z \lesssim 4$  quasar spectra fit with an automated continuum fitting algorithm. It is worth exploring in future work if, analogous to \citet{Garnett17} and \citet{Sun23},  one can use the likelihood in eqn.~(\ref{eqn:lhood_final}) to fit  for both the latent variable model that describes the continua of SDSS/BOSS $2.15 \lesssim z \lesssim 4$  quasar spectra and the astrophysical parameters that govern the low-$z$ Ly$\alpha$ forest.

\section{Quasar Dimensionality Reduction}
\label{sec:DR}

A critical compoment of the formalism presented in the previous section is the representation 
of the intrinsic quasar spectrum, $\bs$, with a DR algorithm. Specifically, before we can compute the likelihood in eqn. (\ref{eqn:lhood_final}) we need to: 1) apply a
DR algorithm to an ensemble of quasar spectra to determine the
function $\bsDR(\boeta)$, 2) demonstrate that the probability distribution
of the relative reconstruction error $\bdelta$ (eqn.~(\ref{eqn:delta})) is well described by a multivariate Gaussian
distribution (eqns.~(\ref{eqn:pdelta}-\ref{eqn:gaussS})), and 3) measure the mean
$\langle\bdelta\rangle$ and covariance $\bDelta$ of this distribution. 

An important question is which DR algorithm to employ. We will compare several different DR approaches in this section. DR methods are commonly divided into linear models and non-linear models. We will start with most widely adopted linear model which is PCA, which will be compared to two non-linear models, namely a Gaussian Process Latent Variable Model (GPLVM), and a Variational Autoencoder (VAE).   Although PCA is provably optimal among linear methods in the sense that it minimizes the average squared reconstruction error for a fixed dimension of the latent space, $n_{\rm latent}$, this optimality does not extend to nonlinear methods. In particular, nonlinear DR algorithms such as the GPLVM and VAE could, in principle, more efficiently capture the structure of the data manifold, and thus achieve smaller  reconstruction errors. For this reason, we assess the performance of each method empirically. First, we will describe the procedure for generating the training data for the DR algorithms  (\S~\ref{sec:traintest}), then we will discuss our implementation of each DR method (\S~\ref{sec:DRmethods}), and finally we quantify and compare their performance for representing quasar spectra (\S~\ref{sec:DRcompare}). Ultimately, we will conclude that a PCA with $n_{\rm DR}=6$ parameters (i.e. $n_{\rm latent}=5$ PCA coefficients plus a normalization parameter, $s_{\rm norm}$) is the best choice for our application.  The reader who is not interested in the details can skip ahead to section \S~\ref{sec:wavegrid}.

\subsection{Training and Test Data}
\label{sec:traintest}

\subsubsection{Automated Continuum Fits of SDSS Quasars}
\label{sec:autofit}

Our training set of quasar spectra is drawn from the  
SDSS-III
Baryon Oscillation Spectroscopic Survey (BOSS) and the SDSS-IV
Extended BOSS (eBOSS) surveys which obtained moderate resolution $(R
\sim 2000)$ spectra of a large sample of $z \gtrsim 2$ quasars
\citep{Dawson13,Dawson16}. Specifically, we consider objects
identified as quasars in the eBOSS DR14 quasar catalog \citep{Paris18}
using the compilation in the \texttt{igmspec} database of public
spectra probing the intergalactic medium, which uses the
\texttt{specdb} database framework \citep{Prochaska17}. Our aim is to
use DR to describe quasars over the rest-frame wavelength range relevant to IGM
damping wing analysis 1170\AA-2040\AA. As the (upgraded) SDSS
spectrograph covers the wavelength range 3580\AA-10,350\AA, the
requirement that our desired rest-frame wavelength range be fully covered restricts the range of
usable quasar redshifts. Adding a small buffer on the
blue end of the spectra to avoid edge effects, and excluding quasars
with $z > 4$ where the Ly$\alpha$ forest transmission is smaller and
harder to correct with automated continuum fitting algorithms (see below), we
arrive at the redshift range $2.149 < z < 4.0$. There are 199,530
quasars in this range in the eBOSS DR14 quasar catalog, of which we removed
13,391 objects that are likely broad absorption line (BAL) quasars given nonzero values of the
BALnicity index characterizing \ion{C}{IV} absorption troughs \citep[see e.g.][]{Paris18},
resulting in a sample of 186,139 quasars. We further require 
that the median signal-to-noise ratio
%% TK be?
${\rm S\slash N} > 10$ within a 5.0\AA\, region
centered at rest-frame 1285\AA. After imposing this  ${\rm S\slash N}$ requirement
and removing a small number of problematic spectra  with large gaps in their spectral coverage,
we are left with a parent sample of 20,201 quasars.

In order to define a continuous smooth spectrum that covers our
desired rest-frame wavelength range, we use automated continuum
fitting algorithms following the approaches adopted
by previous work on quasar continua \citep{Davies18a,Dominika20,Bosman21}.
\citet{Davies18a} and \citet{Bosman20} used the automated
fitting procedure developed by \citet{Young79} and \citet{Carswell82}
as implemented by \citet{Dallaglio08}, which determines a smooth
continuum in the presence of absorption lines and noise both blueward
and redward of the Ly$\alpha$ emission line.  This algorithm
iteratively fits the spectra with a cubic spline with breakpoints
initially spaced by $\sim 1400~{\rm km\,s^{-1}}$ in the Ly$\alpha$
forest (i.e. 20 SDSS pixels, each pixel is $70{\rm km\,s^{-1}}$) and $\sim 1100~{\rm km\,s^{-1}}$ redward of the
Ly$\alpha$ line (16 pixels).  Pixels that lie more than two standard
deviations below the fit are iteratively rejected.
Additional spline breakpoints
are added if the slope between neighboring breakpoints exceeds a
threshold, and spline points are merged if the variations between
neighboring breakpoints are small.  We applied this algorithm to all
20,201 quasars in our target redshift range, which we will henceforth
refer to as the \emph{autofit continua}. For simplicity, our
DR does not attempt to capture luminosity dependent changes in quasar spectral shape.
As such, DR will work best if all quasars are on a common flux scale, and we thus
rescale each of the autofit continua to be unity at a rest wavelength of 1285\AA\, 
and rescale each of the training data spectra by the same factor.

\citet{Dominika20} followed a different approach to
automatically fit continua to SDSS spectra, which we will refer to as
the \emph{QSmooth continua}.  Briefly, they first compute a running
median with a width of 50 SDSS spectral pixels to capture the salient
continuum and emission features in the spectrum. Peak finding is then
performed on the spectrum with the requirement that the peaks lie
above a local threshold set by this running median spectrum, and these
peaks are spline interpolated to define an upper envelope for the
spectrum. After subtracting this envelope from the spectrum, the
RANSAC regression algorithm is applied to the residuals to define
inliers (the continuum level) and outliers (the absorption lines).
The data points that are identified as inliers are interpolated and
smoothed by computing a running median with a bin size of 20 pixels,
resulting in the final automated continuum fit to the spectrum. We
apply a slightly modified version of the publicly available
QSmooth\footnote{\url{https://github.com/DominikaDu/QSmooth}} code to the
20,201 quasars in our target redshift range. Note
that the QSmooth algorithm was applied to the quasar spectra after
rescaling them such that their autofit continua equal unity at 1285\AA,
so we do not independently renormalize the QSmooth continua.

Once we have these two independent estimates of the quasar continuum,
we use them to attempt to further remove problematic objects from the
training data. Associated absorption around the Ly$\alpha$ and
\ion{N}{V} emission line complex resulting from either BAL absorption,
proximate damped Ly$\alpha$ or Lyman limit systems (PDLAs or PLLSs),
or strong metal absorption due to proximate \ion{N}{V} or intevening
absorbers, will result in artifacts in the automated continuum
fits. We remove these cases by simply discarding objects for which
either the autofit or QSmooth continua fall below 0.6 in the
wavelength range 1170\AA~$< \lambda < 1285$\AA (recall that the
spectra are normalized to unity at 1285\AA). In a similar spirit, we
attempt to also exclude BALs by removing any objects which have either
of their automated continua $< 0.1$ in the wavelength range
1285\AA~$< \lambda < 1990$\AA, or which have automated continua $< 0.7$ in the
wavelength range 1300\AA~$< \lambda <$~1570\AA. These thresholds and
wavelength ranges were all determined via trial and error, and collectively
these cuts remove 4433 objects, leaving 15,768 quasars.

\begin{figure*}
  \vskip -0.4cm
  \includegraphics[trim=0 0 0 0,clip,width=0.97\textwidth]{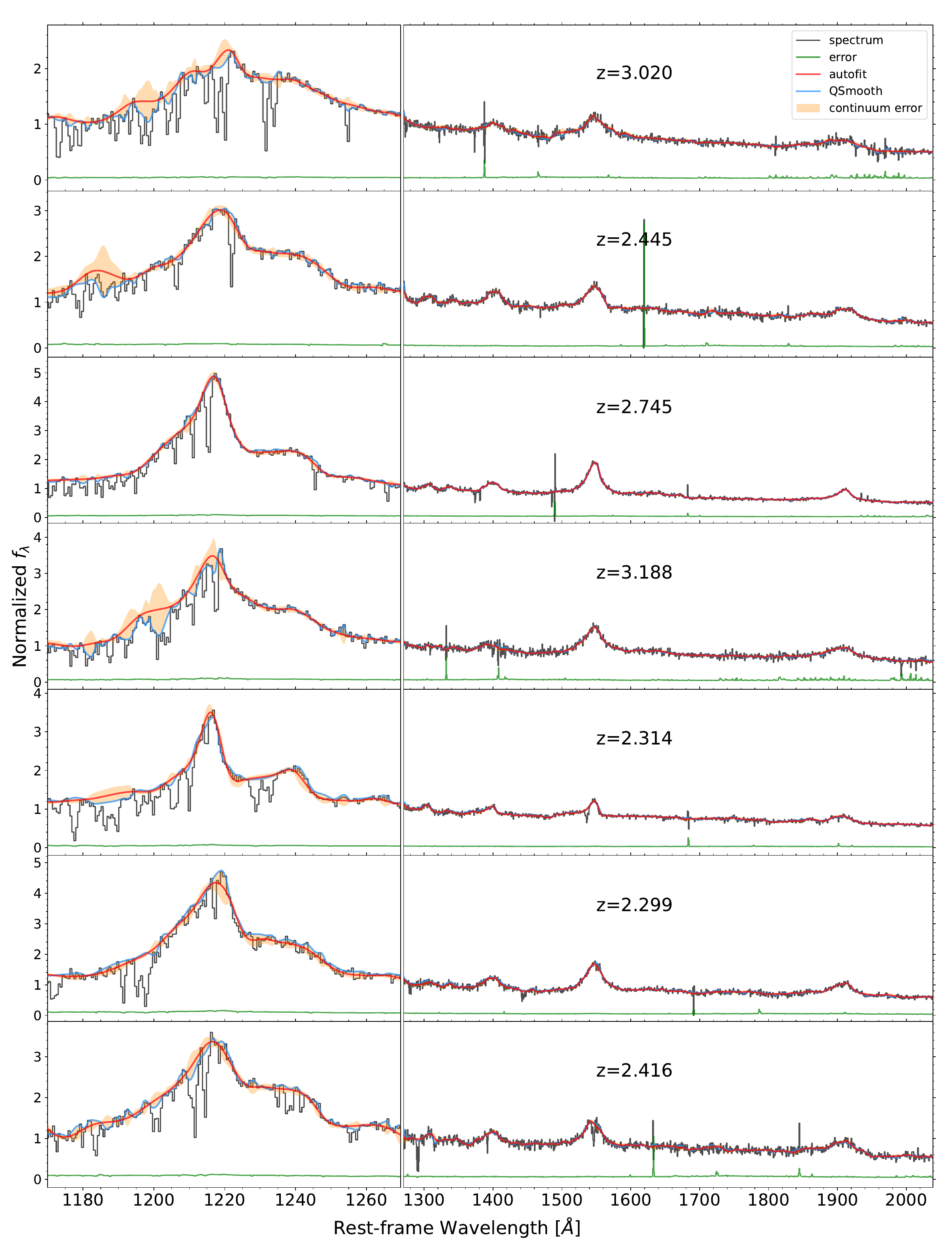}
  \vskip -0.3cm
  \caption{Seven randomly selected quasars from our DR quasar sample. Black and green histograms show the quasar spectrum
    and 1$\sigma$ spectral noise, respectively. Automated continnum fits from the autofit and QSmooth algorithms are shown in blue and red respectively.
    The orange shaded region indicates the error on the continuum fit derived from the differences between the autofit and QSmooth fits, as  defined in eqn.~(\ref{eqn:cont_error}).
    The autofit continua (red) appear to provide good estimates for the continuum and the continuum errors (orange shaded) 
    are generally larger in regions where the continuum is less certain owing to strong 
    absorption. \label{fig:training}}
\end{figure*}

The SDSS spectra are reported on their native observed frame wavelength grid, but the
DR analysis requires a common rest-frame wavelength grid. We convert the observed
wavelengths to rest wavelengths by dividing by $1 + z_{\rm QSO}$, where the quasar redshift, 
$z_{\rm QSO}$ is taken  to be the pipeline redshift, \texttt{Z\_PIPE}, in the DR14 quasar catalog. 
We construct a rest-frame wavelength grid that is linearly spaced in velocity (or equivalently
$\log_{10}{\lambda}$) 
with a pixel size of $\mathrm{d}v_{\rm pix}= 140~{\rm
  km\,s^{-1}}$ (roughly twice the native SDSS pixel scale) spanning
from $1170-2040\,$\AA, resulting in a total of $n_{\rm spec} = 1190$
spectral pixels.
The spectra are rebinned onto this wavelength
grid by averaging the flux values of the spectral pixels that land in
a given wavelength bin. The autofit and QSmooth continua are
interpolated onto this wavelength grid.

A small fraction of objects in
the eBOSS DR14 quasar catalog have incorrect redshifts.  We found that
an effective way to automatically identify and remove these objects is
to analyze their Ly$\alpha$ forest transmission. Specifically, we use
the autofit continua to normalize the rebinned SDSS spectra and
compute the mean Ly$\alpha$ forest transmission of each object in
the wavelength range $1170$\AA~$ < \lambda < 1190$\AA. We compare this
measured mean flux $\langle F \rangle$ to the empirically determined
value $\langle F\rangle_{\rm true}$, where the latter is obtained by
evaluating the fit from \citet{Onorbe17a} at the average Ly$\alpha$
absorption redshifts, $z_{\rm Ly\alpha}$, probed by the quasar
spectrum. For the redshift range we consider the \citet{Onorbe17a} fit
is anchored
by the \citet{Becker13} mean flux measurements. For
each spectrum a flux contrast can then be defined as
$\delta_F =(\langle F\rangle - \langle F\rangle_{\rm true})\slash \langle F\rangle_{\rm true}$.
We then use the median absolute deviation\footnote{Because occasional large outliers, in this
section and in other places in the paper we estimate standard
deviations from the median absolute deviation (MAD) using $\sigma =1.4826\times {\rm MAD}$.} (MAD)
to estimate the effective standard deviation, $\sigma_{\delta_F}$, of $\delta_F$, from all of the
quasars which allows us to define $\chi_{\delta_F} \equiv \delta_F\slash \sigma_{\delta_F}$. Empirically,
we find that large positive fluctuations $\chi_{\delta_F}$ correspond to spurious redshifts,
which are typically lower redshift quasars which  do not exhibit Ly$\alpha$ forest absorption
in their SDSS spectra because another emission line has been incorrectly identified as Ly$\alpha$.
From trial and error we find that applying a cut of $\chi_{\delta_F} > 5$ removes the vast majority
of misidentified redshifts, which
removes 54 objects from our sample. 

Finally, we found that the difference between the autofit continua and
the QSmooth continua provides a reasonable proxy for the continuum
fitting `error'.  If we define ${\Delta s}_{i\lambda} \equiv
s_{{\rm autofit}, i\lambda} - s_{{\rm QSmooth}, i\lambda}$, where $i$ is the index of the quasar in question and
$\lambda$ is the wavelength, then the inverse variance of the continuum fit
can be defined is 
\be \frac{1}{\sigma^2_{s, i\lambda}} \equiv
\frac{1}{{\Delta s}^2_{i\lambda}} + \frac{\left|s_{{\rm autofit},
    i\lambda}\right|^2}{\rm {\rm S\slash N}^2_{\rm clip}}\label{eqn:cont_error},
\ee
where
${\rm S\slash N}_{\rm clip}$ applies a flux dependent floor to the
noise that guarantees that the signal-to-noise ratio ${\rm S\slash N}
\equiv s_{{\rm autofit}, i\lambda}\slash \sigma_{s, i\lambda}$ never
exceeds ${\rm S\slash N}_{\rm clip}$, which we set to be ${\rm S\slash
  N}_{\rm clip} = 30$.
We also define the relative fluctuation $\delta s_{i\lambda} \equiv \Delta s_{i\lambda}\slash s_{{\rm autofit}, i\lambda}$,
and analogous to above, quantify fluctuations relative to the typical scatter via
\be
\chi_{s, i\lambda} \equiv \frac{\delta s_{i\lambda} - \langle \delta s_{\lambda} \rangle}{\sigma_{\delta s, \lambda}}, 
\ee
where the $\langle \delta s_\lambda \rangle$ and $\sigma_{\delta s, \lambda}$ are the wavelength dependent mean and standard deviation
of $\delta s_{i\lambda}$, i.e.  these moments are computed by averaging out the quasar dimension $i$.  
The rms fluctuation of $\chi_{s, i\lambda}$, across  1170\AA~$< \lambda < $~1240\AA,
provides a summary statistic quantifying the relative differences between the two autofit continua for each quasar in the range relevant to IGM damping wings. We require that this
rms fluctuation is less than four, which removes another 155 quasars, or about $1\%$, yielding the final size of
our DR quasar sample of 15,559.

We opt to use the autofit continua since they visually appear to be closer to the truth than the QSmooth continua.
The data thus used to train the DR algorithms are the autofit continua, $s_{\rm autofit, i\lambda}$, and the error $\sigma_{s,{i\lambda}}$.
Fig.~\ref{fig:training} shows seven objects randomly chosen from the DR quasar sample.
Following standard practice in machine learning, we split this DR sample of 15,559 quasars into a
training set of 14,781 objects and a test set of 778 objects, corresponding to a roughly $95\%-5\%$ split.
The training set will be used to train our dimensionality reduction algorithms and the test set, which we refer to as
our autofit test set, will constitute the unseen data that will be used to quantify performance.

\subsubsection{Hand-Fit Continua}
\label{sec:handfit}

The previous section discussed the automated continuum fitting, but it
is also possible (although tedious) to fit quasar continua by
hand. Whether this approach results in continua that are more accurate
than the aforementioned automated algorithms is unclear. But hand fit
continua provide an additional `test' dataset that can be used to
quantify the efficacy of DR algorithms which motivates us to compile a
set for this purpose. \citet{Paris11} hand fit a set of 78 high
signal-to-noise ratio, ${\rm S\slash N} > 14$, quasar spectra in the
redshift range $2.82 < z < 3.0$, where the ${\rm S\slash N}$ was
evaluated at rest-frame 1280\AA. The continua were fit by selecting
cubic spline breakpoints at regular intervals and adjusting the height
and spacing of these breakpoints.  In the Ly$\alpha$ forest, where
this procedure is particularly challenging and subjective, the
breakpoint heights were chosen to follow the peaks of the flux.  We
renormalize the 78 \citet{Paris11} continua to be unity at 1285\AA\,
(for consistency with our autofit continua).  
We augment the
\citet{Paris11} continua, with continua from the XQ-100 survey
\citep{XQ100}, which acquired a set of 100 high signal-to-noise-ratio
(${\rm S\slash N} \sim 33$ at rest-frame
1700\AA) echelle spectra of quasars
at $3.5 \lesssim z \lesssim 4.5$ using the X-shooter instrument on the
Very Large Telescope (VLT).  The public data release of XQ-100
provides hand fit continua for every quasar following an algorithm
that is similar in spirit to that used by \citet{Paris11}. As the
public release of the XQ-100 dataset provided continuum fits only for
the spectra obtained by the individual arms of the X-shooter
spectrograph, we restrict attention to the VIS arm, which covers the
observed frame 5500\,\AA~$< \lambda <$~10200\,\AA.  As our mock high-$z$
quasar spectra with IGM damping wings (see \S~\ref{sec:sims}) cover the rest-frame
wavelength range 1185\AA\,$< \lambda <$~2000\,\AA, the requirement that
this entire range is covered by the VIS arm restricts the number of
quasars to be 43, and the redshift range to be $3.65 \le z \le
4.09$. We similarly renormalize the XQ-100 continua to be unity at
1285\AA. For the quasar redshifts, $z_{\rm QSO}$, we used the values adopted by
\citet{Paris11} and the XQ100 survey. 
Combining the \citet{Paris11}
and XQ-100 spectra we finally arrive at a sample of 121 hand-fit continua, which
we henceforth refer to as our hand-fit test dataset.

\subsection{Dimensionality Reduction Algorithms}
\label{sec:DRmethods}

A DR algorithm transforms data from a high-dimensional space into 
a lower dimensional \emph{latent} space, such that the lower dimensional
representation retains as much information as possible from the higher dimensional process.
In more concrete terms, our training set comprises $n_{\rm train}=14,781$ quasars with
$n_{\lambda} = 1190$ spectral pixels per quasar. The smooth appearance of a quasar continuum (see Fig.~\ref{fig:training}) 
implies these  $n_{\lambda}$ spectral pixels are highly correlated, and hence the bulk of the information
content can be encapsulted by a vector $\boeta$ in a lower dimensional latent space of dimension, $n_{\rm DR}$, 
which parameterizes a DR model $\bsDR(\boeta)$.
Recall from the discussion in \S~\ref{sec:traintest} that we normalized all of our training and test set spectra to unity
at 1285~\AA\, which removes the amplitude degree of freedom from the stochastic process $\bs$. But since
our goal is to eventually fit $\bsDR(\boeta)$ to quasar spectra with arbitrary flux, we add an
additional multiplicative normalization parameter, $s_{\rm norm}$. This implies our DR quasar continuum model 
will have $n_{\rm DR} = n_{\rm latent} + 1$ free parameters, i.e. the normalization and the $n_{\rm latent}$ latent
variables that parameterize a latent variable model for the normalized spectra. We define 
$\bxi$ to be a vector whose elements are the parameters in the $n_{\rm latent}$ dimensional latent space
that describes the normalized spectra, whereas $\boeta \equiv (s_{\rm norm}, \xi_1, \xi_2, ...., \xi_{n_{\rm latent}})$,
is the vector in the $n_{\rm DR} = n_{\rm latent} + 1$ dimensional latent space that describes the non-normalized spectra. 

DR methods are commonly divided into linear and non-linear algorithms, and we consider
both in this study. We start with the most widely adopted linear model, which is PCA.

\subsubsection{Principal Component Analysis}

PCA is a commonly used tool to understand correlations in quasar spectra
\citep[e.g.][]{Boroson92,Francis92,Yip04,Suzuki06}.  \citet{Suzuki05}
first proposed PCA as a method to predict the continuum
absorbed regions of quasar spectra, which was later improved upon by
\citet{Paris11}. This approach, which uses pixels redward of
Ly$\alpha$ to predict the continuum in the Ly$\alpha$ forest, has been
used in many IGM absorption studies
\citep[e.g.][]{Kirkman05,Lee12,Lee13,Lee15,Eilers17a,Eilers17b,Eilers18,Dominika20, Bosman21,Dominika24}. 
Motivated by this technique,
\citet{Davies18a} developed a PCA-based model from a
far larger sample of spectra encompassing a wide range of spectral
properties with an eye toward predicting the quasar
continua for IGM damping wing analysis.
A similar tack was followed by \citet{Bosman21} who modeled a broader spectral region relevant
to the Ly$\alpha$ and Ly$\beta$ forests. While the training data and PCA
models used here are qualitatively very similar to those used by
\citet{Davies18a} and \citet{Bosman21}, the primary difference is that
these works
applied PCA to the blue and red spectral regions independently,
using the latter to predict the former,  whereas here we apply PCA
and other DR algorithms to \emph{the entire spectrum}, since our
inference approach models the IGM and the continuum simultaneously using all spectral
pixels (see \S~\ref{sec:formalism}).

The principal components of a collection of $n_{\rm train}$ quasars that reside in a $n_\lambda$ dimensional spectrum space can be
thought of as a set of unit vectors (with $n_{\lambda}$ components) where the $i$-th vector defines the direction of a line in the  $n_\lambda$
dimensional space that best fits the variations among the spectra while being orthogonal to the first $i-1$ unit vectors. The components are rank ordered
according to the amount of variation in the dataset that they describe, and typically one truncates the latent space at some value $n_{\rm latent} < n_\lambda$,
such that the dimensionality of the process is reduced.

A quasar PCA decomposition can be written as
\be
\bsDR(\boeta) = s_{\rm norm}\left(\langle \bs\rangle + \bxi^T \bA\right)\label{eqn:PCA}, 
\ee
where $\bxi$ is the $n_{\rm latent}$ dimensional latent variable (commonly referred to as the PCA coefficients) and
$\bA$ is the set of $n_{\rm latent}$ PCA vectors or principal components, each of which has $n_\lambda$ spectral pixels, i.e.  $A_{i \lambda}$ is the $n_\lambda$ dimensional PCA vector corresponding to the $i$-th component.

\citet{Davies18a} argued that the dominant mode of variations between quasar spectra are their power-law continua, which are
more naturally described by  PCA decomposition in \emph{log space}, motivating 
\be
\ln \bsDR(\boeta) = \ln s_{\rm norm} + \langle \ln \bs\rangle + \bxi^T \bA, 
\ee
or equivalently
\be
\bsDR(\boeta) = s_{\rm norm}\left[\exp\left(\langle \ln \bs\rangle\right)\exp\left(\bxi^T \bA\right)\right],  
\ee
where $\boeta \equiv (s_{\rm norm}, \xi_1, \xi_2, ...., \xi_{n_{\rm latent}})$. 
We will refer to these two decomposition choices as PCA and lnPCA, respectively.

Given that we have an estimate of the errors for our continuum fits
(eqn.~(\ref{eqn:cont_error})), we also explore whether there is an
advantage to using a weighted PCA\footnote{We use the WPCA implementation available
here: 
\url{https://github.com/jakevdp/wpca}.}. Whereas in standard PCA one
performs a singular value decomposition of the sample covariance,
weighted PCA instead decomposes a weighted covariance matrix to
compute the set of principal component vectors $\bA$ \citep{WPCA}. For our
weighted PCA, we set the weights for each spectral pixel to be
the inverse variance defined in eqn.~(\ref{eqn:cont_error}).

To summarize, there are four different variations of PCA that we
explore: standard (PCA), log PCA (lnPCA), weighted PCA (wPCA), and
weighted log PCA (wlnPCA).

\subsubsection{Gaussian Process Latent Variable Model}

Gaussian processes (GPs) are a supervised learning method for solving regression and classification problems in a powerful
non-parametric probabilistic framework. A Gaussian process latent variable model \citep[GPLVM;][]{GPLVM} uses Gaussian processes for
unsupervised learning tasks like DR or searching for hidden structure in high-dimensional data. In the current context,
the GPLVM will act as a \emph{decoder}, providing a probabilistic mapping from the latent space variables $\bxi$ to the data space
variables $\bs$. The smoothness of this mapping is controlled by kernel functions whose hyperparameters are fit during the training process. This is
analogous to GP regression, where given inputs $X$  and outputs $y$, one chooses a kernel and learns hyperparameters that best describe
the mapping from $X$ to $y$. In the GPLVM, one is not given the latent variables $\bxi$ (i.e. $X$), but is rather only given $\bs$ (i.e. $y$). The
latent variables $\bxi$ representing each example in the training set must be learned along with the kernel hyperparameters. It is in this
sense that the GPLVM acts a decoder, or equivalently a forward mapping from latent space to data space. Once the latent variables that encode each training
instance and the kernel hyperparameters are learned, one can evaluate the model, $\bs_{\rm GPLVM}(\bxi)$, at any location in the latent space. This forward mapping
is governed by GPs which are independently defined for each dimension of the data space, which is to say that each spectral pixel $s_\lambda$ has its own
GP (and associated hyperparameters) which regresses $\bxi$ to produce $s_\lambda$. See \citet{EilersGPLVM22} for a recent application
of GPLVMs for DR of quasar spectra. 

In the canonical formulation of GPLVM \citep{GPLVM}, the covariance
hyperparameters and point estimates for the unknown latent variables
are determined jointly, by optimizing the likelihood of the data. But
it is well known that the optimization required for GP regression
scales as $\mathcal O(n^3)$ for a dataset of size $n$, such that
applying canonical GPLVM to `big data' is computationally
prohibitive. \citet{Hensman13} showed how stochastic variational
inference \citep[SVI;][]{SVI} techniques can be used to apply GPs to
very large datasets by stochastically optimizing over mini-batches of the
dataset. Building upon this, \citet{gplvm_gpytorch} re-formulated the
Bayesian incarnation of the GPLVM \citep{Titsias10} in an SVI
framework by using a structured doubly stochastic lower bound
\citep{Titsias14} which enables training on very large datasets. 

We apply the \citet{gplvm_gpytorch}
implementation\footnote{\url{https://docs.gpytorch.ai/en/latest/examples/045_GPLVM/Gaussian_Process_Latent_Variable_Models_with_Stochastic_Variational_Inference.html}}
of Bayesian GPLVM, built in the
\texttt{gpytorch}\footnote{\url{https://gpytorch.ai}} GP framework
\citep{gpytorch}, to our quasar training set. The loss function
optimized in this formalism exploits the errors on the quasar continua
that were defined in eqn.~(\ref{eqn:cont_error}).  The training set is
passed through a scaler transformation to `whiten' the data. Namely,
the wavelength dependent mean, $\langle s_\lambda \rangle$, and
standard deviation, $\sigma_{s, \lambda}$, of the training set are
computed by averaging over the sample of quasars. We rescale the
spectra $s_{\lambda}$ using the transformation $y_\lambda = (s_\lambda
- \langle s_\lambda \rangle)\slash \sigma_{\rm median}$, and
correspondingly rescale their errors $\sigma_{s, \lambda}$ . Here
$\sigma_{\rm median}$ is a single number which is the median value of
$\sigma_{s, \lambda}$.\footnote{A more common whitening procedure
would be to adopt $y_\lambda = (s_\lambda - \langle s_\lambda
\rangle)\slash\sigma_{s, \lambda}$, where $\sigma_{s, \lambda}$ is the
standard deviation of each feature (wavelength). However, this
transformation changes the shape of the spectra that are fit by the
GPLVM by effectively downweighting larger emission line fluctuations
and upweighting smaller continuum variations. We found that scaling by
a single wavelength independent number, $\sigma_{\rm median}$, yielded
better results.}

\subsubsection{Variational Auto-Encoder}

An autoencoder is an unsupervised learning method that uses an
artificial neural network to learn a latent space representation, or
encoding, of a dataset.  The essential property of autoencoders are
their architecture, which consists of an encoder that maps the input
into the code, and a decoder that maps the code to a reconstruction of
the input. The code or equivalently the latent variables, $\bxi$,
correspond to the output of a specific layer of a multilayer
perceptron (MLP). Typically, the encoder passes the inputs through a
sequence of gradually smaller hidden layers of an MLP until hitting a
bottleneck which has $n_{\rm latent}$ neurons. The decoder then passes
the code back up through gradually larger hidden layers to generate
the reconstruction of the input.  When the latent space has dimensionality
smaller than the data space, autoencoders can be used for DR.
A variational autoencoder \citep[VAE;][]{VAE}
is similar in spirit to an autoencoder, but it essentially replaces the deterministic
decoder, which generates the data from latent variables $\bs(\bxi)$, with a probablistic
decoder $p(\bs  | \bxi)$, and likewise for the encoder  $p(\bxi | \bs)$. One application of the VAE is to address
a limitation of autoencoders, which is that their resulting latent space could have a very complex structure,
i.e., small changes in latent variables could produce large variations in the data, 
and/or the probability distribution
function (PDF) of the latent variables for an ensemble of data 
could be highly multimodal. By virtue of their design, VAEs tend to produce
better behaved latent space PDFs. Specifically, the $\beta$-VAE is an implemenation which
explicitly aims to disentangle the latent space manifold via a tunable parameter $\beta$ \citep{bVAE}, which sets the relative weighting of two competing loss functions. The first
loss term is a standard reconstruction loss (mean squared error; MSE) typically adopted in autoencoders, whereas the second is the Kullback-Leibler (KL) divergence between the conditional encoder
distribution $p(\bxi | \bs)$ and an isotropic diagonal Gaussian distribution with unit variances. 

We implemented a $\beta$-VAE in the \texttt{pytorch} machine
learning framework. The same scaler transformation was applied
to the training data as was used for the GPLVM, and the 
MSE loss was generalized to include our
estimates for the continuum fitting `error' (see
eqn.~(\ref{eqn:cont_error})). 

The design of the autoencoder architecture
is as follows.  The
encoder constitutes an MLP
with two hidden layers. The first layer is
non-linear, and maps the input spectral pixels to 1024 neurons passing
through a SELU activation. Then two distinct linear layers convert
these 1024 outputs into the mean, $\bmu$, and log variance,
$\ln{\bsigma^2}$ (both have size $n_{\rm latent}$) of the normal
distribution which forms the probalistic description of the latent
variables underlying the VAE.  During training or evaluation, a sample
from this normal distribution produces a latent space vector that is
then passed through the decoder. Our decoder constitutes an MLP with
two linear layers, one that maps the latent space vector to 1024
outputs, and another that maps these 1024 outputs to the $n_\lambda$
data vector.

Following standard practice, we optimized to determine the MLP weights
via  stochastic gradient descent  using the Adam optimizer with
weight decay. We adopted a 90-10\% split of the 14,781 training spectra, using 90\%
for training and 10\% for validation, where the validation set was used
to evaluate performance on data unseen during training and mitigate
against overfitting. The loss was computed from a mini-batch size of $N_{\rm batch} = 128$ training set quasars per optimizer step, or
epoch. To prevent overfitting, the validation set loss was also computed each epoch, and the
best model was chosen to be the one that achieved the smallest value of the validation set loss.
A learning rate of $10^{-4}$ was used and we employed
early stopping, which is to say that we stopped training if
the validation loss did not improve after 100 epochs.
Hyperparameters were tuned  via trial
and error and with  a more rigorous grid scan strategy. 
Surprisingly, we found that a very small value of $\beta =
10^{-9}$ produced the best results, which reduces the influence of the
KL loss term. Nevertheless, visual inspection of the latent space
indicated that it was
not significantly multimodal.

\subsection{Comparison of DR Algorithms}
\label{sec:DRcompare}

In this subsection we compare and contrast the performance of the DR
algorithms discussed in the previous section. Since our ultimate goal
is to fit models to quasar spectra, the quantitative metric that we
adopt for evaluating DR algorithm performance will be an estimate of
the variance of the relative reconstrution error $\bdelta$ defined in
eqn.~(\ref{eqn:delta}).  How does one determine the $\bsDR(\boeta)$
representation of the spectrum $\bs$ which appears there?
Whereas for PCA and autoencoders, the inverse
function $\bxi(\bs)$ is tractable and easy to evaluate, the same is
not true for GPLVMs, for which determining the encoding of a new test
data point requires additional assumptions \citep{GPLVM}. Since our
specific application of DR is the construction of a parametric model
to fit quasar spectra, the most natural definition of $\boeta$ is to
fit the function $\bsDR(\boeta)$ to the spectrum $\bs$. That is, for
each spectrum in our test set, we determine the value $\boeta_{\rm true}$ that
minimizes the mean square error
\be
   {\rm MSE} = \sum_\lambda \left(s_{\lambda}-s_{{\rm
         DR},\lambda}(\boeta)\right)^2,  \label{eqn:MSE}
\ee
where the `true' subscript denotes that this represents the best DR representation
of the quasar continuum for the hypothetical case where we directly fit a noiseless
spectrum with no IGM absorption.  Note that the sum in eqn.~(\ref{eqn:MSE}) weights each spectral pixel equally in the mean-square error 
loss computation.  Indeed, the \emph{formally correct} choice for the relative weighting of the terms in the equation
above is rather subtle\footnote{Our formalism in
\S~\ref{sec:formalism} suggests that one should take the covariant
relative reconstruction errors into account when fitting for $\boeta$
(see eqn.~(\ref{eqn:gaussS})), which would amount to maximizing a
likelihood implied by eqn.~(\ref{eqn:gaussS}) which would differ
markedly from the uniform weighting in the MSE in
eqn.~(\ref{eqn:MSE}).  There is thus a chicken-and-egg problem in
that, formally the likelihood one should maximize to fit for $\boeta$
depends on the mean, $\langle \bdelta\rangle$, and covariance,
$\bDelta$, of the relative reconstruction error (eqn.~(\ref{eqn:gaussS})), which can only be determined by analyzing the
distribution of residuals $\bdelta$ of an ensemble of such fits. For
this reason we adopt the simple approach of fitting for $\boeta$ with uniform unit
`errors' as in eqn.~(\ref{eqn:MSE}) and define $\bdelta$ to be the relative reconstruction error
resulting from those fits.} and here we adopt a uniform weighting for
simplicity.

\begin{figure*}
  \vskip -0.4cm
  \includegraphics[trim=0 0 0 0,clip,width=0.99\textwidth]{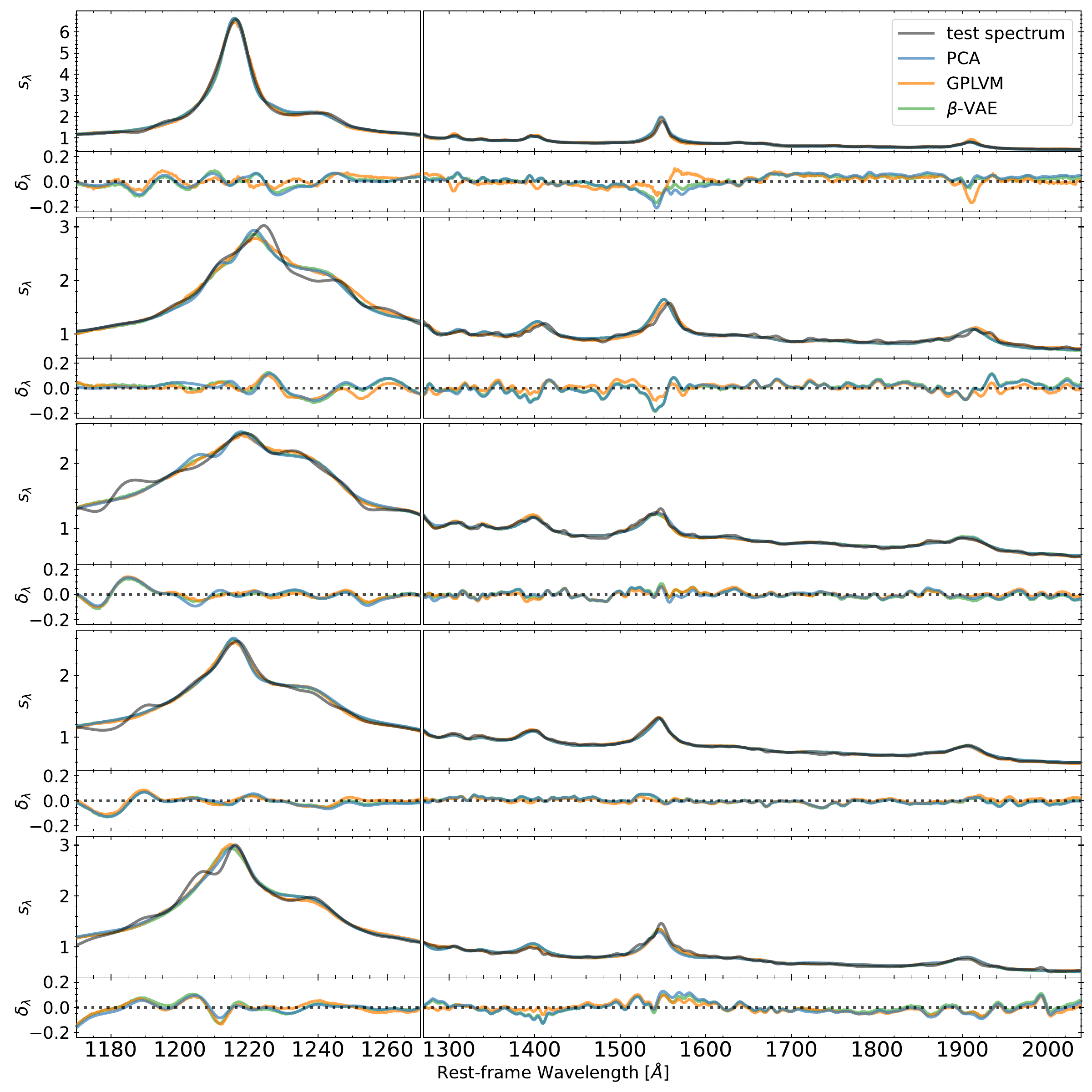}
  \vskip -0.3cm
  \caption{DR reconstructions of five randomly selected quasars from our the SDSS test set for $n_{\rm latent} = 5$. Upper panels show the quasar continuum
    $s_{\lambda}$ where the black curves are the autofit continuua and the blue, yellow, and green curves are the DR reconstructions for the
    PCA, GPLVM, and $\beta$-VAE, respectively. The lower panels show the relative reconstruction errors, $\delta_\lambda$, defined in eqn.~(\ref{eqn:delta}),
    where the horizontal dotted line indicates the zero level. \label{fig:dr_examples}}
\end{figure*}

As previously discussed, the hyperparameter governing the dimensionality of the latent space, $n_{\rm latent}$, sets
the number of free parameters $n_{\rm DR} = n_{\rm latent}+1$, and generally the variance of $\bdelta$ (eqn.~(\ref{eqn:delta})) will decrease with
increasing $n_{\rm latent}$ because the DR model is more flexible. We trained each of the aforementioned DR algorithms on our training
set data for $n_{\rm latent}$ from 1 to 25 in
unit steps from 1-15 and then in steps of two from 15-25. After training, we fit the models to both the 778 spectra in our
autofit test set (see \S~\ref{sec:autofit}) and the 121 hand-fit continua in our hand-fit test set (see \S~\ref{sec:handfit}).
Fig.~\ref{fig:dr_examples} shows examples of our fits from the PCA, GPLVM, and $\beta$-VAE algorithms for $n_{\rm latent} = 5$
for five randomly selected quasars from the autofit test set, where the lower set of panels shows
$\bdelta$.

\begin{figure*}
  \hskip -0.09cm
  \includegraphics[trim=0 0 0 0,clip,width=0.50\textwidth]{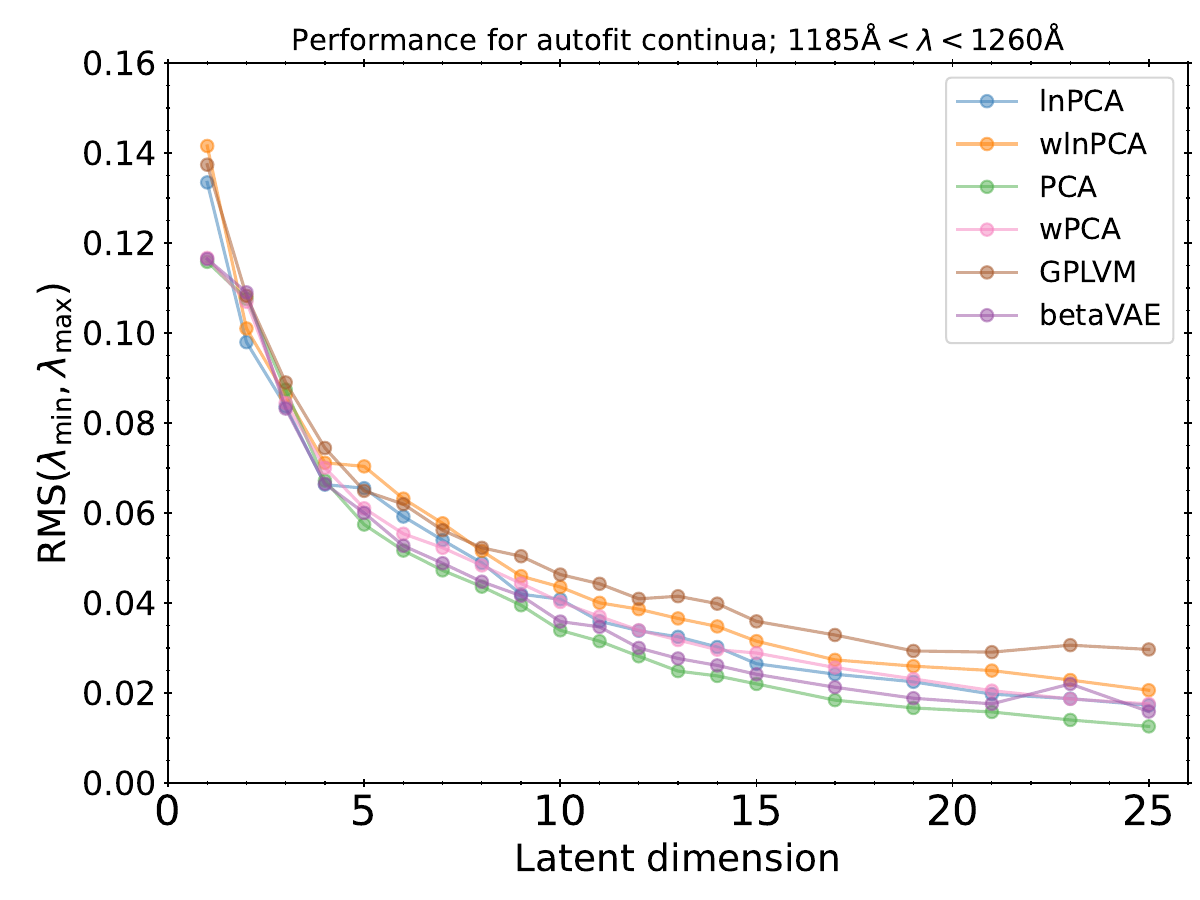}
  \includegraphics[trim=0 0 0 0,clip,width=0.50\textwidth]{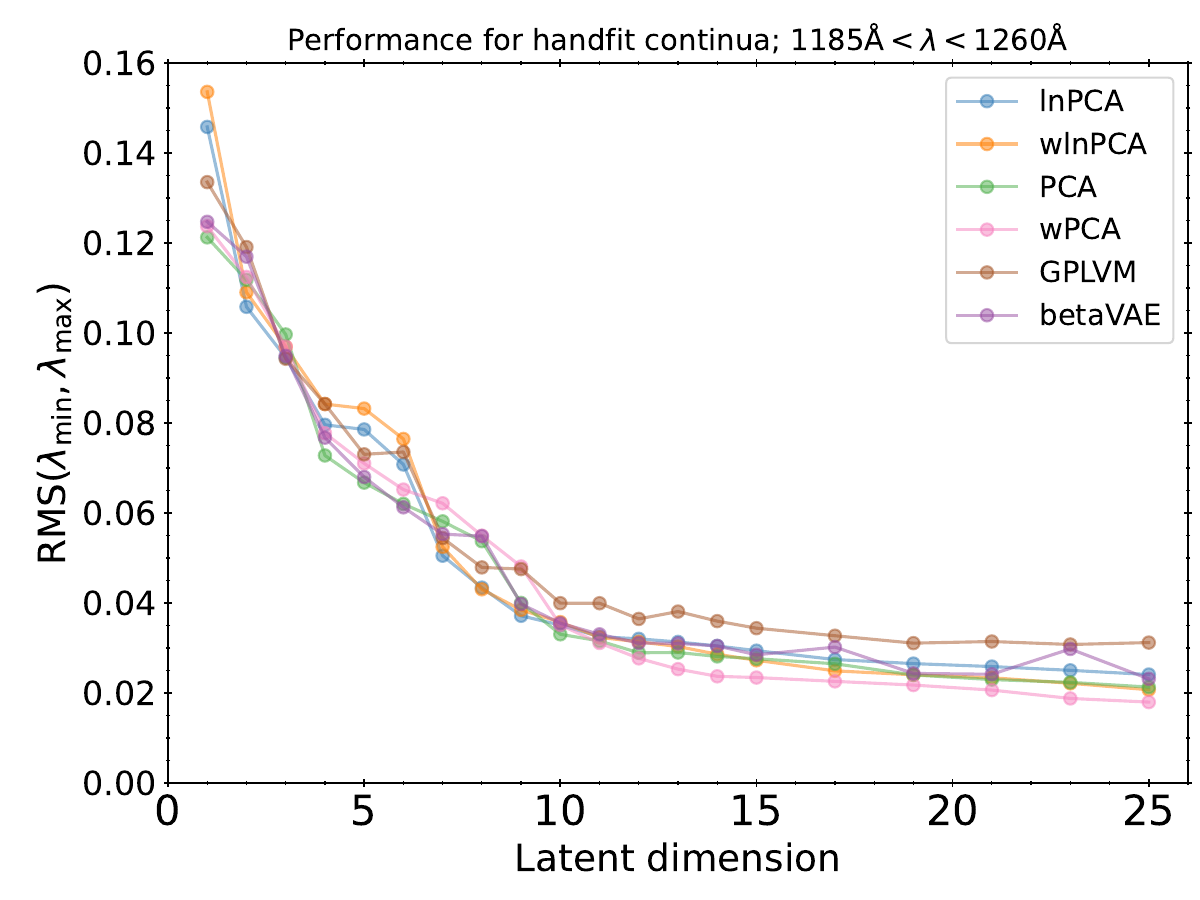}
  \vskip -0.1cm
  \caption{Comparison of DR algorithm performance for `blue' pixels in the wavelength range $1185\,$\AA~$ < \lambda <
    1260\,$\AA\, for the six DR algorithms considered. The curves show the summary
    statistic ${\rm RMS}(\lambda_{\rm min},\lambda_{\rm max})$ (see eqn.~(\ref{eqn:rms})), for the
    relative reconstruction error, $\delta_\lambda$ (see
    eqn.~\ref{eqn:delta} or the lower panels of the examples in
    Fig.~\ref{fig:dr_examples}) as a function of latent
    dimension. Left panels show results when the DR algorithms
    were fit to the 778 autofit continua in the SDSS test dataset,
    whereas the right panel shows the same for the 121 hand fit
    continua.  Comparison of the left and right panels indicates that
    the behavior of the DR algorithms does not depend on the test set
    used. Furthermore, the fact that the level of relative
    reconstruction error as a function of latent dimension is very
    nearly the same for autofit and hand fit continua, suggests that
    the automated algorithms do not add a significant amount of extra
    `noise' to the continua\label{fig:dr_performance_blue}.}
\end{figure*}

We define a simple summary statistic for the purposes of comparing the
different DR algorithms. The covariance, $\bDelta$, of the relative reconstruction error  (see eqn.~(\ref{eqn:gaussS}))
is defined by
\be
\Delta_{\lambda \lambda^\prime} = \langle (\delta_\lambda - \langle \delta\rangle_{\lambda})(\delta_{\lambda^\prime} - \langle \delta\rangle_{\lambda^\prime})\rangle\label{eqn:covardelta},
\ee
where the angle brackets denote an average over the members of the test set. 
The variance of $\bdelta$  as a function of wavelength
is given by the diagonal elements of the covariance matrix $\Delta_{\lambda\lambda}$. 
As our metric for comparing DR aglorithms, we adopt the root-mean-square variation per spectral pixel
with the mean computed over the $N_\lambda$ spectral pixels in the wavelength interval  $[\lambda_{\rm
  min}, \lambda_{\rm max}]$
\be
{\rm RMS}(\lambda_{\rm min},\lambda_{\rm max}) \equiv \left(\frac{1}{N_\lambda} \sum_{\lambda_{\rm min}}^{\lambda_{\rm max}}  \Delta_{\lambda\lambda}\right)^{1\slash2}\label{eqn:rms}. 
\ee
We consider a blue region
$[\lambda_{\rm min}, \lambda_{\rm max}]=$[1185\,\AA, 1260\,\AA] covering the continuum relevant to the proximity zone and IGM damping wing, and a red
region $[\lambda_{\rm min}, \lambda_{\rm max}]=$[1260\,\AA, 2000\,\AA] constituting the rest of the
spectrum.

\begin{figure*}
  \hskip -0.09cm
  \includegraphics[trim=0 0 0 0,clip,width=0.50\textwidth]{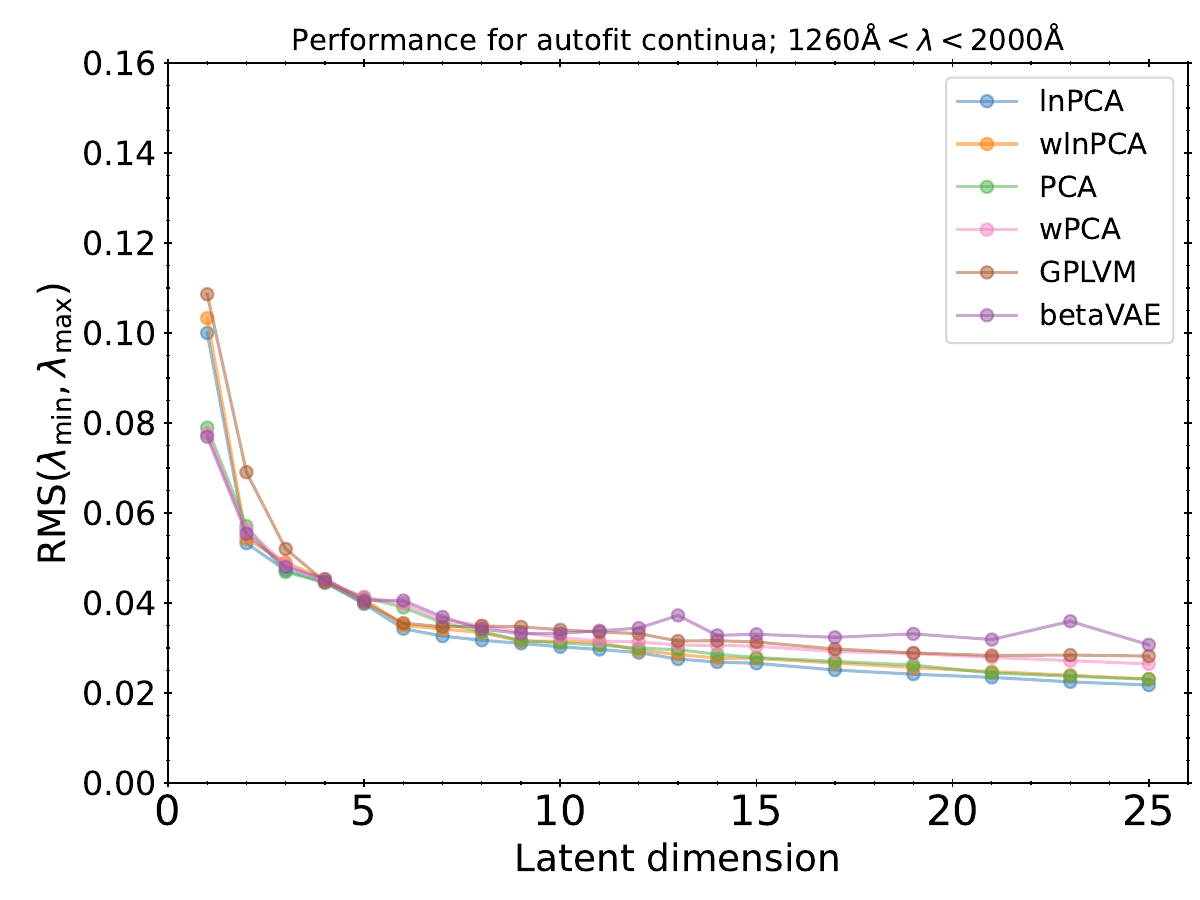}
  \includegraphics[trim=0 0 0 0,clip,width=0.50\textwidth]{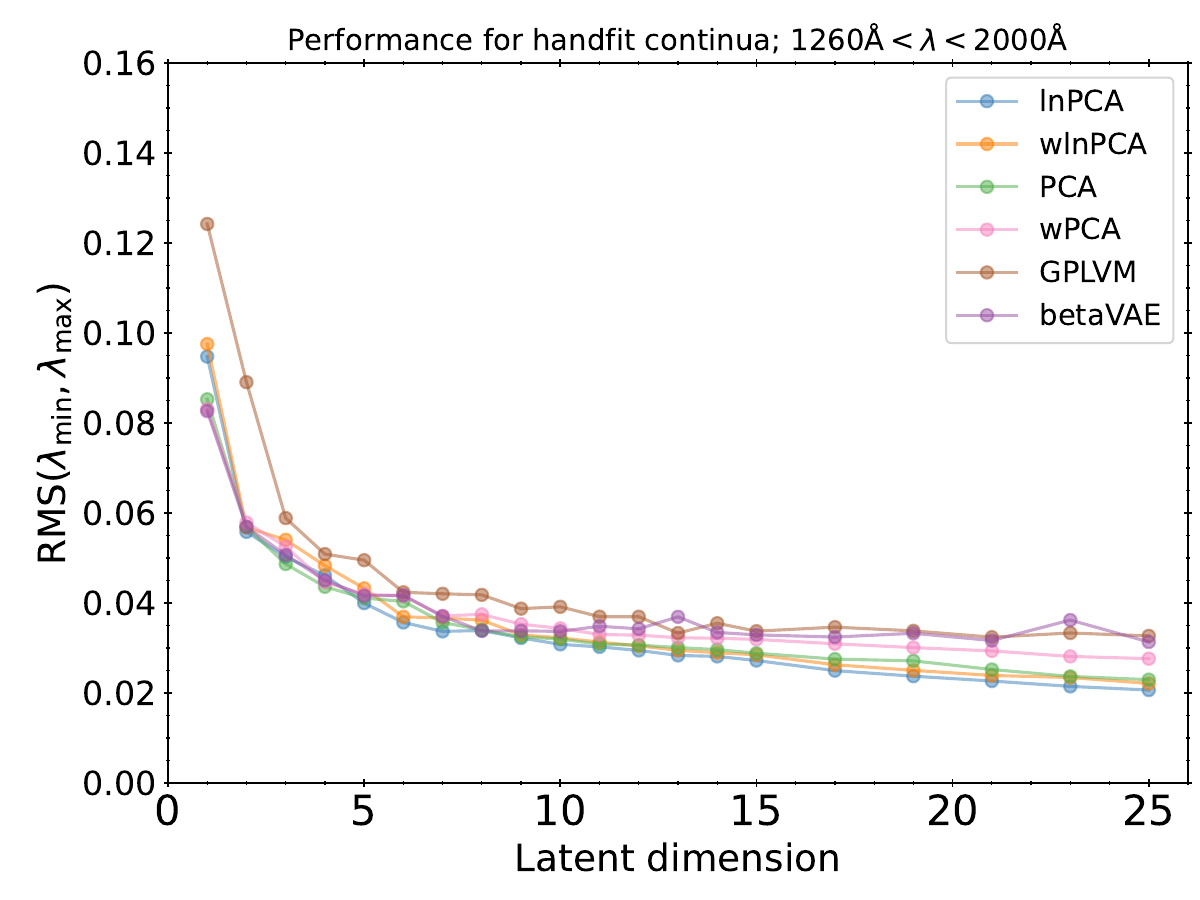}
  \vskip -0.1cm
  \caption{Comparison of DR algorithm performance for `red' pixels in the rest-frame wavelength range $1260~$\AA $< \lambda < 2000$~\AA\,
  for the six DR algorithms considered. Curves and panels are the same as Fig.~\ref{fig:dr_performance_blue}. The difference in amplitude as compared to Fig.~\ref{fig:dr_performance_blue} likely occurs because the ratio of pixels
    in emission lines to pixels in the continuum is smaller in the $1260$~\AA $< \lambda < 2000$\,\AA\, wavelength range, as compared to $1185$~\AA $< \lambda < 1260$\,\AA\, where
    the Ly$\alpha\,\lambda 1216$\AA, \ion{N}{V}\,$\lambda 1240$\AA, and \ion{Si}{II}\,$\lambda 1260$\AA\, emission lines constitute a large fraction of the pixels and inflate the variation\label{fig:dr_performance_red}.}
\end{figure*}

Fig.~\ref{fig:dr_performance_blue} shows DR algorithm performance as
a function of $n_{\rm latent}$ for the autofit test set (left) and the
hand-fit test set (right) for the blue wavelength range, and
Fig.~\ref{fig:dr_performance_red} is the analogous plot for the red
wavelengths.  These plots generalize the canonical explained variance
vs. number of components plots that one constructs in applications of PCA that are used to 
determine which dimensionality to compress down to. All of the
DR algorithms show the expected trend of decreasing
${\rm RMS}(\lambda_{\rm min},\lambda_{\rm max})$
with increasing $n_{\rm latent}$
-- smaller reconstruction errors will always result from a more
flexible model. The most striking conclusion that one draws from
Figs.~\ref{fig:dr_performance_blue} and \ref{fig:dr_performance_red}
is that simple linear DR, i.e. standard
PCA or lnPCA, performs as well or better
than the more sophisticated ML based non-linear DR techniques.
Furthermore, this conclusion holds for both the autofit (left panels)
and handfit (right panels) test sets, and the comparable values of
${\rm RMS}(\lambda_{\rm min},\lambda_{\rm max})$ achieved indicates that
both continuum test data sets have comparable levels of noise. 
Finally, there appears to be no obvious advantage to the variations of PCA that
we discussed, standard PCA performs just as well as lnPCA
or weighted PCA. These results
motivate us to adopt PCA as our DR algorithm, and to use the larger
autofit test set continua for the construction of mock data in the rest of
this work.

How should we choose the value of the $n_{\rm latent}$ hyperparameter?
As is often the case with DR algorithms, the rate of decrease of the
reconstruction error for a unit increment of $n_{\rm latent}$ varies
with $n_{\rm latent}$. All curves decrease steeply around $n_{\rm
  latent}$ of a few, and all flatten out at the largest values of
$n_{\rm latent}$.  Such behavior is expected on physical grounds --
the steep decrease occurs as one approaches the number of parameters
required to capture the salient features of quasar spectra, whereas
the flattening occurs when this number is significantly exceeded. At
the largest values of $n_{\rm latent}$, incremental decreases in
reconstruction error likely result from fitting noise in the autofit
or handfit continua, or `intrinsic noise' in the quasar continuum
stochastic process, marginally better with a more flexible
model. Often in the DR literature, one sets the $n_{\rm latent}$
hyperparameter to the location of a `knee' in such plots where the
slope begins to flatten.  Whereas for red wavelengths
Fig.~\ref{fig:dr_performance_red} shows the reconstruction error
starts to saturate around 2.5-3\% around $n_{\rm latent} \simeq 6$,
for blue wavelengths the curves in
Fig.~\ref{fig:dr_performance_blue} exhibit a more gradual trend with
$n_{\rm latent}$ with 2.5-3\% errors reached around $n_{\rm latent}
\simeq 10$. It is possible that the red wavelengths compress harder
because the approximately power-law continuum of quasar spectra
contributes relatively more pixels to this wavelength range, whereas
the blue wavelengths are dominated by strong emission lines complexes
from Ly$\alpha$ 1216\,\AA, \ion{N}{V} 1240\,\AA, and \ion{S}{II} 1260\,\AA\, that exhibit more
intrinsic variation in shape. Another possibility is that IGM
absorption blueward of Ly$\alpha$ results in more noise in the
continuum estimates, making it harder to describe and compress the
blue wavelengths. Our companion paper quantifies the variation in the precision with which
the two IGM damping wing parameters (volume averaged IGM neutral fraction $\langle x_{\rm HI}\rangle$ and the quasar
lifetime $t_{\rm Q}$) can be recovered as a function of $n_{\rm latent}$ \citep{Kist25a}. There, adopting the same autofit continuum test dataset used here for the mock spectra (see \S~\ref{sec:mocks}), it is found that parameter precision does not increase significantly
beyond $n_{\rm latent}> 5$, which motivates us to use $n_{\rm latent} = 5$ as our fiducial value for
the rest of this paper, corresponding to $n_{\rm DR} = 6$ after adding the additional normalization parameter, 
$s_{\rm norm}$.

\subsection{Rebinning onto Coarser Wavelength Grids}
\label{sec:wavegrid}

As discussed in \S~\ref{sec:autofit} our DR test and training set
spectra cover from $1170-2040\,$\AA\, with $\mathrm{d}v= 140~{\rm km\,s^{-1}}$
pixels resulting in $n_\lambda = 1190$ total spectral pixels. However,
these continua are smooth and can be interpolated onto a finer grid to
construct mock observations of quasars with a finer spectral
sampling. Whereas a spectrum of a high-$z$ quasar 
might have
fixed spectral resolution (given by the instrument FWHM in 
${\rm km}\,s^{-1}$) and
spectral sampling (the number of pixels per FWHM), for the purposes of
IGM damping wing analysis, there are several reasons why rebinning the
spectrum onto a coarser wavelength grid is advantageous.
Note that here rebinning refers not to interpolation, but rather to
averaging the flux values
of finer grid pixels that land within a given low resolution pixel.
First, at a
typical resolution of FWHM=$100\,\kms$ a uniform velocity grid would
result in $\gtrsim 10^3$ pixels, making the matrix compuations in the
likelihood in eqn.~(\ref{eqn:lhood_final}) costly to evaluate during
HMC sampling. A fine velocity grid is only required blueward of
rest-frame Ly$\alpha$ where there are narrow Ly$\alpha$ forest
absorption lines, whereas redward of the Ly$\alpha$ line the smooth
damping wing signature and the broad emission lines imply that a
coarse velocity grid will suffice. Second, as we discuss further in
\S~\ref{sec:inference}, adopting a Gaussian form
(eqn.~(\ref{eqn:gaussT})) for the transmission PDF, $P(\bt |
\btheta)$, in quasar proximity zones is an approximation which
compromises the fidelity of the inference with the likelihood in
eqn.~(\ref{eqn:lhood_final}).  However, rebinning a high-resolution
spectrum onto a coarser wavelength grid, ameilorates the problems with
the inference, because it follows from the central limit theorem that 
the averaging involved in rebinning
Gaussianizes the non-Gaussian Ly$\alpha$ forest transmission
stochastic process, making the Gaussian approximation more valid.

Motivated by these considerations, one can consider `hybrid' wavelength grids which  have 
a distinct velocity grid with size $\mathrm{d}v_{\rm
  blue}$, for blue wavelengths $\lambda <
\lambda_{\rm blue-red}$ where it may be necessary to resolve narrow absorption features in the proximity zone, 
concatenated with a coarse uniform grid with pixel size $\mathrm{d}v_{\rm red}$ for red wavelengths $\lambda > \lambda_{\rm
  blue-red}$ for which the damping wing absorption and intrinsic quasar spectrum are smooth.  Whereas such 
  hybrid grids are considered in our companion paper \citep{Kist25a} to assess the impact of spectral resolution 
  on the resulting measurement precision, in this paper 
  we simply adopt $\mathrm{d}v = \mathrm{d}v_{\rm
  blue} =  \mathrm{d}v_{\rm red}=500\,{\rm km\,s^{-1}}$ throughout.

\subsection{Properties of the PCA Decomposition}
\label{sec:PCA}

\begin{figure*}
  \vskip -0.4cm
  \includegraphics[trim=0 0 0 0,clip,width=0.98\textwidth]{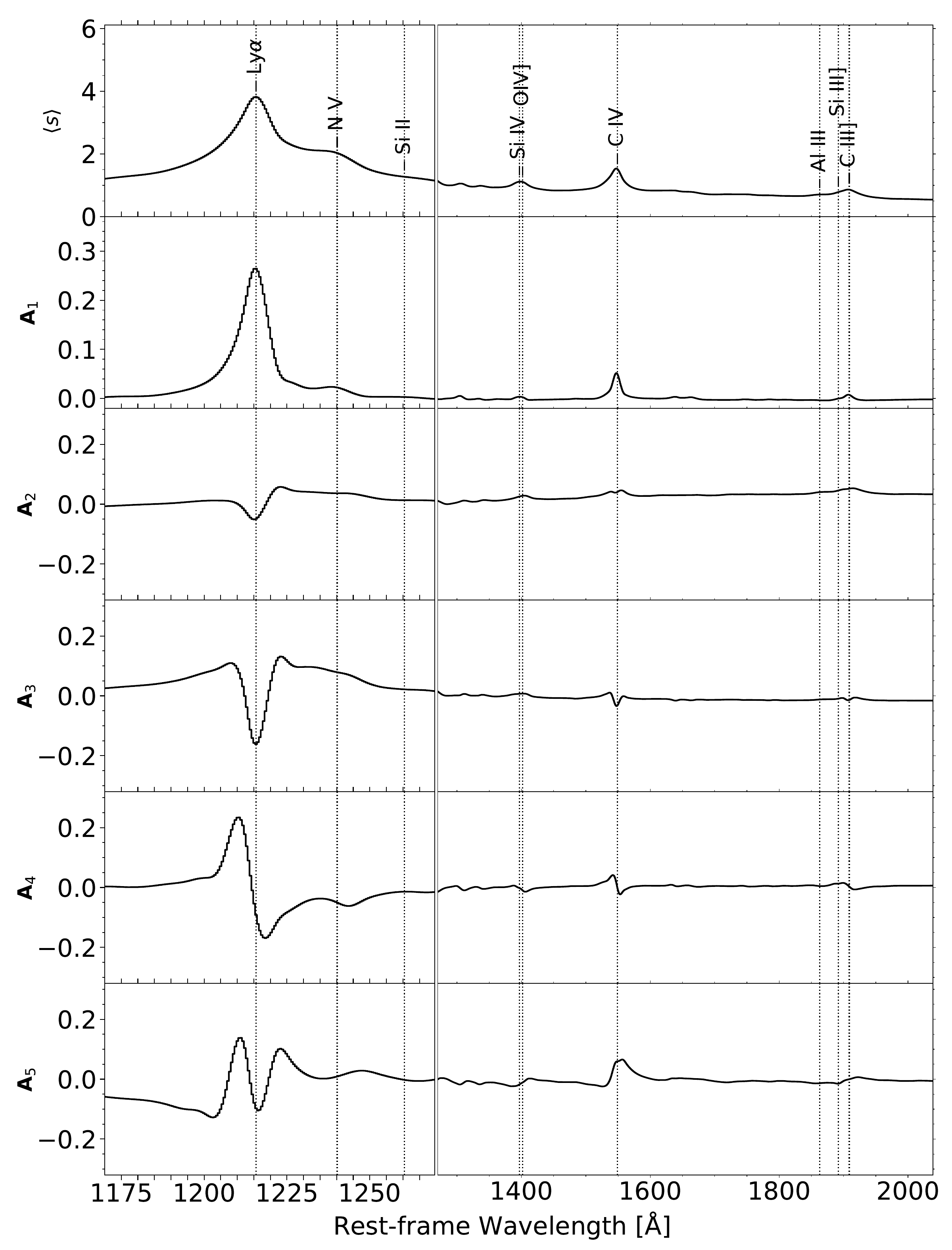}
  \vskip -0.3cm
  \caption{PCA vectors from the PCA decomposition of  14,781 SDSS quasar spectra for $n_{\rm latent} = 5$. The upper panel labeled $\langle s\rangle$ reprensets the mean spectrum, while the lower panels show the principal components ordered in terms of the amount of explained variance from top to bottom. Strong broad emission lines in the quasar spectra are indicated by the vertical dashed lines and labled in the upper panel. \label{fig:pca_vectors}}
\end{figure*}

Having arrived at PCA with $n_{\rm latent}=5$ as our preferred DR
algorithm we now briefly describe and quantify the properties
of the decomposition. Fig.~\ref{fig:pca_vectors} shows the mean spectrum $\langle
s_\lambda \rangle$ and the $n_{\rm latent}=5$ PCA basis
spectra. Recall that our approach differs from previous work using PCA
to predict quasar continua
\citep[e.g.][]{Suzuki05,Paris11,Eilers17a,Davies18a,Dominika20,Bosman21,Dominika24} in that
we do not construct separate red-side and blue-side PCA
decompositions, but rather a single PCA deomposition of the entire
spectral range (the blue-red split in Fig.~\ref{fig:pca_vectors} is
only to better visualize the blue side).  PCA basis vectors are sorted
in terms of the amount of variance they explain, and we see that the
first component $\bA_1$ describes correlated variation in the
strengths of various broad emission lines, with the subsequent
components encoding more subtle correlated variations in shape.

\begin{figure*}
  \vskip -0.4cm
  \includegraphics[trim=0 0 0 0,clip,width=0.99\textwidth]{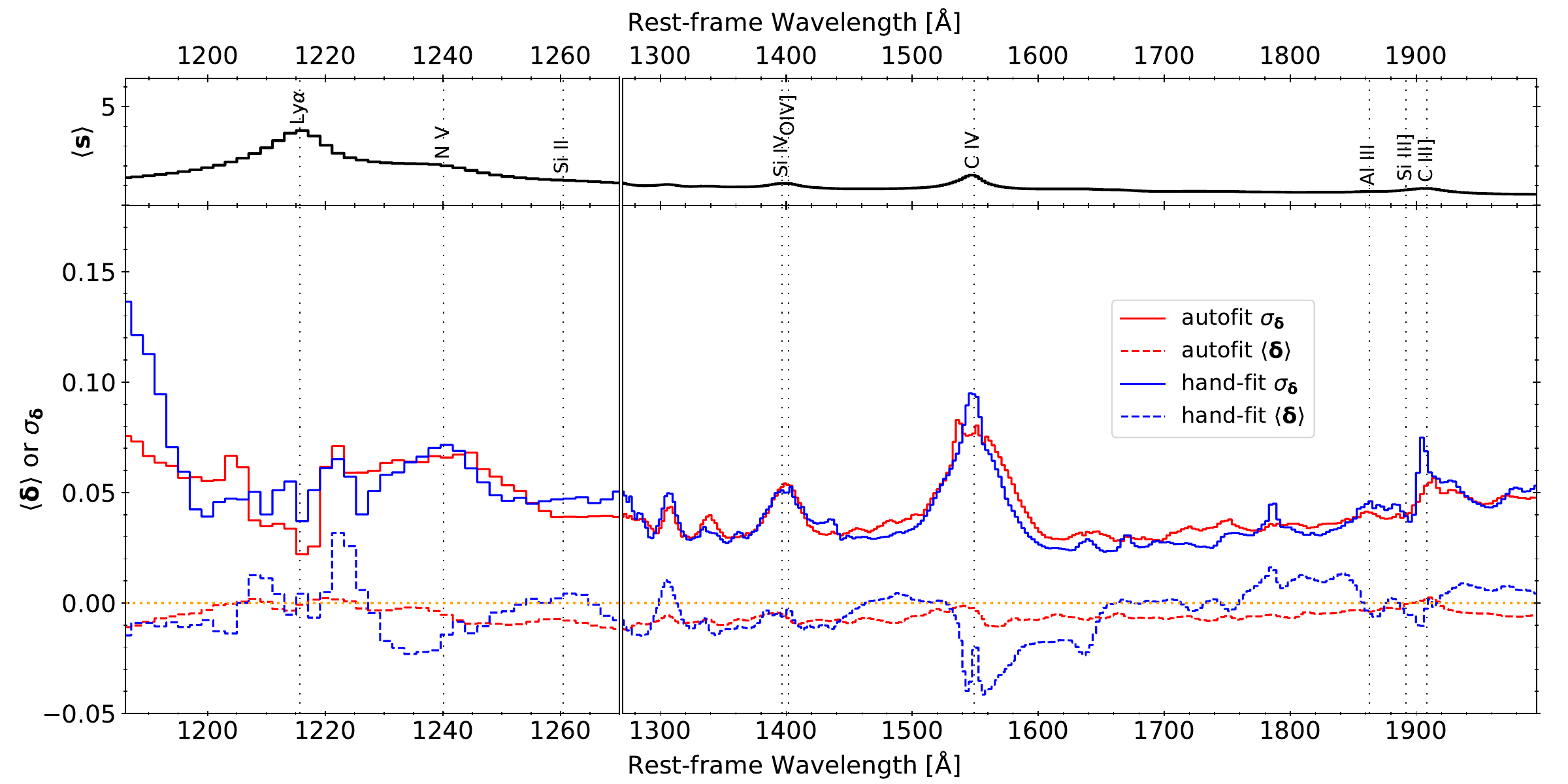}
  \vskip -0.3cm
  \caption{Moments of the relative reconstruction error as a function of wavelength for the PCA model with $n_{\rm latent}=5$. The lower panels show the
    mean $\langle \bdelta\rangle$ (dashed) and standard deviation $\sigma_{\bdelta} \equiv
\Delta_{\lambda\lambda}^{1\slash 2}$  (solid) of the 
    relative reconstruction error evaluated from the  778 spectra in the autofit (red) and 121 spectra in the hand-fit (blue) test sets, respectively.  
    For reference, the upper panel shows the mean quasar spectrum $\langle \bs\rangle$ constructed from the autofit test set with
    prominent emission lines labeled.\label{fig:DR_resid}}. 
\end{figure*}

We now compute the mean, $\langle \bdelta\rangle$, and covariance,
$\bDelta$, of the relative reconstruction error (see
eqn.~(\ref{eqn:gaussS})) required to evaluate the likelihood in
eqn.~(\ref{eqn:lhood_final}). In general, these must be computed from
the test set data on the wavelength grid adopted (i.e. see the rebinning discussion in \S~\ref{sec:wavegrid}) for the IGM damping wing fits.  Here we show results for the wavelength grid used for the mock quasar spectra generated in this paper (see \S~\ref{sec:mocks}), which extends from $1185-2000\,$\AA\, with a uniform pixel size of $\mathrm{d}v=500~{\rm km\,s^{-1}}$, resulting in $n_\lambda=313$ pixels. 
To obtain the $\bdelta$
field we interpolate each test set spectrum $\bs$ onto this wavelength grid
using a cubic spline and fit for $\bsDR(\boeta)$. 
As we discussed in the previous section, we fit for $\boeta_{\rm true}$ by
minimizing the mean-square error loss function in 
eqn.~(\ref{eqn:MSE}). Adopting the uniform weighting of the spectral pixels
that we discussed there\footnote{Note we use a uniform grid with $dv=500~{\rm km\,s^{-1}}$ in this paper 
so each pixel receives the same weight. However, on a hybrid wavelength grid 
with different blue and red pixel scales, a naive uniform weighting implies
the more numerous finer blue side pixels would receive a higher relative weight.
To equally treat all spectral regions, one would then need to instead upweight the red pixels by the
factor  $\mathrm{d}v_{\rm red}\slash \mathrm{d}v_{\rm blue}$.},   
we fit the PCA with $n_{\rm DR} = 6$ (i.e. $n_{\rm latent}=5$ plus the normalization parameter $s_{\rm norm}$) 
to each member of the autofit test set and the hand-fit test set.

\begin{figure}[b!]
  \vskip -0.4cm
  \includegraphics[trim=0 0 0 0,clip,width=0.50\textwidth]{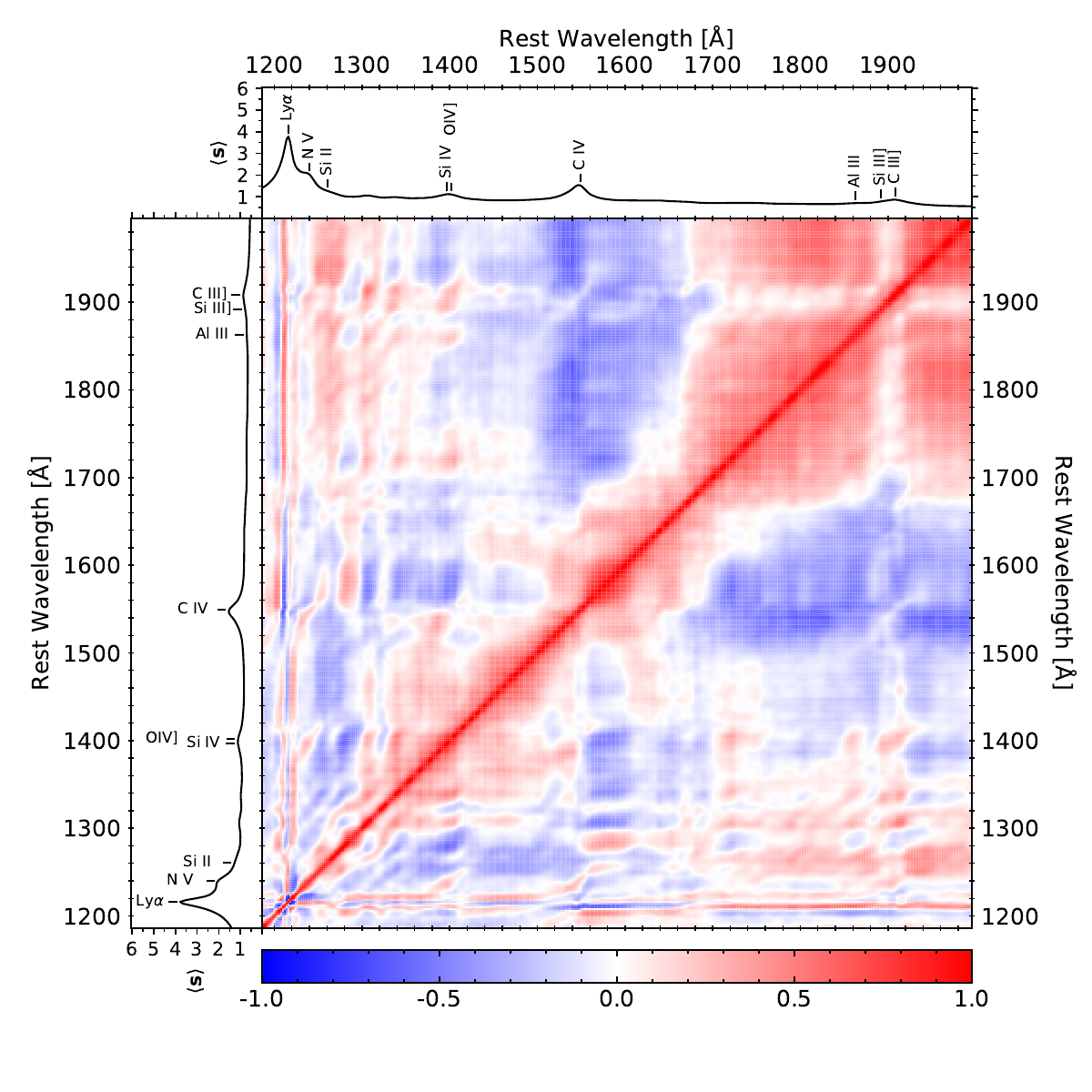}
  \vskip -0.3cm
  \caption{Correlation matrix of the relative reconstruction errors for the PCA model with $n_{\rm latent}=5$
    evaluated from  778 autofit test set spectra. 
    The mean quasar spectrum $\langle \bs\rangle$ constructed from the autofit test set is shown
    for reference and prominent emission lines are labeled.\label{fig:resid_corr}}
\end{figure}

Fig.~\ref{fig:DR_resid} shows the mean $\langle \bdelta\rangle$
(dashed curves) and standard deviation $\sigma_{\bdelta} \equiv
\Delta_{\lambda\lambda}^{1\slash 2}$ (solid curves; see eqn.~(\ref{eqn:covardelta})) of the relative
reconstruction error evaluated from the autofit (red) and hand-fit
(blue) test sets. For the autofit test set we see that the fits are
unbiased -- the mean is consistent with zero to within $\sim 0.1\%$.
The factor of $\sim 2-4$ larger variations in the mean for the hand-fit
test set results from the factor of $\sim 6$ times fewer quasars and
as a result possibly increased sensitivity to outliers. The results
for $\sigma_{\bdelta}$ are consistent between the two sets and
indicate typical relative reconstruction errors of $\sim 4-5\%$, which are relatively independent of wavelength, 
although larger errors occur in the vicinity of the strong emission lines, 
particularly the \ion{C}{IV} $\lambda 1549$ line. 

Finally, we can
visualize the full covariant structure of $\bdelta$ by computing the
correlation matrix
\be
   {\rm Corr}(\bDelta)_{\lambda \lambda^\prime} \equiv
\frac{\bDelta_{\lambda \lambda^\prime}}{\sqrt{\bDelta_{\lambda
      \lambda}\bDelta_{\lambda^\prime\lambda^\prime}}}\label{eqn:corr},
\ee
which is shown in Fig.~\ref{fig:resid_corr} computed from the autofit test set.  
One clearly sees that the fit residuals are
highly correlated for neighboring pixels as is also apparent from the
lower panels showing the relative reconstruction error, $\bdelta$, in Fig.~\ref{fig:dr_examples}.  However, whereas
correlations are typically positive in continuum regions, in the
vicinity of the stronger emission lines like Ly$\alpha$ and
\ion{C}{IV}, one sees oscillation between correlation and
anti-correlation at smaller wavelength separations. Such behavior is
expected since in general there is more small-scale wavelength
structure in the emission lines than in the smoother continuum.  As
for the correlations of residuals between different emission lines,
the structure appears rather complex. For example, residuals around
the peak of \ion{C}{IV} are anti-correlated with the
\ion{Al}{III}-\ion{Si}{III}]-\ion{C}{III}] complex at $\sim$1900\,\AA, whereas
for pairs of pixels from the \ion{C}{IV} and Ly$\alpha$-\ion{N}{V}
emission line complex, one observes both correlated
and anti-correlated residuals.

\begin{figure*}
  \vskip -0.4cm
  \includegraphics[trim=0 0 0 0,clip,width=0.99\textwidth]{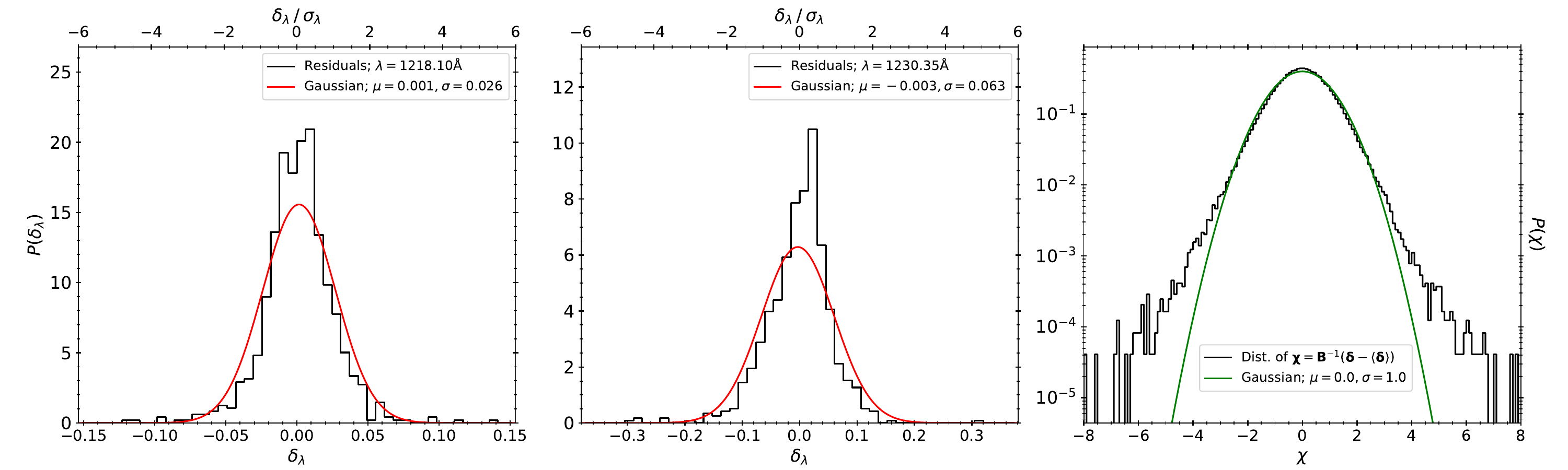}
  \vskip -0.3cm
  \caption{Distribution of relative reconstruction errors determined
    from fits to 778 autofit test set spectra. {\it Left and Center:}
    Histograms show the distribution of $\delta_\lambda$ for two
    different wavelengths relevant to studies
    of IGM damping wings, $\lambda = 1217.90\,$\AA\, (left) and $\lambda
    = 1230.35\,$\AA\, (center). Red curves show a normal distribution
    $\mathcal{N}(\delta_\lambda | \mu, \sigma)$ with mean and standard deviation set
    to that measured from the distribution. The histograms appear consistent with
    Gaussianity but with stronger tails. {\it Right:} The histogram shows the distribution of
    transformed residuals $\bchi$ (see eqn.~(\ref{eqn:chi})) for the
    full set of spectral pixels and quasars. The green curve shows a
    zero-centered unit variance normal distribution $\mathcal{N}(\chi | 0,
    1)$ for comparison. If the relative reconstruction errors
    $\bdelta$ follow a multivariate Gaussian distribution
    (eqn.~(\ref{eqn:pdelta})), then the distribution of $\bchi$ should
    follow the green curve. Similar to the left and center panels,
    the histogram appears consistent with Gaussianity but with stronger tails.
    \label{fig:resid_hist}}
\end{figure*}

Fig.~\ref{fig:resid_hist} allows us to visualize the reliability of
our approximation that the distribution of relative reconstruction
errors is a multivariate Gaussian (see eqn.~(\ref{eqn:pdelta})). The
left and center panels show the distribution of $\delta_\lambda$ for
two different wavelengths relevant to studies of IGM damping wings,
$\lambda = 1217.90\,$\AA\, (left) and $\lambda = 1230.35\,$\AA\, (center)
for the 778 spectra in our autofit test set. The right panel shows the
distribution of the quantity 
\begin{equation}
\bchi = \mathbf{B}^{-1}(\bdelta -
\langle \bdelta\rangle)\label{eqn:chi}, 
\end{equation}
where $\mathbf{B} = \mathbf{V} \cdot \diag{\sqrt{\blambda}}$, is the
product of the matrix of eigenvectors, $\mathbf{V}$, of the covariance matrix
$\bDelta$ and the diagonal matrix, $\diag{\sqrt{\blambda}}$, formed from the square root of the
vector of eigenvalues $\blambda$ of $\bDelta$. For a stochastic variable
described by a general multivariate Gaussian distribution
(i.e. eqn.~(\ref{eqn:pdelta})), the transformed variable
$\bchi$ will be a draw from $\mathcal{N}(\bchi;
\mathbf{0}, \mathbf{I})$ where $\mathbf{0}$ is a vector of zeros and
$\mathbf{I}$ is the identity matrix.   The histogram shows the distribution of 
$\mathbf{B}$ for  all 313 spectral pixels for the 778 autofit test set spectra. 
Inspection of the distribution
$\bchi$ marginalized over the spectral pixel dimension,
generalizes the intuitive method of using $(x-\mu)\slash \sigma$ to
assess the Gaussianity of a stochastic variable $x$, to the case of
a multivariate Gaussian distribution with a non-diagonal covariance matrix.  The
panels in Fig.~\ref{fig:resid_hist} all paint a similar picture
for the distribution of relative reconstruction errors. Namely, the distribution
generally follows a Gaussian shape, but has stronger tails. These strong tails
result in larger variance estimates than implied by the values in the `core'
of the distribution.

\section{Simulating Damping Wing Observations}
\label{sec:sims}

This section describes our procedure for generating mock $z > 7$ quasar spectra with IGM damping wing absorption. First, we introduce the IGM damping wing simulations used to generate transmission skewers, then describe how these skewers are combined with our autofit continuum test set and a noise model to create realistic mock spectra. While we adopt a specific physical model to generate the reionization topology and use a hydrodynamical simulation to describe the IGM opacity, we note that our approach does not explore uncertainties in the underlying reionization model or the impact of galaxy formation physics on the IGM opacity.  We will discuss these limitations further when we summarize our conclusions in \S~\ref{sec:summary}.

\subsection{Damping Wing Simulations}
\label{sec:DWsims}

We generate IGM transmission skewers following the procedure developed in
\citet{Davies18b} to simulate spectra of quasar proximity
zones with IGM damping wing absorption and quasar lifetime
effects. Here we briefly summarize the most important elements and
refer the reader to \citet{Davies18b} for additional details. Owing
to the complexity of the problem, this is a hydrid model that
combines three main components: 1) a high-resolution density field from a large volume cosmological hydrodynamical
simulation, 2) a reionization topology generated from a semi-numerical
model of reionization, and 3) one-dimensional ionizing radiative
transfer to treat the impact of the quasar's ionizing radiation on its
surrounding IGM.

Density, velocity, and temperature skewers are taken from the $z=7.0$
output of a Nyx hydrodynamical simulation \citep{Almgren13,Lukic15} in
a box with side equal to $100~h^{-1}{\rm cMpc}$ run with a $4096^3$
baryon grid and $4096^3$ dark matter particles. The skewers were
extracted along an axis of the simulation domain and were chosen to
originate on the 200 most massive dark matter halos, $M_{\rm halo}
\gtrsim 2\times 10^{11} M_\odot$, identified with a custom halo
finder adapted specifically to grid hydro codes \citep[see][for
  additional details]{Sorini18}.

To compute a realistic topology of reionization around massive
quasar-hosting halos we use a modified version of the
\texttt{21cmFast}\footnote{https://github.com/andreimesinger/21cmFAST}
code \citep{Mesinger11}, with an improved treatment of the ionizing
photon mean free path \citep{Davies22}. As the Nyx simulation volume
is too small to characterize the distribution of ionized/neutral
regions around the rare massive halos hosting quasars, we compute the
ionization field in a larger volume, 400 cMpc on a side. The
semi-numerical model starts with cosmological initial conditions using
the Zel'dovic approximation \citep{Zeldovich70} generated on a
$2048^3$ grid, and then produces evolved density and ionization fields
output at a lower resolution of $512^3$. We keep all parameters that
govern the reionization topology fixed except the ionizing efficiency,
$\zeta$, which sets the total number of ionizing photons emitted per
collapsed baryon.  Increasing $\zeta$ decreases the fraction of the
volume that is neutral and vice-versa. By tuning $\zeta$, we generate
ionization fields with global volume-averaged neutral fractions
$\langle x_{\rm HI}\rangle = 0.05-0.95$, in steps of $\Delta \langle x_{\rm
  HI}\rangle=0.05$. The \texttt{21cmFast} code constructs dark matter halos
from the initial conditions following the approach of
\citet{Mesinger07}.  Starting at the location of the 500 most massive
halos $M_{\rm halo} = 3\times 10^{11}$ in the 400 cMpc domain, we randomly
sampled them to extract 1200 randomly oriented skewers of the neutral fraction, $x_{\rm HI}$, for
each value of the parameter $\langle x_{\rm HI}\rangle$. 
We trivially add a completely ionized model, $\langle x_{\rm HI}\rangle=0.0$, and a
completely neutral model, $\langle x_{\rm HI}\rangle=1.0$, by setting $x_{\rm
  HI}$ to a constant everywhere to arrive at 21 total reionization
models spanning the range $\langle x_{\rm HI}\rangle=0.0-1.0$.

To model the impact of the quasar radiation on the IGM, we perform
ionizing radiative transfer (RT) using an udpated version of the
one-dimensional RT code described in
\citet{Davies16a}, which computes the time-dependent evolution of the
ionized fractions of hydrogen and helium, as well as the gas
temperature.  This time evolution is governed by the quasar's
radiative history, and we assume a so-called `light bulb' lightcurve
parameterized by the quasar lifetime, $t_{\rm Q}$, whereby the quasar
turned on at some point $t_{\rm Q}$ in the past, and has been shining
at constant luminosity ever since.  The ionizing photon output from
the quasar was computed using the \citet{Lusso15} spectral energy
distribution normalized to agree with the photometric measurements of
the quasar. In this paper we use models constructed for the quasar
ULAS J1342$+$0928 at $z=7.54$, which has an  apparent
$J$-band AB magnitude of $J_{\rm AB}=20.3$ \citep{Banados18}.

To generate IGM transmission skewers originating from quasars, we use
the density and temperature fields from the Nyx simulations and the
$x_{\rm HI}$ skewers from the semi-numerical ionization topology as
initial conditions for the radiative transfer code.  For ionized
regions which have values of $x_{\rm HI} = 0$ in the semi-numerical
ionization skewers, we initialize to $x_{\rm HI}\sim 10^{-3}$ by
adopting a fixed amplitude of the ultraviolet background (UVB)
photoionization rate, $\Gamma_{\rm HI}$.  The IGM temperature was
initialized to the values from the Nyx simulation in ionized regions,
whereas we assume a temperature of $T = 2000\,{\rm K}$ inside neutral
regions. For each value of the global IGM neutral fraction $\langle x_{\rm HI}\rangle$ parameterizing the reionization topology, we ran RT calculations using 1200 distinct random
Nyx skewers (six per halo for 200 Nyx halos), and associated them with each of the 1200 semi-numerical $x_{\rm HI}$
skewers.  The RT was computed on a uniform logarithmic
grid of quasar lifetimes spanning from $\logtq = 3-8$ in steps
of $\Delta \logtq = 0.1$ dex \citep[this grid of RT computations were generated in][]{DaviesBH19} and IGM transmission
skewers were computed over the velocity range $-10,000\,{\rm
  km\,s^{-1}} \le \Delta v \le 10,000\,{\rm km\,s^{-1}}$ on a uniform
velocity grid with 5000 pixels of size $\mathrm{d}v = 4\,{\rm km\,s^{-1}}$.
This was
done by performing the optical depth velocity integral using the
neutral hydrogen density and temperature outputs from the RT, the peculiar velocity field from the Nyx simulation, and
weighting by the Voigt line profile that describes the frequency
dependence of the Ly$\alpha$ absorption cross section. Because the
Nyx simulations do not have star or galaxy formation sub-grid recipes,
hydro grid elements in collapsed structures can evolve to have unrealistically
high densities, which would not be present in a simulation with galaxy formation
prescriptions. These high overdensities are problematic for several reasons. First, the
RT can get stuck modeling self-shielding from such high-density systems,
because the convergence criteria push the adaptive timestep to very short values. Second,
when analyzing IGM damping wings, objects with strong proximate absorbers (PDLAs or PLLSs), are
excluded from analysis, so it is sensible to do the same for the simulations. 
To mitigate the undesired effects of unphysically large densities, we omit the first $35~{\rm pkpc}$
from consideration when extracting the
density, temperature, and velocity fields from the Nyx simulation, and we clip the gas overdensity to always be below 200, i.e. roughly
the virial overdensity. Nevertheless, even for the combination of model parameters where proximate DLAs are least expected ($x_{\rm HI} = 0.0$ and 
$\logtq = 8$),  we still found that a small fraction (17 of 1200) of the simulated IGM transmission spectra exhibited proximate DLA absorption\footnote{These were identified as skewers with transmission deviating by more than $2\%$ from unity at a pixel $1000~{\rm km\,s^{-1}}$ redward of
the quasar redshift for the model with $x_{\rm HI} = 0.0$ and $\logtq = 8$. This model choice is conservative, since it is least likely to
exhibit damping wing absorption both because the IGM is reionized and the longest quasar lifetime maximizes the likelihood of photoevaporating proximate absorbers.},
which we chose to exclude from consideration for all models. Thus, the final
output of our IGM transmission simulations is a set of 1183
transmission spectra at each location of a coarse $21\times 51$ grid
corresponding to the two IGM parameters, $\btheta = (\langle x_{\rm
  HI}\rangle, \logtq)$. Figures illustrating our numerical simulation procedure can be found in
\citet{Davies18b}.

\begin{figure}[b!]
  \vskip -0.4cm
  \includegraphics[trim=0 0 0 0,clip,width=0.50\textwidth]{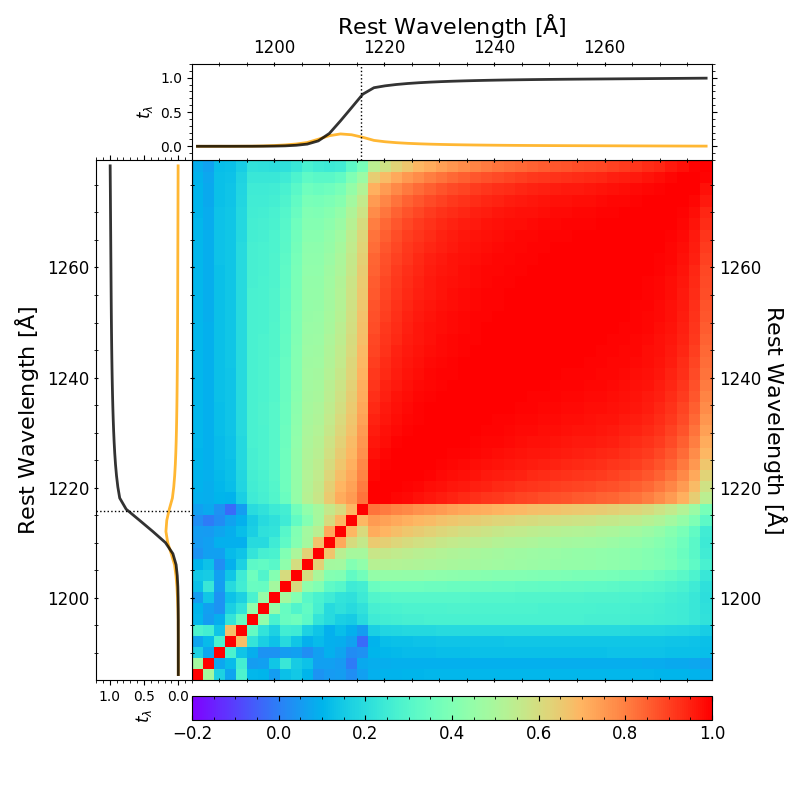}
  \vskip -0.3cm
  \caption{Mean transmission and covariance structure of the proximity
    zone Ly$\alpha$ transmission field $\bt$ evaluated from the 1183
    transmission spectra for a model $\langle x_{\rm HI}\rangle =
    0.50$ and $\logtq = 6$ on a velocity
    grid with $\mathrm{d}v = 500~{\rm km s^{-1}}$.  Top and left panels show
    the mean transmission $\langle \bt\rangle$ (black) and its
    1$\sigma$ variation $\sqrt{\diag{\bC_{\bt}}}$ (orange), whereas
    the 2d image shows the correlation matrix with the level of
    correlation indicated by the color bar. The wavelength of
    Ly$\alpha$ at line center, $1215.67\,$\AA, is shown by the vertical
    dotted line. The smooth coherent IGM damping wing signature
    results in highly correlated transmission fluctuations redward of
    Ly$\alpha$; whereas, for $\lambda \lesssim 1215.67\,$\AA\, the
    correlations are smaller owing to Ly$\alpha$ forest fluctuations in the proximity zone, 
    although they are not insignificant.
    \label{fig:trans_corr}}
\end{figure}

To evaluate the likelihood in eqn.~(\ref{eqn:lhood_final}) we require
the mean, $\langle \bt(\btheta)\rangle$, and covariance,
$\bC_{\bt}(\btheta)$ of the proximity zone Ly$\alpha$ transmission as
a function of model parameters
$\btheta = (\langle x_{\rm HI}\rangle, \logtq)$. Fig.~\ref{fig:trans_corr} illustrates these quantities at the wavelengths 
  relevant for IGM damping wings on our velocity grid with $\mathrm{d}v = 500~{\rm km
  \,s^{-1}}$ pixels for a 
model with $\langle x_{\rm HI}\rangle = 0.50$ and $\logtq = 6$. 
The top and left panels show the mean transmission $\langle \bt\rangle$
(black) and its 1$\sigma$ variation  $\sqrt{\diag{\bC_{\bt}}}$ (orange).
The covariant structure of the transmission can be visualized
via the correlation matrix shown in the 2d image, computed from
$\bC_{\bt}$ (see the definition in eqn.~(\ref{eqn:corr})). The smooth coherent IGM damping wing signature results in highly
correlated transmission fluctuations redward of Ly$\alpha$; whereas, for $\lambda \lesssim 1215.67\,$\AA\, the correlations
    are smaller owing to  Ly$\alpha$ forest fluctuations, although not insignificant.

\subsection{Creating Mock Quasar Spectra}
\label{sec:mocks}

In this section we describe our procedure for generating realistic
mock quasar spectra with IGM damping wings. As described in
\S~\ref{sec:DWsims}, we focus on models of the quasar ULAS
J1342$+$0928 at $z=7.54$ with $J_{\rm AB}=20.3$. 
The simulated spectra
cover the rest-frame (observed-frame) wavelength range
$1185-2000\,$\AA\, ($10120-17080\,$\AA). As discussed in \S~\ref{sec:wavegrid}, it is
advantageous to rebin the spectra from the native sampling of the simulations onto
a  coarser wavelength grid with $\mathrm{d}v = 500~\kms$. 
We assume observations at a spectral resolution described by a Gaussian line-spread function (LSF) with
FWHM=$100\,\kms$. To forward model the impact of spectral resolution and rebinning onto the coarse
$\mathrm{d}v = 500~\kms$ grid,  we
convolve our ensemble of $1183\times21\times 51$ IGM transmission
skewers (1183 skewers, $21\times 51$ models) defined on the simulation spectral grid $\mathrm{d}v_{\rm sims}=4\,\kms$, 
with this Gaussian LSF and then rebin them onto the coarse wavelength grid. Mock quasar spectra are
generated by randomly selecting a continuum from our 778 autofit test
set, cubic spline interpolation of the continuum onto the
coarse wavelength
grid, and then multiplying by a randomly selected resolution convolved
and rebinned IGM transmission skewer\footnote{Note that the two
operations, 1) multiplication of the full resolution IGM transmission
with the 
continuum, and 2) convolution of the resulting spectrum
with the LSF and rebinning do not formally commute. The correct order
is to multiply by the continuum first and then convolve and rebin, but
the differences are negligible. We choose to convolve and rebin the IGM
transmission skewers first, and then multiply by the continuum afterward
because the expensive convolution and rebinning operations can then be
performed only once in pre-processing, dramatically speeding up all
downstream computations.}. 

\begin{figure*}
  \vskip -0.4cm
  \includegraphics[trim=27 0 23 0,clip,width=0.99\textwidth]{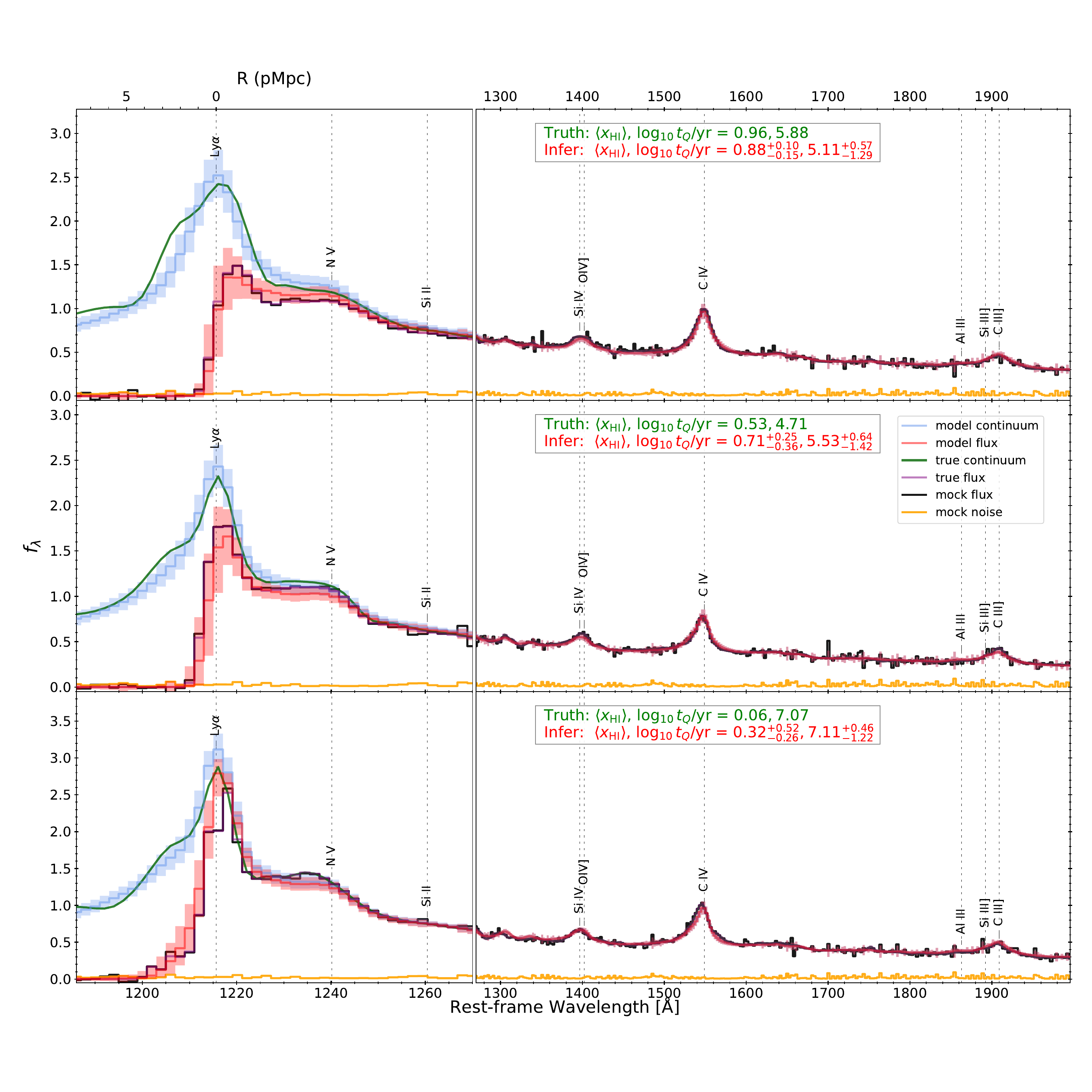}
  \vskip -1.4cm
  \caption{Examples of inference performed on mock spectra for for
    three different combinations of $\langle x_{\rm HI}\rangle $ and $\logtq$. Black shows the noisy mock quasar spectra, orange
    the $1\sigma$ spectral noise level, purple the true noiseless flux, green the true underlying continuum,
    blue the median inferred continuum model, and red the median inferred damping wing model flux. 
    Shaded regions indicate the 16th and 84th percentiles of the  continuum model (blue) and  model flux (red). 
    For the continuum model the median and shaded regions incorporate parameter variations,  continuum reconstruction errors,
    and spectral noise, whereas for the model flux the incorporate these effects as well as IGM transmission fluctuations (see \S~\ref{sec:inf_examples} for details).  True model parameters are annotated in green, whereas median inferred parameters and 16th and 84th percentile credibility intervals (determined from marginalized posteriors) are annotated in red. 
    The inferred constraints are those determined after the marginal coverage correction weights (which slightly dilate the contours) have been applied (see \S~\ref{sec:reweighting}).  A corner plot illustrating the full eight dimensional posterior for the mock with  $\langle x_{\rm HI}\rangle =0.96$ and $\logtq = 5.88$  in the upper panel is shown in Fig.~\ref{fig:mock_corner_full} (also coverage corrected). 
    \label{fig:mock_example}}. 
\end{figure*}

We simulate observational data collected with a ground based 8m class telescope with a hypothetical
instrument with a constant throughput of 30\%. To correctly model the noise, we used the
\texttt{skycalc\_ipy}\footnote{https://github.com/AstarVienna/skycalc\_ipy}
python package to generate realistic models of the sky background and atmospheric transmission,
which are used to construct a full spectrum simulator
including telluric absorption and noise contributions from object
photons, sky background, and detector read noise (assumed to be four
electrons per  $\mathrm{d}v = 500~\kms$ spectral pixel). We tuned a hypothetical exposure time to
achieve a median signal-to-noise ratio of ${\rm S\slash N}=10$ per
$100~\kms$ velocity interval, computed over the telluric absorption
free observed frame (rest frame) wavelength range $11750-13300\,$\AA\,
($1376-1557\,$\AA). This in turns allows us compute the correct relative 
contributions of photon counting and detector read noise to the noise
budget, resulting in a realistic noise vector $\bsigma$. 
Multiplying this noise vector into a random draw from a unit variance Gaussian distribution,
generates a realistic realization of heteroscedastic noise (due to OH sky lines and
telluric absorption features), which is then added to the mock quasar
spectrum. Three examples of mock quasar spectra generated via this procedure 
are shown in Fig.~\ref{fig:mock_example}.

\section{Inference Results}
\label{sec:inference}

In this section we present the results from statistical inference performed on the mock EoR quasar spectra
that were introduced in \S~\ref{sec:mocks} using the new likelihood we derived in \S~\ref{sec:formalism} (see eqn.~(\ref{eqn:lhood_final})). First we discuss how we use HMC to sample the posterior distribution, then we describe the coverage test we perform to assess the reliability of our statistical inference. Our inference turns out to be  overconfident and thus does not pass the coverage test, but we describe a procedure that reweights the HMC samples to remedy this problem and thus perform reliable inference. After showing examples of our inference at work, we build intuition for why our inference fails the coverage test. We conclude by quantifying how well we recover the underlying quasar continuum and compare the accuracy of our reconstructions with past work.

\subsection{Hamiltonian Monte Carlo}
\label{sec:HMC}
HMC \citep{HMC1} is a powerful variant of the traditional Markov Chain Monte Carlo (MCMC) algorithm for sampling
probability distributions in high dimensions \citep[see e.g.][for a
  review]{HMC2}.  Based on a powerful analogy with Hamiltonian
dynamics, HMC introduces auxiliary momentum variables that interact
with the target variables representing the samples from the
distribution of interest.  HMC numerically integrates the equations of Hamiltonian dynamics, and
then interleaves Metropolois-Hastings steps to accept or reject the
proposed target states ensuring that detailed balance is
satisfied. Compared to MCMC, HMC capitalizes on the gradients of the
log-likelihood, making it more efficient for sampling in
high-dimensional spaces and handling correlated target variables. This
advantage enables HMC to generate distant proposals more effectively,
leading to reduced autocorrelation in samples, better performance for multi-modal distributions, 
and faster convergence. 

The main challenge of using HMC is that it requires gradients of the log-likelihood function with respect to the model parameters, which can be challenging or costly to compute. 
Gradients can be computed in several ways: by numerical approximation using finite differences, through symbolic manipulation of analytic expressions, or via automatic differentiation \citep[AD; e.g.][]{autograd}, a programmatic application of the chain rule that yields exact derivatives at machine precision by tracing the operations in a computation. In our case, we rely on AD, which yields exact gradients for arbitrary compositions of differentiable operations and scales efficiently even in high-dimensional parameter spaces. Modern AD frameworks automate this process, making it straightforward to obtain derivatives even for complicated likelihood functions involving matrix operations, linear algebra, and complex control flow. The introduction of automatic differentiation environments in \texttt{Python} such as \texttt{JAX} \citep{JAX}, makes exploiting AD to compute the reqiured gradients for HMC straightforward.

\begin{figure*}
  \vskip -0.4cm
  \includegraphics[trim=10 0 0 10,clip,width=0.49\textwidth]{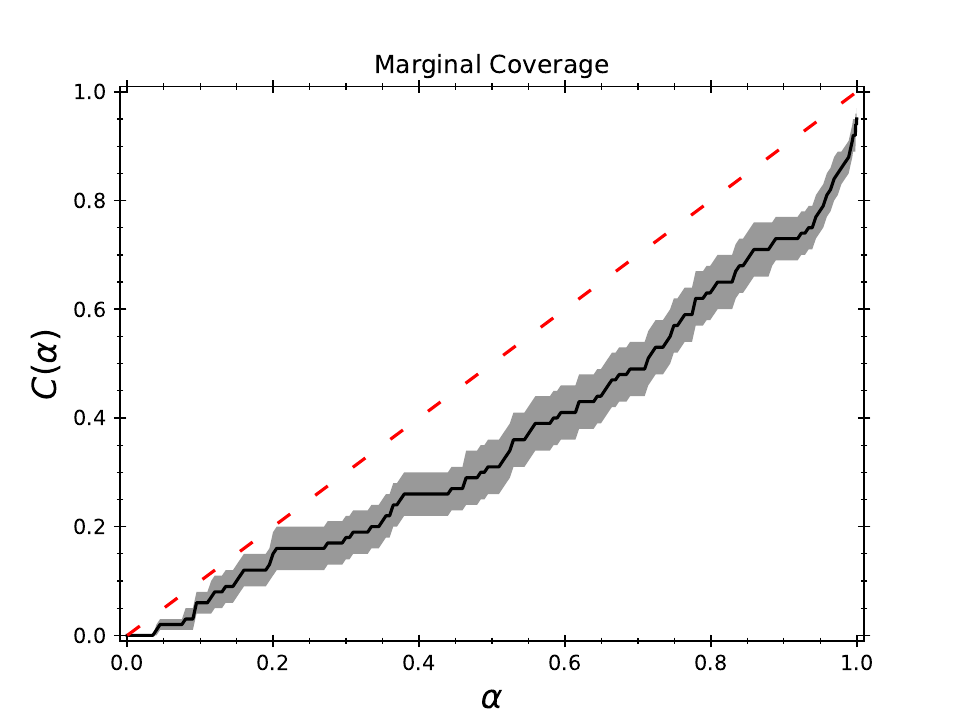}
  \includegraphics[trim=10 0 0 10,clip,width=0.49\textwidth]{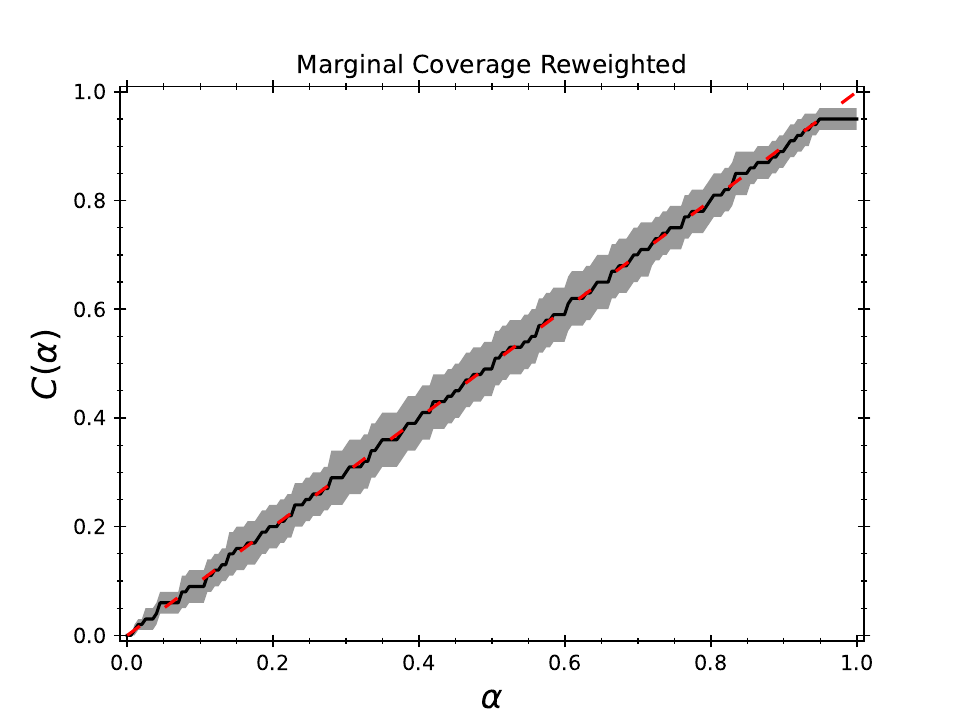}
  \vskip -0.3cm
  \caption{Coverage test results determined from $N=100$ mock spectra. 
    {\emph Left:} The black line shows the coverage $C(\alpha)$ of the marginal posterior distributions
    $P(\btheta | \bflam_j,\bsigma_j)$, which emprically quantifies how often the true
    astrophysical parameters $\btheta_{\rm true} = (\langle x_{\rm HI}\rangle , \logtq)$
    lie within the $\alpha$-th credibility contour.  The red dashed line shows the expected
    $y=x$ curve that one would obtain for a perfect inference. The gray shaded region shows the effective $1\sigma$ error range
    on $C(\alpha)$ determined from 16th and 84th percentile ranges of the binomial distribution.
    Given that $C(\alpha)$ lies everywhere below the red-dashed line, we see that our
    pipeline systematically delivers overconfident inference. {\emph Right:} Coverage test results after enlarging
    the credibility contours by reweighting the HMC samples. By construction, $C(\alpha) =\alpha$, indicating that on
    average the inference pipeline delivers reliable credibility contours\label{fig:marginal_coverage}.}
\end{figure*}

We developed \texttt{Python} software to compute an automatically
differentiable version of the 
likelihood in eqn.~(\ref{eqn:lhood_final})
in \texttt{JAX}.  For the HMC, we use the No-U-Turn Sampler
\citep[NUTS;][]{NUTS} variant of HMC implemented in the
\texttt{JAX}-based \texttt{NumPyro} \citep{pyro,numpyro} probabilistic
programming package in the \texttt{Python} module
\texttt{numpyro.infer.NUTS}. The NUTS sampler was run with four chains of
1000 samples, each with a warmup phase of 1000 samples, using the
``vectorized'' chain option.  The NUTS algorithm builds a binary tree
that is used to take forward/backwards `directional' steps to explore
the target posterior using gradients to guide it towards the highest
probability regions.  The \texttt{max\_tree\_depth} parameter, which
sets the size of this binary tree, was set to ten (i.e. up to a
maximum of 1024 steps for each iteration). As alluded to in
\S~\ref{sec:wavegrid}, the run time is a strong function of the number
of pixels, $n_{\rm pix}$, in the quasar spectrum, as this sets the
dimensionality of the matrices in eqn.~(\ref{eqn:lhood_final}), and
several of the required matrix operations scale as $\mathcal{O}(n_{\rm
  pix}^3)$. In general, there are order unity variations in the
runtime which depend on details of the warmup phase that constructs
the `mass matrix' determining the Hamiltonian dynamics in HMC, which
in turn will depend on the parameter dependent shape of the underlying
posterior distribution. The wavelength grid adopted in this paper
has $n_\lambda = 313$.  Running on
a single thread of an AMD EPYC 7763 2.45 Ghz processor\footnote{These
processors have 64 cores, 128 threads, and $256\,{\rm MB}$ of L3 Cache.}, the
typical runtimes were $\sim 2.5$ hours. Running on
a single thread of an Intel Xeon Gold 6126 2.6GHz processor\footnote{These
processors have 12 cores, 24 threads, and $19.25\,{\rm MB}$ of L3 Cache.} equipped with a NVIDIA GeForce RTX 2080TI GPU (with 11 GB of memory), the
typical runtimes were $\sim 15$ minutes.

\subsection{Coverage Testing}
\label{sec:cov_test}
Let $\bTheta$ be the parameter vector of interest for a Bayesian inference problem that will be applied to experimental data $\bx$. In the current 
context of IGM damping wings, $\bTheta = \{\btheta, \boeta\}$, where $\btheta$ are the astrophysical parameters and $\boeta$ are the DR parameters, and
the data $\bx$ are the quasar spectrum $\bflam$ and its associated noise vector $\bsigma$. 
Following Bayes theorem, the posterior distribution for $\bTheta$, given experimental data $\bx$ is
\be
P(\bTheta | \bx) = \frac{L(\bx | \bTheta)P(\bTheta)}{P(\bx)}\label{eqn:bayes}, 
\ee
where $L(\bx | \bTheta)$ is the likelihood of the data given the model, $P(\bTheta)$ reflects our prior knowledge
of the parameters, and $P(\bx)$ is known as the evidence, which is interpreted as a normalization constant since it is independent of the parameters $\bTheta$.
If the inference procedure is reliable, one expects that the probability $\alpha$ obtained by integrating the posterior probability density in eqn.~(\ref{eqn:bayes}) over a volume $V$ of parameter space $\bTheta$,
\be
\alpha = \int_{V_\alpha} P(\bTheta | \bx) \mathrm{d}\bTheta\label{eqn:Valpha}
\ee
corresponds to a true probability. 
In other words, that the parameter space volume enclosed by the 68 percent credibility contour
contains the true parameters 68 percent of the time under repetitions of the experiment, and analogously for all the other credibility levels. However, this need not always be the case. Imperfections in an inference procedure can cause the posterior distribution to exclude the true
parameters more or less often than indicated by the nominal credibility contours, because the posterior is shifted, or
too narrow and hence overconfident, or too broad and hence underconfident, or combinations thereof. 
Such imperfections can result from bugs in an inference pipeline, or from an inappropriate choice of the prior, or by adopting an approximate
likelihood that fails to accurately capture the statistical properties of the measurement process.
In the current context, we know that the likelihood in eqn.~(\ref{eqn:lhood_final}) is only approximate, 
as discussed in \S~\ref{sec:formalism}, which motivates us to explore its coverage.

The \emph{coverage probability}, $C(\alpha)$, of a posterior credibility
level $\alpha$ is the fraction of the time that the true parameters
lie within the volume enclosed by the corresponding credibility
contour under repetitions of the experiment \citep[see
  e.g.][]{Sellentin19}.  It provides a mathematically rigorous method
to quantify whether a posterior distribution delivers reliable
probabilities. If an inference procedure is robust, then the coverage
probability should equal to the posterior credibility for every level.
This approach of comparing coverage probabilities to posterior
credibility levels is referred to
as an `inference test' or a `coverage
test'.

We conduct a coverage test following the algorithm described in more detail in Appendix~\ref{appendix:coverage}
\citep[see also][]{Wolfson23},
which we now summarize: 
\begin{enumerate}
\item Draw $N=100$ astrophysical parameter vectors $\btheta_{{\rm true},j}$ from uniform priors defined by
  $\langle x_{\rm HI}\rangle \sim \text{Uniform}(0, 1)$ and $\logtq \sim \text{Uniform}(3, 8)$.
  These are the `true' parameters that generate the mock datasets used to perform the coverage test.

\item Generate realizations of mock quasar spectra $\{\bflam_j, \bsigma_j\}$
  from these `true' parameters following the approach
  described in \S~\ref{sec:mocks}.  

\item Perform HMC inference on each dataset as described in \S~\ref{sec:HMC} resulting in a set of 2000 
samples for the astrophysical parameters $\btheta$ and continuum nuisance parameters $\boeta$, from each of the $N=100$
  posterior distributions $P(\btheta, \boeta | \bflam_j,\bsigma_j)$. 
  
\item Consider a set of $M$ credibility contour levels $\alpha \in [0,1]$.  For each value $\alpha$ and each mock, 
  test whether the true astrophysical parameter values,  $\btheta_{{\rm true},j}$, reside within the volume $V_{\alpha}$ enclosed by $\alpha$-th contour. 
  For each $\alpha$, the  coverage probability $C(\alpha)$ is the fraction of the $N$ mock datasets for which the
  true values lie within the volume $V_{\alpha}$ defined by eqn.~(\ref{eqn:Valpha}). 
\end{enumerate}

As described in Appendix~\ref{appendix:coverage},
a coverage test can be performed for
the entire parameter vector, here $\bTheta = \{\btheta, \boeta\}$ using the full
posterior $P(\btheta, \boeta | \bflam_j,\bsigma_j)$, as well as for the
astrophysical (i.e. non-nuisance) parameters $\btheta$ using their
marginal posterior $P(\btheta | \bflam_j,\bsigma_j)$. Obviously, the
coverage of the marginal posterior $P(\btheta | \bflam_j,\bsigma_j)$
is most critical, since nuisance parameters will be marginalized out, which is
what we focus on in what follows.

For $\alpha$, we consider a vector of $200$ 
linearly spaced values in the
range $\alpha \in [0,0.994]$ concatenated with a vector of 101
linearly spaced values in the range $\alpha \in [0.995,1.0]$, resulting in 
$M=301$ values of $\alpha$.  As will
be apparent in the next section, the steep $C(\alpha)$ vs $\alpha$
curve near $\alpha \sim 1$ (see Fig.~\ref{fig:marginal_coverage}) motivates
adopting a more finely spaced grid as $\alpha$ approaches unity.

\subsection{Marginal Coverage Test Results}
\label{sec:cov_test_results}

In Fig.~\ref{fig:marginal_coverage} we show the marginal coverage test results
determined from the approach described in \S~\ref{sec:cov_test} and Appendix~\ref{appendix:coverage} for $N=100$ mock quasar
spectra.  The black line shows the coverage $C(\alpha)$ of the marginal posterior distributions
$P(\btheta | \bflam_j,\bsigma_j)$, which empirically quantifies how often the true
astrophysical parameters $\btheta_{\rm true} = (\langle x_{\rm HI}\rangle, \logtq)$
lie within the $\alpha$-th credibility contour.  The red dashed line shows the expected
$y=x$ curve that one would obtain for a perfect inference pipeline in
the limit $N\rightarrow \infty$. An overconfident inference pipeline
will yield a curve $C(\alpha)$ versus $\alpha$ that lies
systematically below the line $y=x$, whereas for an underconfident
inference procedure the $C(\alpha$) will lie above $y=x$.  As
discussed in Appendix~\ref{appendix:coverage} \citep[see also][]{Sellentin19}, since
$C(\alpha)$ counts how often the true parameters fall inside the
$\alpha$-th contour, it is the number of successes in a sequence of $N$
independent experiments each asking a yes-or-no question. Hence $C(\alpha)$ follows
the Binomial distribution $B(N, C(\alpha))$,
which we use to assign errors. The gray shaded
region shows the effective $1\sigma$ error range\footnote{Note however
that the different values of $\alpha$ are clearly correlated since the
same ensemble was used to calculate all of them.} on $C(\alpha)$
determined from 16th and 84th percentile ranges of $B(N, C(\alpha))$.

Given that $C(\alpha)$ lies everywhere below the red-dashed line, we see that our
pipeline systematically delivers overconfident
inference. For example, for the effective $1\sigma$ credibility level
$\alpha=0.68$, i.e. the 68th percentile credibility contour, the
coverage is actually just $C(0.68) = 0.48$, indicating that on average
the contours include the true $\btheta_{\rm true} = (\langle x_{\rm HI}\rangle,
\logtq)$ only $48\%$ of the time.  Similarly, for the effective $2\sigma$
credibility level of $\alpha=0.95$ the coverage is $C(0.95) = 0.79$.

\subsection{Reweighting Samples to Pass a Coverage Test}
\label{sec:reweighting}

The coverage test for the marginal posterior $P(\btheta | \bflam,\bsigma)$ distribution shown in
left panel of Fig.~\ref{fig:marginal_coverage}. indicates that $C(\alpha)$ lies everywhere below the red-dashed line $y=x$,
indicating that our pipeline systematically delivers overconfident inference.
How can we nevertheless perform statistically reliable inference in
light of this overconfidence? 

In Appendix~\ref{appendix:coverage} we introduce a novel procedure whereby the HMC posterior samples can be assigned
weights, such that the reweighted samples produce reliable inference which passes a coverage
test by construction \citep[see also][]{Wolfson23}. The mathematics underlying this procedure is
described in detail in Appendix~\ref{appendix:coverage}, but it can be understood
heuristically as follows. \citet{Sellentin19} advocate that one simply relabel the credibility contours to reflect the fact that
inference is overconfident.  In other words, since the coverage plot in Fig.~\ref{fig:marginal_coverage} indicates
that the 68th percentile contour only contains the true value 48\% of the time,  we will simply refer to this contour as the 48th percentile rather than the 68th. The real 68th
percentile contour containing the true parameters 68\% of the time
under the inference test actually corresponds to the value $\alpha=0.85
= C^{-1}(0.68)$ contour for our original approximate inference. In
other words, we can systematically expand all of the
credibility contours of the original inference by the right amount, such that they contain the
true model the empirically correct fraction of the time. In general,
as we show in Appendix~\ref{appendix:coverage}, this remapping of the
credibility levels $\alpha$ into true coverage probabilities
$C(\alpha)$ can be achieved by solving for the set of weights for the HMC
samples from the original posterior, which by construction
guarantees that we will pass a coverage test. 

The purpose of HMC (or
MCMC) samples from a posterior is
to estimate credibility intervals
on parameters, perform marginalization integrals, and compute
`moments' of the posterior via Monte Carlo integration. If we can
determine the set of weights that corrects the imperfect inference
such that it passes a coverage test, these weights can then be
used in all of the downstream computations that one performs with the
samples, guaranteeing the reliability of our statistical inference.

The right panel of
Fig.~\ref{fig:marginal_coverage} shows the coverage
of the  marginal posterior $P(\btheta | \bflam,\bsigma)$ distribution
after the HMC samples have been reweighted.
Samples from the core of the distribution with higher $P(\btheta| \bx)$ are
downweighted, whereas samples in the outskirts of the distribution
with lower $P(\btheta| \bx)$ are upweighted, such that the net effect is
to grow the credibility contours. The agreement of $C(\alpha)$ with the
red dashed $y=x$ line indicates that we now achieve perfect coverage,
which as explained in Appendix~\ref{appendix:coverage}, occurs by
construction because we solve for the set of weights that guarantees
this outcome. 

Note that although we compute the weights to guarantee that the reweighted
HMC astrophysical parameter samples, $\btheta_j$, will pass a coverage test for
their marginal posterior distribution, we nevertheless apply these weights to
the entire parameter vector $\bTheta = \{\btheta, \boeta\}$. In other words, the contours
for all parameters including the nuisance parameters will also be dilated. In practice
this implies that we pass the coverage test perfectly for the astrophysical parameters  (by construction, as shown in the left panel of Fig.~\ref{fig:marginal_coverage}), $\btheta$, 
whereas our inference is slightly underconfident for the full parameter vector, $\bTheta$, which includes the nuisance parameters, $\boeta$ (i.e. the contours are slightly too large).  Since underconfident inference is more conservative,  and as this applies only to the nuisance parameters, we have demonstrated that our full  pipeline passes a coverage test and delivers reliable statistical inference.

\subsection{Inference Examples}
\label{sec:inf_examples}

\begin{figure*}
  \vskip -0.4cm
  \includegraphics[trim=0 0 0 0,clip,width=0.99\textwidth]{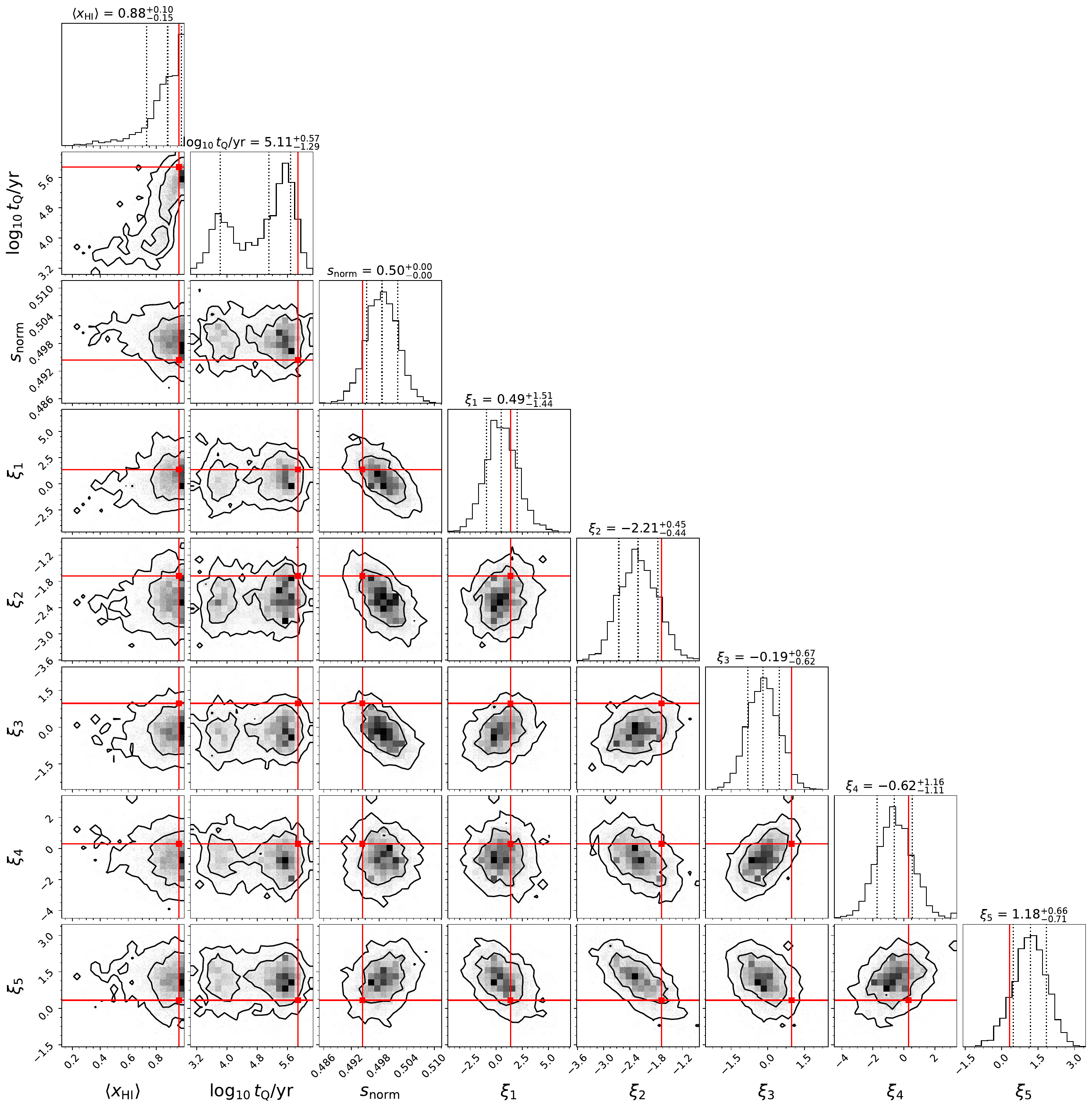}
  \vskip -0.3cm
  \caption{Corner plot illustrating the full eight dimensional posterior distribution resulting from our statistical inference procedure applied to the mock quasar spectrum in the top panel of Fig.~\ref{fig:mock_example} 
  with $x_{\rm HI}=0.96$ and $\logtq = 5.88$. The red square and the horizontal and vertical red lines indicate the true parameter values. The posterior distribution shown here is the result  after the marginal coverage correction weights described in \S~\ref{sec:reweighting} have been applied, which slightly dilates the contours relative to the original HMC inference\label{fig:mock_corner_full}.} 
\end{figure*}

Examples of our inference applied to mock
spectra (see \S~\ref{sec:mocks})  with reweighted HMC samples 
(see \S~\ref{sec:reweighting}) 
are shown for three different
combinations of $\langle x_{\rm HI}\rangle$ and $\logtq$ in
Fig.~\ref{fig:mock_example}. 
The black histograms show the noisy mock quasar
spectra, orange the $1\sigma$ spectral noise levels, green the true
underlying continuua, blue the median inferred continuum models, and
red the median inferred damping wing model flux profiles.
True model parameters are annotated in green, whereas median inferred parameters
and 16th and 84th percentile credibility intervals (determined from
marginalized posteriors after the coverage correction reweighting) are 
annotated  in red.

Our approach for visualizing the quality of the fits, and hence the solid lines and shaded regions in Fig.~\ref{fig:mock_example},  
warrants  further discussion. A realization of the model of the spectrum can be determined by
evaluating the model of the spectra (e.g. the DR quasar continuum, the proximity zone transmission profile, or the product of the two) at each HMC
sample, which has an associated weight. As discussed in
\S~\ref{sec:reweighting}, median model curves and model confidence
intervals can be determined by computing weighted percentiles of these
model curves.  However, according to the likelihood in
eqn.~(\ref{eqn:lhood_final}), there are multiple sources of scatter that
govern how well a model curve will fit the data, namely spectral
noise, IGM fluctuations, and continuum reconstruction errors. A naive
visual data-model comparison that does not take all these sources of
stochasticity into account can be misleading. As a concrete example,
if one compares the median and $1\sigma$ interval of the PCA continnum
to the quasar spectrum redward of the Ly$\alpha$ region, it will not
appear to be a good fit relative to the spectral noise alone, since
this ignores the continuum reconstruction error budget. Thus in the current
context, choosing sensible `error-bars' for visual data-model
comparison is rather subtle.

To generate the median and 16th and 84th percentile model intervals in
Fig.~\ref{fig:mock_example} we proceed as follows. For each parameter vector
$\bTheta = \{\btheta, \boeta\}$ from the HMC posterior Markov chain, we:
\begin{itemize}
\item Evaluate the PCA DR model $\bsDR(\boeta)$ via
  eqn.~(\ref{eqn:PCA}).
\item Draw a realization of the relative continuum reconstruction
  error $\bdelta$ from the Gaussian distribution in
  eqn.~(\ref{eqn:pdelta}), and compute $\bs = \bsDR(\boeta)\ccirc (\mathbf{1} + \bdelta)$.
\item Draw a random IGM transmission skewer $\bt$ for the value of
  $\btheta$ from the set of simulated skewers, allowing us to compute
   $\bflam = \bs \ccirc \bt$.
\item Draw a realization of Gaussian spectral noise $\bsigmatilde$
  consistent with the noise vector $\bsigma$ and compute $\bstilde = \bs +
  \bsigmatilde$ and $\bflamtilde = \bflam + \bsigmatilde$
  which we refer to as the 
  noisy model continuum and the noisy model flux. 
\end{itemize}

The blue histogram and shaded region in Fig.~\ref{fig:mock_example} represent
the weighted median and 16th and 84th weighted percentiles of the noisy model continuum, 
$\bstilde$, respectively, which reflects parameter uncertainty,
continuum reconstruction errors, and spectral noise. The red histogram
and shaded region are the weighted median and the same
percentiles of the noisy model flux, $\bflamtilde$, reflecting parameter uncertainty,
continuum reconstruction errors, IGM transmission fluctuations, and
spectral noise.  Note that IGM transmission fluctuations are intrinsically accounted for by drawing
IGM transmission skewers from the model for a choice of astrophysical parameters $\btheta$, whereas
we are explicitly adding the continuum reconstruction errors and spectral noise by drawing
realizations from their respective Gaussian distributions.

\begin{figure}
  \vskip -0.4cm
  \includegraphics[trim=0 0 0 0,clip,width=0.99\columnwidth]{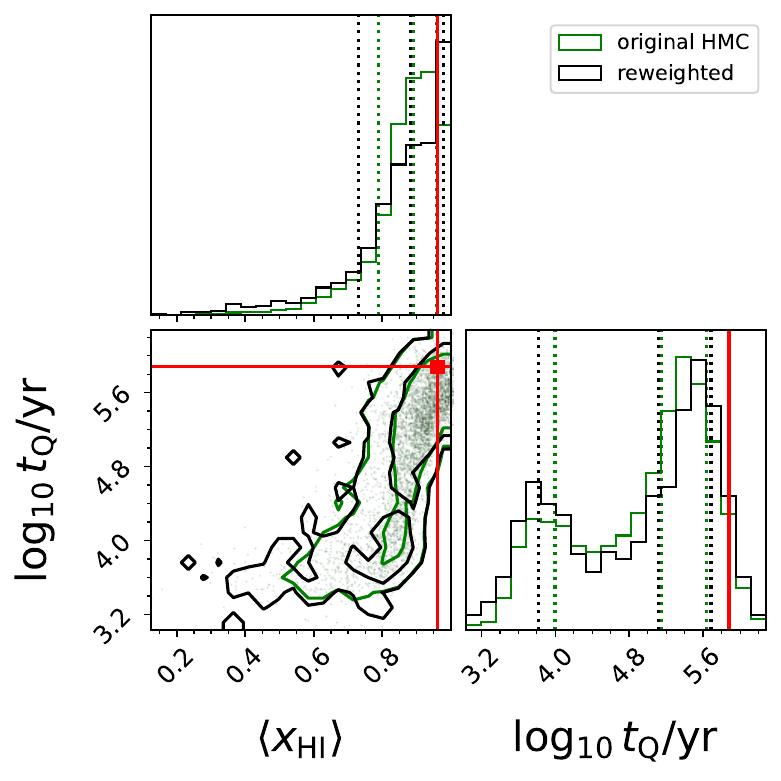}
  \vskip -0.3cm
\caption{Comparison of the 2D marginal posterior distribution in the astrophysical parameters,  $\langle x_{\rm HI}\rangle$ and $\logtq$ (i.e. the upper left panel of the full posterior in Fig.~\ref{fig:mock_corner_full}), before (original HMC; green) and after (reweighted; black) applying the coverage correction weights to the samples from the HMC chain (see \S~\ref{sec:reweighting}). The red square and the horizontal and vertical red lines indicate the true parameter values. Vertical dotted lines in the marginal posterior panels indicate the 16-50-84 percentile ranges.  It is evident that reweighting the samples broadens the posterior, correcting for the overconfidence  (see Fig.~\ref{fig:marginal_coverage}) of the original marginal posterior distribution as discussed in \S~\ref{sec:reweighting}.\label{fig:mock_corner_marg}}
\end{figure}

Fig.~\ref{fig:mock_corner_full} shows a corner plot illustrating the full eight dimensional posterior for the mock spectrum in the top panel of Fig.~\ref{fig:mock_example}, which has true model parameters $\langle x_{\rm HI}\rangle=0.96$ and $\logtq = 5.88$. The red square and the horizontal and vertical red lines indicate the true parameter values. Note that Fig.~\ref{fig:mock_corner_full} shows the posterior distribution after applying the marginal
coverage correction weights described in \S~\ref{sec:reweighting}. Fig.~\ref{fig:mock_corner_marg} compares the nuisance parameter marginalized posterior distributions for the two astrophysical parameters, $\langle x_{\rm HI}\rangle $ and $\logtq$ for the same mock spectrum 
in the top panel
of Fig.~\ref{fig:mock_example} (i.e. the upper left panel of the full posterior in Fig.~\ref{fig:mock_corner_full}), before (original HMC; green) and after (reweighted; black) applying the coverage correction weights to the samples from the HMC chain (see \S~\ref{sec:reweighting}). It is evident that this reweighting broadens the posterior, correcting for the overconfidence of the original marginal posterior distribution (see Fig.~\ref{fig:marginal_coverage}), as
discussed in \S~\ref{sec:reweighting}. Qualitatively, the shape of the posterior in the $\langle x_{\rm HI}\rangle-\logtq$
plane resembles the shape of the posteriors recovered in the analysis of real $z > 7$ quasar
spectra by \citet{Davies18b}. Specifically, the well known degeneracy \citep[see e.g.][]{Bolton11,Davies18b} between IGM neutral fraction and quasar lifetime in determining the shape of the proximity zone and IGM damping wing profile is apparent.

\begin{figure*}
  \vskip -0.4cm
  \includegraphics[trim=0 0 0 0,clip,width=0.49\textwidth]{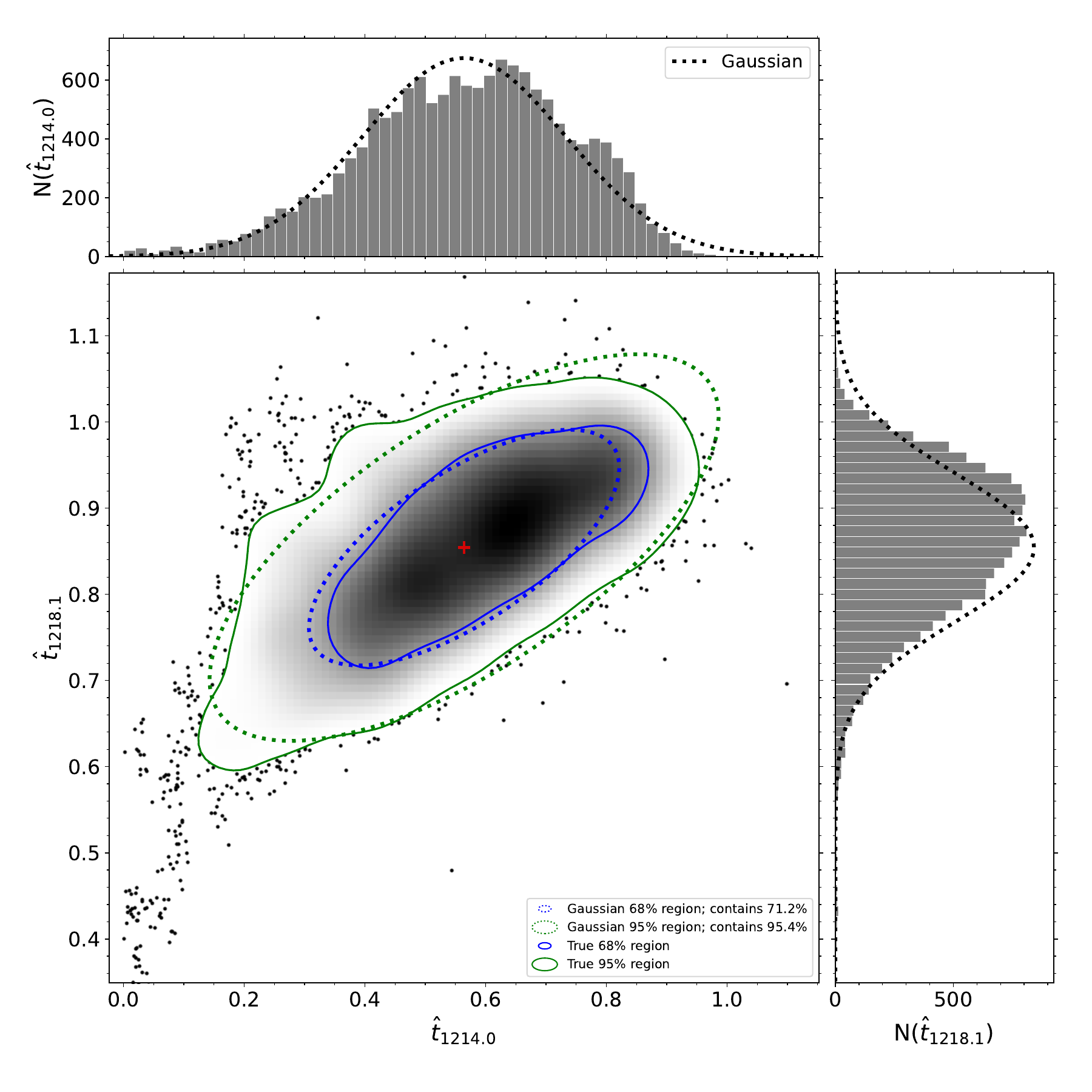}
  \includegraphics[trim=0 0 0 0,clip,width=0.49\textwidth]{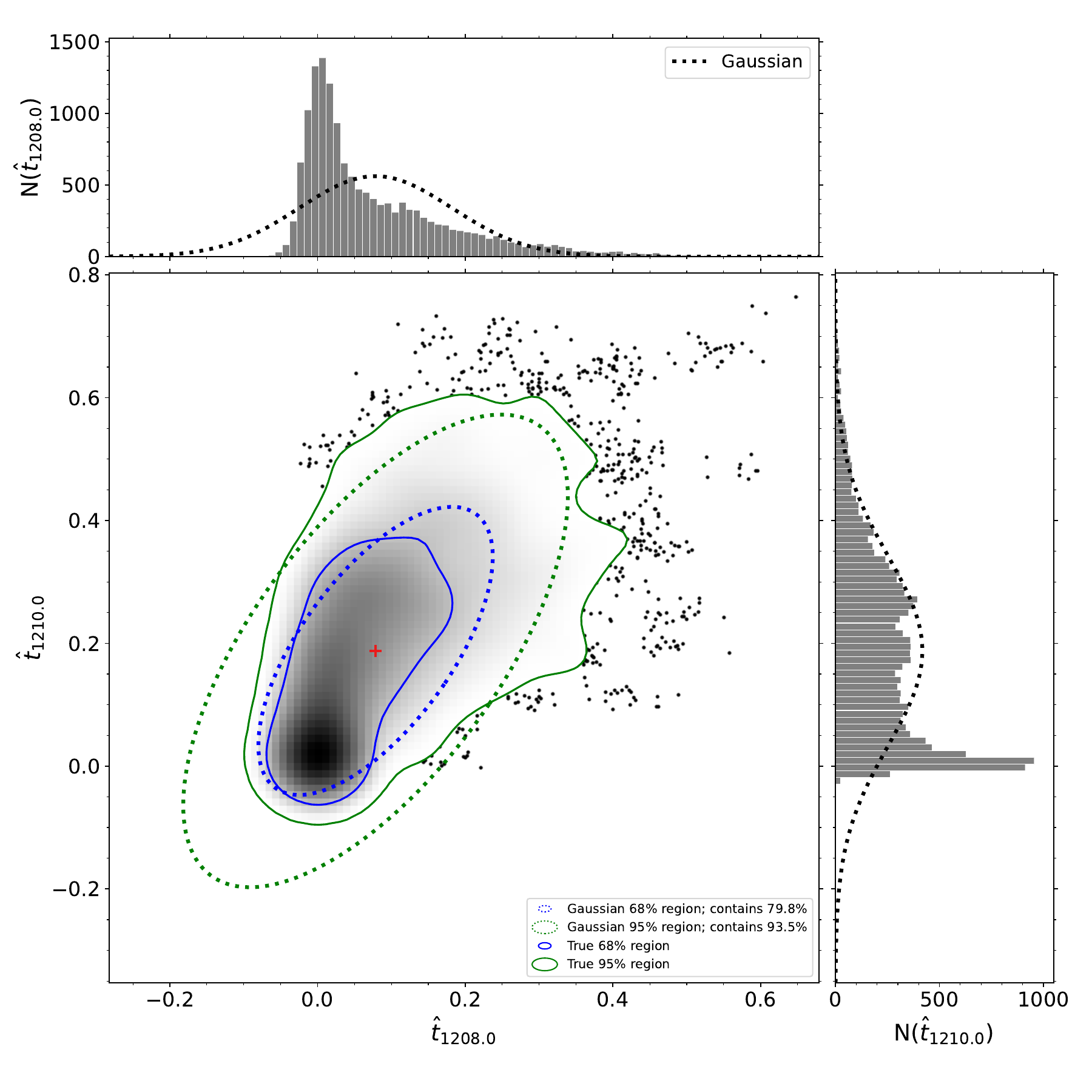}
  \vskip -0.3cm
  \caption{{\emph Left:} Two dimensional slice of the multivariate distribution of the pseudo transmission field  ${\hat t}_\lambda$
    for two rest-frame wavelengths $\lambda = 1214.0\,$\AA\, ($x$-axis) and $\lambda = 1218.1\,$\AA\, ($y$-axis) in the quasar proximity zone,
    corresponding to mean transmission values of $\langle t_{1214.0}\rangle = 0.56$ and $\langle t_{1218.1}\rangle = 0.85$,  which is indicated by the
    red cross. The model considered has $\langle x_{\rm HI}\rangle =
    0.50$ and $\logtq = 6$. Shading indicates the probability density with a logarithmic stretch. 
    Solid lines show the 68\% (blue) and 95\% (green) probability density contours determined
    from the distribution of $\bthat$ samples, whereas the dotted lines show the corresponding probability density contours for the Gaussian distribution in
    eqn.~(\ref{eqn:contour}). The legend indicates the percentage of the samples that are within the Gaussian probability density contours. Histograms show
    the one dimensional marginal distributions where black dotted lines show the Gaussian prediction based on  eqn.~(\ref{eqn:contour}). 
    {\emph Right:} Same as left except for a different set of rest-frame wavelengths $\lambda = 1208.0\,$\AA\, ($x$-axis) and $\lambda = 1210.0\,$\AA\, (y-axis),
     corresponding to lower values of the mean transmission $\langle t_{1208.0}\rangle = 0.08$ and $\langle t_{1210.0}\rangle = 0.19$. 
    \label{fig:that_dist}}
\end{figure*}

\subsection{Understanding the Poor Coverage}
\label{sec:poor_coverage}

The overconfidence of our inference arises from the approximate form of
the analytic likelihood adopted in
eqn.~(\ref{eqn:lhood_final}). Specifically, the problematic
approximation was the assumption of a Gaussian form for the Ly$\alpha$
forest transmission PDF, $P(\bt| \btheta)$, in
eqn.~(\ref{eqn:gaussT}). We have explicitly demonstrated this by generating
mocks where the simulated $\bt$ transmission skewers are repalaced
with samples from the multivariate normal distribution in
eqn.~(\ref{eqn:gaussT}). We carried out a coverage test on $N=100$
such \emph{Gaussianized mocks} for the same set of mock quasar
continua, and find that $C(\alpha) = \alpha$ within the binomially distributed counting
errors, indicating that the assumption of a Gaussian form for $P(\bt| \btheta)$ is
responsible for the failure of the coverage test.

To better visualize
how this non-Gaussianity manifests, one could compare the actual
distribution of $\bt$ from the simulated proximity zone skewers to the
Gaussian adopted in eqn.~(\ref{eqn:gaussT}). However, this would not
be the entire story since the observable is not the noiseless
transmission $\bt$, but rather the flux $\bflam$. Notwithstanding the
non-Gaussianity of $P(\bt| \btheta)$, the resulting distribution of
$\bflam$ could still be close to the Gaussian adopted in
eqn.~(\ref{eqn:lhood_final}), because convolution with the Gaussian
spectral noise (the $\bSigma$ term in eqn.~(\ref{eqn:lhood_final}))
and the very nearly Gaussian (see right panel of
Fig.~\ref{fig:resid_hist}) relative reconstruction errors (the
$\langle \bT \rangle \bC_{\bs} \langle \bT\rangle$ term in
eqn.~(\ref{eqn:lhood_final})) could nevertheless
Gaussianize the
distribution of $\bflam$.

However, it is easier to visualize the distribution of a continuum normalized quantity than the flux $\bflam$, and this would allow us to aggregate mocks with different intrinsic continua $\bs$ on a single plot. Motivated by this, we consider the quantity $\bthat \equiv \bflam \ccirc \langle \bs(\boeta_{\rm true})\rangle^{-1}$, which we will refer to as the pseudo transmission. It is akin to the real transmission $\bt$, but the flux is divided by  $\langle \bs(\boeta_{\rm true})\rangle$ instead of the true continuum. Here $\boeta_{\rm true}$ is the latent variable obtained by fitting the DR model to the mock continuum with no noise or IGM absorption as discussed in \S~\ref{sec:DRcompare}. 
Recalling that $\langle \bs(\boeta)\rangle = \bsDR(\boeta)\ccirc(\mathbf{1} + \langle \bdelta \rangle)$, 
we see that $\bthat$ is the flux normalized by the product of  our best-fit DR estimate to the continuum, $\bsDR(\boeta_{\rm true})$, and the average bias $(\mathbf{1} + \langle \bdelta \rangle)$ of these
estimates. Thus the distribution of $\bthat$ will include scatter arising from both spectral noise and the imperfection of the DR continuum model.
We can recast the likelihood in eqn.~(\ref{eqn:lhood_final}) as a PDF for $\bthat$, which gives
\be
P\left(\bthat | \bsigma, \btheta, \boeta)\right) = \mathcal{N}\left(\bthat; \langle \bt\rangle, \langle \bS\rangle^{-1}\bSigma\langle\bS\rangle^{-1} + \bC_{\bt}  +  \frac{\langle \bT\rangle}{\mathbf{1} +\langle \bdelta \rangle}\bDelta\frac{\langle \bT\rangle}{\mathbf{1} +\langle \bdelta \rangle}\right)
\label{eqn:pre_contour}, 
\ee
where we recall that  $\langle\bS\rangle \equiv \diag{\langle \bs(\boeta_{\rm true})\rangle}$ and hence $\langle \bS\rangle^{-1} \equiv \diag{\langle \mathbf{1}\slash \bs(\boeta_{\rm true})\rangle}$, and where we define $\frac{\langle \bT\rangle}{\mathbf{1} +\langle \bdelta \rangle} \equiv {\rm diag} \left(\frac{\langle \bt\rangle}{\mathbf{1} +\langle \bdelta \rangle}\right)$, where as
before the division of one vector by another is performed elementwise. 

To further simplify the expression in eqn.~(\ref{eqn:pre_contour}), we assume the spectral noise vector can
be written as $\bsigma = \langle \bs(\boeta_{\rm true})\rangle \slash \mathrm{\bf snr}$, where $\mathrm{\bf snr}$ is a vector of spectral
signal-to-noise ratio values and again division of two vectors is performed element wise. By choosing the noise to be proportional to the continuum level $\langle \bs(\boeta_{\rm true})\rangle$, the dependence on
$\langle \bS\rangle$ cancels out of the $\langle \bS\rangle^{-1}\bSigma\langle\bS\rangle^{-1}$ term of the covariance in eqn.~(\ref{eqn:pre_contour}),
yielding an expression that is independent of $\bs$ and $\boeta$
\be
 P\left(\bthat | \bsigma, \btheta\right)   =  \mathcal{N}\left(\bthat; \langle \bt\rangle, \mathrm{\bf SNR}^{-2} + \bC_{\bt}  +  \frac{\langle \bT\rangle}{\mathbf{1} +\langle \bdelta \rangle}  \bDelta  \frac{\langle \bT\rangle}{\mathbf{1} +\langle \bdelta \rangle}\right)\label{eqn:contour}, 
 \ee
 where $\mathrm{\bf SNR}^{-2} \equiv \diag{\mathbf{1}\slash \mathrm{\bf snr}^2}$. This expression 
allows us to express the distribution of $\bthat$ generated from many different mock spectra (with different $\bs$) via a single PDF, which can be easily visualized. The form of the PDF in eqn.~(\ref{eqn:contour}) is intuitive -- as expected
$\bthat$ is distributed about the mean IGM transmission $\langle \bt\rangle$, and the total covariance is a sum of three matrices, the first  $\mathrm{\bf SNR}^{-2}$  quantifying  spectral noise, the second $\bC_{\bt}$ quantifying fluctuations of the IGM transmission field $\bt$, and the third quantifying
continuum reconstruction errors.

To generate realizations of $\bthat$ to compare to the Gaussian
distribution in eqn.~(\ref{eqn:contour}), we start by generating mock
IGM damping wing spectra $\bflam$ using the procedure described in
\S~\ref{sec:mocks} for a model with $\langle x_{\rm HI}\rangle =
0.50$ and $\logtq = 6$. The continuum signal-to-noise ratio of these mocks
is $\bs \slash \bsigma$  (where $\bsigma$ is
the noise from our simulator described in \S~\ref{sec:mocks}). We then
set $\mathrm{\bf snr}$ to be the median value (i.e. median taken over
the quasar dimension), thus adopting a single but representative value
of the spectral signal-to-noise ratio as a function of wavelength.  We then regenerated the mocks
with the noise $\bsigma = \langle \bs(\boeta_{\rm true})\rangle \slash
\mathrm{\bf snr}$, with $\mathrm{\bf snr}$ set to this median value, and divided the fluxes $\bflam$ by $\langle
\bs(\boeta_{\rm true})\rangle$,  yielding realizations of $\bthat$.
Whereas throughout
this work we restricted to the 778 test set
autofit continua for constructing mock IGM damping wing spectra, here for the
sole purpose of visualization, we instead use the larger training set of
14,781 autofit continua to increase the number of samples\footnote{As
we only have 1183 IGM transmission skewers for each parameter value
$\btheta$, we randomly assigned the autofit continua to an IGM
transmission skewer with replacement.}.

To determine how well 
the PDF in eqn.~(\ref{eqn:contour})
approximates the true distribution of $\bthat$, we plot slices of this
distribution in the 2D ${\hat t}_{\lambda}-{\hat t}_{\lambda^\prime}$
plane, where $\lambda$ and $\lambda^\prime$ are the wavelengths of two different
spectral pixels.  Examples for two distinct pairs of transmission values are
shown in Fig.~\ref{fig:that_dist}.  The left panel of
Fig.~\ref{fig:that_dist} indicates that at rest-frame wavelengths where the
mean proximity zone transmission is high, $\langle t_\lambda\rangle \sim
0.5-0.9$,  the Gaussian approximation is decent, as indicated by: 1) the
similarity of the 2D probability density contours for the
${\hat t}_{\lambda}-{\hat t}_{\lambda^\prime}$ samples (solid lines) and the
contours of the approximate analytical Gaussian PDF 
(dotted lines; eqn.~\ref{eqn:contour}), 2) the same
comparison for the 1D marginal distribution (i.e. gray marginal histograms
compared to black dotted lines for the Gaussian PDF), and 3) 
the fact that the percentage
of samples (indicated in the legend) within the 68\% (71.2\%; dotted blue) and 95\% (95.4\%; dotted green) analytical Gaussian probability density contours are very close to the expected values for each contour.

However, the right panel of Fig.~\ref{fig:that_dist} indicates that
for bluer wavelengths where the mean proximity zone transmission is
lower, $\langle t_\lambda\rangle \sim 0.1-0.2$, the Gaussian 2D contours
and 1D marginal PDFs poorly approximate the distribution of the ${\hat
  t}_{\lambda}-{\hat t}_{\lambda^\prime}$ samples. Whereas the Gaussian
PDF is symmetric about the mean $\langle t_\lambda\rangle$, the
samples are significantly skewed to positive transmission values.
This disagreement is easily understood -- whereas a Gaussian centered
at a low $\langle t_\lambda\rangle\sim 0.1-0.2$ must be symmetric by
construction and thus predicts a significant probability for negative
values, the transmission is an inherently positive
quantity. Besides the positivity constraint,  additional skewness of the
$t_\lambda$ field PDF results from the
reionization topology and density fluctuations in the IGM.  At intermediate and low values of the average IGM
neutral fraction, $\langle x_{\rm HI}\rangle\lesssim 0.5$, the
distribution of distances from quasar host halos to the first patch of
neutral gas exhibits a strong tail to large distances (see Fig.~2
of \cite{Davies18a}). The corresponding reionized regions near the quasar
will be further photoionized by the quasar's radiation resulting in 
transmissive proximity zone regions. Thus tails in the distance to the nearest neutral patch, 
combined with the strong tails in the in Ly$\alpha$ forest transmission PDF at the relevant optical depths
\citep[see e.g.][]{Davies18ABC}, manifest as a tail towards high
transmission, $t_\lambda$, for bluer proximity zone wavelengths where
the average profile has $\langle t_\lambda\rangle \sim 0.1-0.2$.  These heavy positive
tails bias the Gaussian transmission covariance, $C_{\bt}$ high, relative
to the width of the core of the distribution of samples (see marginal
histograms in the right panel of Fig.~\ref{fig:that_dist}). As a result  
the percentage of samples within the 68\% Gaussian
contour (79.8\%; dotted blue) deviates significantly from the Gaussian
expectation.

We conclude that despite the Gaussianizing effects of Gaussian spectral noise and approximately Gaussian continuum reconstruction errors, our approximate Gaussian form  for the likelihood in eqn.~(\ref{eqn:lhood_final}) is nevertheless a poor approximation 
at low average transmission values  $\langle t_\lambda\rangle\sim 0.1-0.2$ because of the strong underlying non-Gaussianity of the IGM transmission PDF $P(\bt| \btheta)$.

\begin{figure*}
  \vskip -0.4cm
  \includegraphics[trim=0 0 0 0,clip,width=0.99\textwidth]{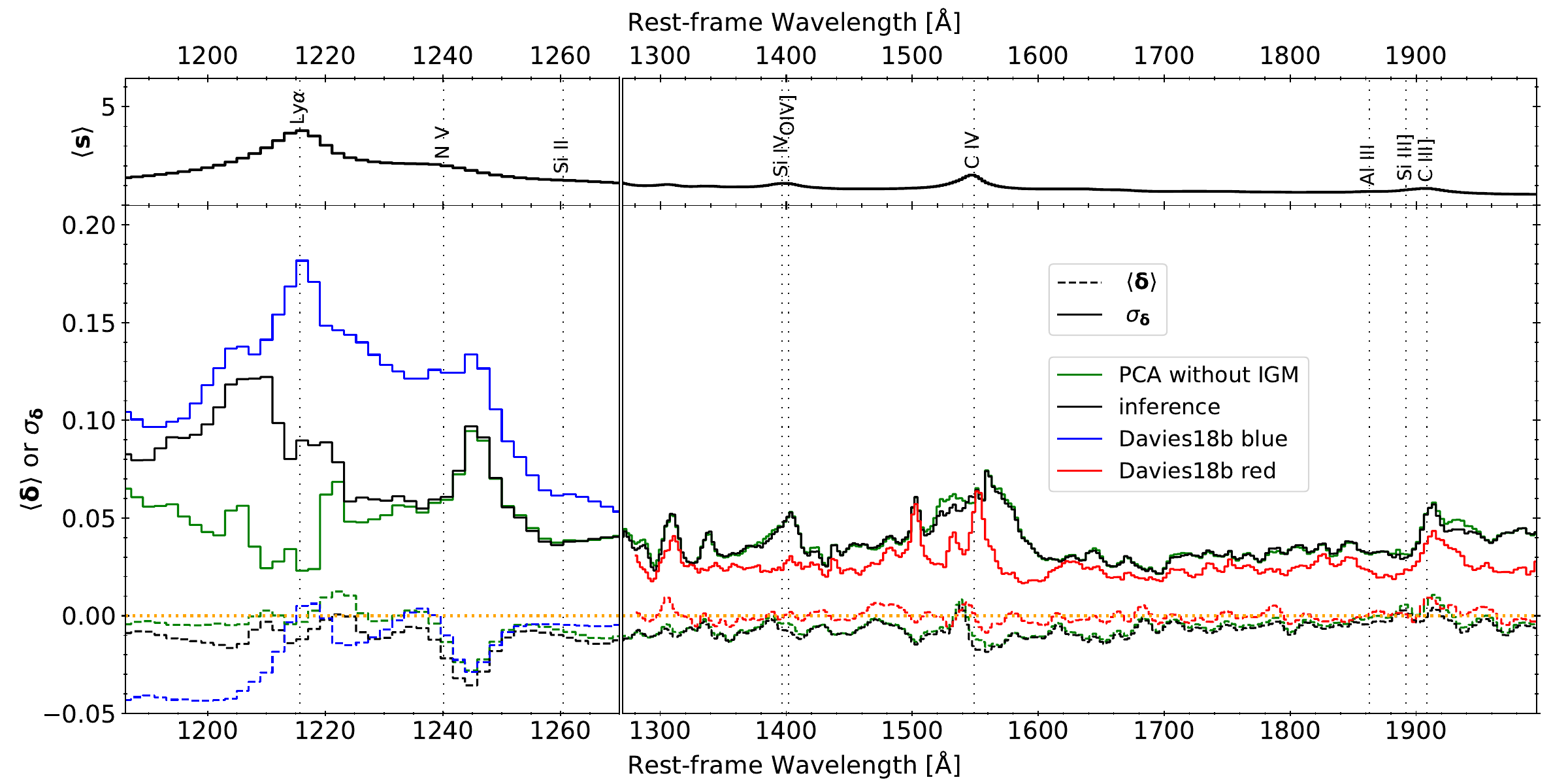}
  \vskip -0.3cm
  \caption{Efficacy of quasar continuum reconstruction algorithms as quantified by the 
  moments of the relative reconstruction error $\bdelta$. 
  The lower panels show the mean $\langle \bdelta\rangle$ (dashed) and standard deviation $\sigma_{\bdelta} \equiv
\Delta_{\lambda\lambda}^{1\slash 2}$  (solid) of the
  relative reconstruction error evaluated from $N=100$ mock spectra on which we performed statistical inference (see  Fig.~\ref{fig:mock_example} and \S~\ref{sec:cov_test_results}).  Black shows these moments evaluated using the \emph{inferred quasar continuum} (see eqn.~(\ref{eqn:delta_inf})) that results from our joint statistical inference of the PCA coefficients and the astrophysical parameters. Green shows the moments arising from the intrinsic imperfections of the DR alone, i.e. for the case where we fit the PCA basis to the same set of spectra over the entire spectral range with no noise or IGM absorption
   (see eqn.~(\ref{eqn:delta_noigm})).  For   $\lambda > 1220\,$\AA, the inferred quasar continua (black) achieve the intrinsic limiting precision of the DR itself (green, i.e. PCA fits without noise or IGM absorption), whereas at bluer wavelengths degeneracy with and censorship by IGM absorption is seen to increase the errors as quantified by $\sigma_{\bdelta}$.  For comparison, we show the moments of the relative reconstruction error of the red-side continuum fit (red, right) and the blue-side prediction (blue, left) obtained by applying the \citet{Davies18b} red-blue prediction algorithm to the same set of $N=100$ mock spectra. Over the wavelength range $1216\,$\AA~$< \lambda < 1240\,${\rm \AA} most critical for measuring IGM damping wings, our inference  significantly outperforms the \citet{Davies18b} algorithm, yielding $\sigma_{\bdelta}$ a factor of 1.7-2.5 lower, with an average reduction of $\sigma_{\bdelta}$ over this range of a factor of $2.1$. 
  For reference, the upper panel shows the mean quasar spectrum $\langle \bs\rangle$ constructed from our test set with prominent emission lines labeled. \label{fig:inf_DR_resid}}
\end{figure*}

\subsection{Continuum Reconstruction Recovery and Comparison to Previous Work}

The main advantage of the statistical inference method at the heart of this paper is that it constructs a generative model for the entire quasar spectrum, including absorbed pixels in the proximity zone, to perform parameter inference. This contrasts with the red-blue prediction approach that has been adopted in most past work \citep[e.g.][]{Davies18b,Dominika20,Fathi20,Reiman20,Chen22,Greig24b} modeling IGM damping wings, whereby only pixels redward of some cutoff (typically $\lambda > 1280$\AA) are used to predict the blue-side ($\lambda < 1280$\AA) continuum and its error, which are then used to perform inference. Our approach is generically expected to perform better, given that it uses all of the information available, and specifically the spectral pixels $\lambda  < 1280$\AA\, where much of the information about the intrinsic quasar continuum shape around Ly$\alpha$, and all of the information about the IGM damping wing are manifest. However, quantifying the improvement in precision on the astrophysical parameters $\langle x_{\rm HI}\rangle$ and $\logtq$ and directly comparing to previous work would be challenging, given the heterogeneity of the different modeling approaches that have been employed to date.  Furthermore, such a comparison might not be fair given that the current study is the only one to  investigate (and explicitly correct for) the coverage of the inference: if the inference pipelines used in past work were overconfident, a comparison to our results would not be a fair one.  

In our companion paper  \citet{Kist25a} we quantify the precision on the astrophysical parameters yielded by our inference procedure in 
detail by analyzing an ensemble of 1000
mock spectra spanning the full astrophysical parameter range ($0.0 \le \langle x_{\rm HI}\rangle \le 1.0\,$; $3 \le \logtq \le 8)$.  We find that \citep[see Table~1 and Fig.~9 of][]{Kist25a} 
even for the hypothetical case of perfect knowledge of the quasar continuum, the median $\sim 1\sigma$ precision on $\langle x_{\rm HI}\rangle$ 
is $\sim 15\%$ and the median precision on $\logtq$ is $\sim 0.55$ dex. For inference on full mocks generated according to the procedure in \S~\ref{sec:mocks}, where we model and marginalize over the unknown  quasar continuum, the typical measurement error increases to  $\sim 28\%$ on $\langle x_{\rm HI}\rangle$ and $\sim 0.80$\,dex for $\logtq$. This analysis suggests that continuum reconstruction errors contribute roughly an equal amount to the error budget as the other sources of stochasticity in the problem, namely the distribution of distances to the to the nearest patch of neutral hydrogen \citep[see Fig.~2 of][]{Davies18a} and fluctuations in the location of the quasar ionization front due to the distribution of sinks along the sightline.  Thus significant gains in astrophysical parameter precision should result from an algorithm that better reconstructs the quasar continuum. 

Motivated by this, we perform a careful comparison of the precision of our continuum reconstructions to the PCA based red-blue continuum prediction method introduced in \citet{Davies18b}.
While this is only one of the many algorithms that have been 
used for quasar continuum reconstruction to date, \citet{Greig24b} recently conducted a detailed comparison 
of  all ten of the 
quasar continuum prediction pipelines in existence, and found that while they all yield roughly comparable precision, the \citet{Davies18b} pipeline consistently performed among the best for the various samples and metrics considered. Hence comparing the performance of our continuum reconstruction approach to the \citet{Davies18b} algorithm should, broadly speaking, constitute a sufficient comparison to the diverse set of existing continuum prediction algorithms in existence. 

To perform this comparison we considered the $N=100$ mock quasar spectra on the $\mathrm{d}v = 500~\kms$ wavelength grid that we conducted inference on (see Fig.~\ref{fig:mock_example} for examples) to arrive at the coverage test results described in \S~\ref{sec:cov_test_results} and shown in Fig.~\ref{fig:marginal_coverage}. We define the \emph{inferred quasar continuum} for a given mock spectrum to be the weighted median (using the marginal coverage weights) of the PCA DR models, $\bsDR(\boeta_i)$, evaluated at each of the HMC samples $\boeta_i$ for each mock spectrum, which we denote by $\bsDRmed$. However,  the DR model itself is not perfect and will result in some relative reconstruction error, $\bdelta$, even if the PCA is fit to a spectrum with no noise or IGM absorption (see \S~\ref{sec:DRcompare} and Fig.~\ref{fig:DR_resid}). Thus a proper analysis requires comparing the moments of the relative reconstruction error of  the inferred continua
\be
\bdelta_{\rm inf} = \frac{\bs - \bsDRmed}{\bs} \label{eqn:delta_inf},
\ee
to 
the moments of the relative reconstruction error
\be
\bdelta = \frac{\bs - \bsDR(\boeta_{\rm true})}{\bs} \label{eqn:delta_noigm},
\ee
obtained by fitting the PCA to the same mock quasar spectra, but without noise or IGM absorption. As described in \S~\ref{sec:DRcompare}, we determine the best-fit PCA DR parameters, $\boeta_{\rm true}$, by minimizing the 
MSE loss given in eqn.~(\ref{eqn:MSE}), and Fig.~\ref{fig:DR_resid} shows the moments of the relative reconstruction error, $\langle \bdelta \rangle$ and  $\sigma_{\bdelta} \equiv  \Delta_{\lambda\lambda}^{1\slash 2}$ (see eqn.~(\ref{eqn:covardelta})), 
computed from the spectra in our continuum test sets.  In 
the current context,  we will compare the moments
of $\bdelta_{\rm inf}$ directly to the moments of $\bdelta$, both computed from the
same 100 mock spectra for which we performed statistical inference. Since the DR model always results in reconstruction errors (even in the absence of noise and IGM absorption),  the best continuum reconstruction (in the presence of noise and IGM absorption) would be one for which $\bdelta_{\rm inf}$ is very close to $\bdelta$.  

To compare the accuracy of our continuum reconstructions to those from the \citet{Davies18b} method,  we applied 
their red-blue prediction algorithm to the same set of 100 mock spectra. Specifically, we followed the \citet{Davies18b} approach and fit their  red-side PCA vectors  to the red spectral pixels ($\lambda < 1280$\AA) of our mocks yielding ten  PCA coefficients for each mock, and then we used their transformation matrix to transform each set of ten red-side PCA coefficients into six blue-side PCA coefficients, which finally yields a predicted blue side continuum ($\lambda < 1280$\AA) for each mock. 

The efficacy of the quasar continuum reconstruction algorithms is illustrated in Fig.~\ref{fig:inf_DR_resid}. In the lower panel, the dashed lines show the mean $\langle \bdelta\rangle$ of the relative reconstruction errors, whereas the solid lines show the standard deviation $\sigma_{\bdelta} \equiv \Delta_{\lambda\lambda}^{1\slash 2}$ (see eqn.~(\ref{eqn:covardelta})).  Black shows these moments evaluated using the \emph{inferred quasar continuum}  that results from our joint statistical inference of the PCA coefficients and the astrophysical parameters. Green shows the moments arising from the intrinsic imperfections of the DR, i.e. for the case where we fit our PCA basis (see Fig.~\ref{fig:pca_vectors}) to the same set of spectra over the entire spectral range with no noise or IGM absorption.  For   $\lambda \gtrsim 1220\,$\AA,
the inferred quasar continua (black) achieve the intrinsic limiting precision of the DR itself (green, i.e. PCA fits without noise or IGM absorption), whereas at bluer wavelengths, degeneracy with and censorship by IGM absorption is seen to increase the errors as quantified by $\sigma_{\bdelta}$.  For comparison, we show the moments of the relative reconstruction error of the red-side continuum fit (red, right) and the blue-side prediction (blue, left) obtained by applying the \citet{Davies18b} red-blue prediction algorithm to the same set of $N=100$ mock spectra. Over the wavelength range $1216\,$\AA~$< \lambda < 1240\,$\AA\, most critical for measuring IGM damping wings, our inference  significantly outperforms the \citet{Davies18b} algorithm, yielding $\sigma_{\bdelta}$ a factor of 1.7-2.5 lower, with an average reduction of a factor of $2.1$ taken over this wavelength range.

To appreciate the significance of this reduction, consider that for a model with $\langle x_{\rm HI}\rangle = 1.0$ and $\logtq=6$ the mean IGM transmission varies from  $\langle \bt\rangle = 0.49-0.91$ over this same range $1216\,$\AA~$< \lambda < 1240\,$\AA.  Heuristically, the signal-to-noise ratio of IGM damping wing absorption for a single spectral pixel is ${\rm S\slash N}\sim (1-\langle \bt\rangle)\slash \sigma_{\bdelta}$. Averaging this quantity over the range $1216$\,\AA $\,< \lambda < 1240$\,\AA, we find that our inference yields an average statistical significance of 2.9 to be compared to 1.4 for the \citet{Davies18b} red-blue prediction algorithm. This heuristic ${\rm S\slash N}$ likely underestimates the actual improvement, since one optimally combines all the spectral pixels near Ly$\alpha$ 
when performing a fit, although  correlations of the transmission field $\langle \bt\rangle$ and the covariance of the relative reconstruction error $\bdelta$ make it difficult to quantify the improvement more rigorously. But it is abundantly clear that our approach of jointly fitting for the astrophysical parameters governing the IGM absorption and the latent variables that describe the continuum yields far more accurate continuum reconstructions than red-blue prediction.   

\section{Summary and Conclusions}
\label{sec:summary}

In this paper we introduced a new approach for analyzing the IGM damping wings that are imprinted on the proximity zones of EoR quasars.  Whereas past work has typically forgone the additional constraining power afforded by the blue side continuum  ($1216$\AA~$\lesssim \lambda \lesssim 1280$\AA) and opted not to model the large correlated IGM transmission fluctuations in the proximity zone
($\lambda \lesssim 1216$\AA),  we derived a single Bayesian likelihood for the entire spectrum allowing us to fit all of the spectral pixels and thus jointly model the fluctuating transmission in the proximity zone, the smooth IGM damping wing signature, and the underlying quasar continuum \emph{simultaneously}. The latter constitutes a nuisance stochastic process from the standpoint of constraining the average IGM neutral fraction, $\langle x_{\rm HI}\rangle$, and quasar lifetime, $t_{\rm Q}$, that govern the IGM transmission. A key aspect of our approach is the use of DR to 
describe the quasar continuum with a small number of latent variables and then designate the imperfections of this model, 
which we refer to as relative  reconstruction errors, as a source of irreducible correlated noise.  Using a large sample of $15,559$ SDSS/BOSS quasars at $z \gtrsim 2.15$ we trained and quantified the performance of six distinct  DR methods, including machine learning techniques like GPLVMs and VAEs, and find that a six parameter PCA model 
(five PCA coefficients $\bxi$ plus a normalization $s_{\rm norm}$) performs best \citep[see also][]{Kist25a}, with complex machine learning methods providing no advantage. Fitting this PCA model to a subset of 778 spectra which were unseen by the training process provides an  empirical  calibration of  the relative reconstruction errors, which are an important ingredient of the likelihood we derived.  Following our approach, all sources of error -- the stochasticity induced by the reionizing IGMs ionization topology, the unknown quasar lifetime $t_{\rm Q}$ and location of the corresponding ionization front,  continuum reconstruction errors, and spectral noise -- are accounted for by a generative probabilistic model, which enables us to marginalize out nuisance parameters
in a principled manner. 

The only drawback of the Gaussian likelihood that we derive in this paper is that it is approximate, because the true likelihood is analytically intractable.  We used HMC to conduct statistical inference on an ensemble of 100 realistic mock EoR quasar spectra to determine the \emph{coverage} of our inference with this approximate likelihood, which quantifies the validity of the credibility contours that we obtain for  $\langle x_{\rm HI}\rangle$ and $\logtq$ from this new method. We find that our inference is overconfident, which is to say that the 68\% credibility contour contains the true astrophysical parameters ($\langle x_{\rm HI}\rangle$ and $\logtq$) just 48\% of the time, and the 95\% credibility contour contains the true parameters just 79\% of the time (see Fig.~\ref{fig:marginal_coverage}).  We show convincingly that this overconfidence results from the non-Gaussianity of the IGM transmission field, $\bt$, at proximity zone pixels $\lambda$ where the mean transmission takes on low values,  $\langle t_\lambda\rangle \sim 0.1-0.2$, owing both to the fact that, physically, the transmission must be positive, and because of an intrinsic strong tail towards higher transmission values (see Fig.~\ref{fig:that_dist}). Although the HMC posterior samples from our approximate likelihood yield biased inference, we introduced a procedure whereby the HMC samples can be assigned weights, such that the reweighted samples produce reliable inference which passes a coverage test by construction. This reweighting procedure, which amounts to a small dilation of the credibility contours of the original inference, finally yields a state-of-the-art Bayesian inference pipeline that uses all of the spectral pixels to reliably measure the cosmic reionization history and quasar lifetime from quasar spectra. 

The accuracy of the quasar continuum reconstructions afforded by this new method are unprecedented. For  $\lambda > 1220\,$\AA, we find that our inferred quasar continua  achieve the intrinsic limiting precision of the DR model itself, in other words, they are as good as fits to `perfect' spectra without noise or IGM absorption.  At bluer wavelengths, as expected, degeneracy with and censorship by IGM absorption degrades our ability to reconstruct the underlying quasar continuum. But we nevertheless achieve far more accurate reconstructions than the red-blue prediction algorithms that have been adopted in previous IGM damping wing measurements.  Over the wavelength range $1216\,$\AA~$< \lambda < 1240\,${\rm \AA} most critical for such measurements, our continuum reconstructions have a factor of 1.7-2.5 smaller error than red-blue prediction, which increases the statistical significance of a putative IGM damping wing per spectral pixel to $\sim 2.9$ compared to $\sim 1.4$ for red-blue prediction. 

In our companion paper \citet{Kist25a}, we quantify the precision with which IGM damping wings analyzed with this new inference
approach can measure  the astrophysical parameters, $\langle x_{\rm HI}\rangle$ and $\logtq$, and the dependence of this precision on the location in  parameter space, the  dimensionality of the DR latent variable model, as well as on the spectral resolution, signal-to-noise ratio, and spectral coverage of the quasar spectra that are analyzed. After performing a battery of tests on 1000 mocks, \citet{Kist25a} find that the precision is highest when running this new pipeline with a six-parameter DR model   (five PCA coefficients $\bxi$ plus a normalization $s_{\rm norm}$) on $\mathrm{S}/\mathrm{N} \sim 10$ spectra, rebinned to a $\sim 500\,\mathrm{km}/\mathrm{s}$ velocity pixel scale, and extending at least out to the \ion{C}{IV} $\lambda 1549\,$\AA\, emission line.  With this configuration, \citet{Kist25a} find that a single EoR quasar spectrum constrains the IGM neutral fraction, $\langle x_{\rm HI}\rangle$,  to $28.0_{-8.8}^{\hspace{.056em}+\hspace{.056em}8.2}\,\%$ and the quasar lifetime, $\logtq$, to $0.80_{-0.55}^{\hspace{.056em}+\hspace{.056em}0.22}\,\mathrm{dex}$,
where the error bars indicate the 16 and 84 percentile ranges, and 
where the constraints improve on both parameters for spectra with a stronger IGM damping wing signature.   

Higher precision constraints on $\langle x_{\rm HI}(z)\rangle$  can of course be achieved by averaging over statistical samples of EoR quasars samples. An ambitious program to obtain sensitive JWST spectra of the sample of hundreds of EoR quasars that will be delivered by the ESA/Euclid satellite would revolutionize the study  of IGM damping wings towards quasars  and constrain the cosmic reionization history to unprecedented precision. Averaging over $\sim 30$ quasars in a redshift bin would deliver a precision of $\sim 5\%$ on $\langle x_{\rm HI}\rangle$ at that redshift, which, when performed as a function of redshift
across the EoR would measure $\langle x_{\rm HI}(z)\rangle$ far more precisely than the CMB. Furthermore, such an analysis would 
also yield, as a byproduct, the distribution of quasar lifetimes  \citep[see e.g.][]{Khrykin21} providing novel constraints on the buildup of supermassive black holes (SMBHs) in the young Universe. 

Given that the formal precision achievable on $\langle x_{\rm HI}(z)\rangle$ is so high, a natural question arises:
will modeling uncertainties due to poorly understood galaxy formation physics eventually limit the precision 
with which we can constrain reionization? How sensitive are our IGM transmission models (see \S~\ref{sec:sims}) to galaxy formation, which regulates both the ionizing photon sources (via Lyman continuum escape) and sinks (via Lyman limit systems) that determine the reionization topology and the size of the ionized bubble powered by the quasar itself? While galaxy formation physics ultimately determines the reionization topology, we emphasize that  precision constraints on $\langle x_{\rm HI}(z)\rangle$ do not require that this topology can be predicted from first principles. Specifically, we adopted a parameterized semi-numerical \texttt{21cmFast} model 
\citep{Mesinger11,Davies22}  in which the source and sink prescriptions are governed by a handful of tunable sub-grid parameters. Although we fix these parameters in the present study, yielding a fiducial 
reionization topology as a function of a single parameter, $\langle x_{\rm HI}(z)\rangle$, 
an important direction for future work would be to vary and marginalize over the full-suite of sub-grid source/sink parameters to assess their impact on the inferred $\langle x_{\rm HI}(z)\rangle$ constraints. Furthermore, as discussed below, it is likely that IGM damping wing measurements have the potential to contstrain the reionization topology (or the sub-grid parameters that govern it) as well (Kist et al., in preparation).

Along similar lines, while our 1D radiative transfer of the quasar's radiation currently neglects dense absorbers in the quasar environment (Lyman limit systems and DLAs), which are not captured by the Nyx hydrodynamical simulations, such absorbers can be directly identified in high-${\rm S\slash N}$ quasar spectra via associated metal-line absorption systems  \citep[e.g.][]{DaviesPisco23} and excluded from analysis -- analogous to how  "gold samples" are selected in supernova cosmology.  Moreover, if optically thick absorbers introduce additional opacity in quasar proximity zones, this can be empirically tested using existing high-resolution spectra of $z\sim 5-6$ quasars (for which the global IGM is expected to be highly ionized), for example by comparing the flux PDF of their proximity zones to our models. Any residual disagreement could be addressed by introducing a simple sub-grid opacity parameter to govern Lyman limit systems in the 1D radiative transfer modeling \citep[e.g.][]{Khrykin15,Davies16a}. Thus to summarize, while the impact of uncertain galaxy formation physics  on the reionization topology and the presence of dense absorbers in the quasar environment are important considerations that may require adding additional nuisance parameters, they do not constitute a fundamental limitation of our approach. Precision constraints on reionization do not rest upon a full {\it ab initio} solution to galaxy formation. Rather, the impact of galaxy formation physics can 
be treated via a small set of empirically calibratable and marginalizable modeling uncertainties. This situation is closely analogous to precision weak lensing analyses, where the impact of baryonic physics on the matter power spectrum is accounted for using parametric transfer functions derived from hydrodynamical simulations, which are then marginalized over to recover unbiased cosmological constraints \citep[e.g.][]{KIDS21,DES22,Schaller24}. We therefore argue that IGM damping wings towards quasars provide a viable and powerful method for precision constraints on reionization  and SMBH growth, but modeling and marginalizing over uncertainties
due to galaxy formation constitutes an important direction for future work.

In addition to addressing these modeling uncertainties, there are several other promising avenues for improving upon the results presented here. First and foremost, our paper  argued that an optimal analysis algorithm must construct a fully generative probabilistic model for the entire spectrum, however our likelihood is not optimal because it is only an approximation to the true intractable likelihood. As a result, we had to dilate our credibility contours which degraded the precision of our parameter constraints. Hence, an obvious priority for the future is to attack this problem in the simulation based inference framework \citep[see e.g.][for a review]{Cranmer20} and use machine learning to obtain an expression for the intractable likelihood that we here approximated as a Gaussian \citep[e.g.][]{Chen23b}.  This would surely result in higher precision parameter constraints, both  because  coverage correction would not be needed and because only the true likelihood can achieve the true optimal precision. Finally, our companion paper \citet{Kist25a} finds that roughly half of the error budget on $\langle x_{\rm HI}\rangle$ originates from variations of IGM damping wing strength arising from the stochastic distribution of the line-of-sight neutral column density resulting from the topology of reionization itself \citep{Davies18a}. Several studies have recently noted that IGM damping wing transmission profiles are actually well described by a single parameter \citep{Chen24} which is effectively this line-of-sight neutral column density $N_{\rm HI}$ \citep[][Kist et al., in preparation]{Keating24}.  It follows that models of IGM damping wings can be parameterized in two distinct ways --- either one elects to measure the volume averaged neutral fraction $\langle x_{\rm HI}\rangle$, which in turn governs the stochastic distribution of the line-of-sight $N_{\rm HI}$ (via an assumed reionization topology),  or one can measure an $N_{\rm HI}$ \emph{for each quasar individually}, and use ensembles of quasars to map out the distribution of $N_{\rm HI}$ empirically as a function of redshift.  Whereas this paper adopted the former formulation, an upcoming study explores the latter using the inference machinery that we developed here (Kist et al., in preparation).  
The great advantage of this latter approach is that it opens up the 
exciting possibility of using ensembles of quasars to actually determine the distribution of a $\sim 100~{\rm cMpc}$ \emph{one dimensional moment} through the Universe's $n_{\rm HI}$ field, which would not only measure the Universe's reionization history, $\langle x_{\rm HI}(z)\rangle$, but also possibly constrain its topology. These are the primary objectives of cosmological studies of reionization in general and 21cm experiments in particular, and the methodology that we have presented here paves the 
way for achieving them with EoR quasar spectra.

%The extraordinary potential of IGM damping wings towards quasars as a probe of reionization and SMBH growth motivates  future studies that could improve upon the results presented here, and there are several interesting directions for future work.  

\section*{Acknowledgements}

JFH acknowledges helpful discussions with Daniel Foreman-Mackey about Hamiltonian Monte Carlo, Elena Sellentin about coverage tests, and Anna-Christina Eilers about GPLVMs. The authors 
also wish to thank Elia Pizzati, Silvia Onorato, and Linda Jin for comments on an early version  of the manuscript, and the ENIGMA group at UC Santa Barbara and Leiden University for valuable feedback. Finally, the authors are grateful to the anonymous referee for their valuable comments and recommendations,  and for identifying a mathetmical mistake in the original version, all of which impproved the clarity and presetation of the manuscript.
This work made use of \texttt{NumPy} \citep{Numpy2020},  \texttt{SciPy} \citep{Scipy2020}, \texttt{JAX} \citep{Jax2018}, \texttt{NumPyro} \citep{NumPyro2019a, NumPyro2019b}, \texttt{sklearn} \citep{sklearn2011}, \texttt{Astropy} \citep{Astropy2013, Astropy2018, Astropy2022}, \texttt{SkyCalc\_ipy} \citep{SkyCalc_ipy2021}, \texttt{h5py} \citep{h5py2013}, \texttt{Matplotlib} \citep{Matplotlib2007}, \texttt{corner.py} \citep{Corner2016}, and \texttt{IPython} \citep{IPython2007}, and \texttt{PypeIt} \citep{PypeIt}. Computations were performed using the compute resources from the Academic Leiden Interdisciplinary Cluster Environment (ALICE) provided by Leiden University. TK and JFH acknowledge support from the European Research Council (ERC) under the European Union’s Horizon 2020 research and innovation program (grant agreement No 885301),  and JFH from the National Science Foundation under Grant No. 2307180.

\section*{Data Availibility}

The derived data generated in this research will be shared on
reasonable requests to the corresponding author.

%%%%%%%%%%%%%%%%%%%%%%%%%%%%%%%%%%%%%%%%%%%%%%%%%%


\bibliographystyle{mnras}
\bibliography{allrefs, pyrefs}

\begin{thebibliography}{}
\makeatletter
\relax
\def\mn@urlcharsother{\let\do\@makeother \do\$\do\&\do\#\do\^\do\_\do\%\do\~}
\def\mn@doi{\begingroup\mn@urlcharsother \@ifnextchar [ {\mn@doi@}
  {\mn@doi@[]}}
\def\mn@doi@[#1]#2{\def\@tempa{#1}\ifx\@tempa\@empty \href
  {http://dx.doi.org/#2} {doi:#2}\else \href {http://dx.doi.org/#2} {#1}\fi
  \endgroup}
\def\mn@eprint#1#2{\mn@eprint@#1:#2::\@nil}
\def\mn@eprint@arXiv#1{\href {http://arxiv.org/abs/#1} {{\tt arXiv:#1}}}
\def\mn@eprint@dblp#1{\href {http://dblp.uni-trier.de/rec/bibtex/#1.xml}
  {dblp:#1}}
\def\mn@eprint@#1:#2:#3:#4\@nil{\def\@tempa {#1}\def\@tempb {#2}\def\@tempc
  {#3}\ifx \@tempc \@empty \let \@tempc \@tempb \let \@tempb \@tempa \fi \ifx
  \@tempb \@empty \def\@tempb {arXiv}\fi \@ifundefined
  {mn@eprint@\@tempb}{\@tempb:\@tempc}{\expandafter \expandafter \csname
  mn@eprint@\@tempb\endcsname \expandafter{\@tempc}}}

\bibitem[\protect\citeauthoryear{{Abbott} et~al.,}{{Abbott}
  et~al.}{2022a}]{DESWL22}
{Abbott} T.~M.~C.,  et~al., 2022a, \mn@doi [\prd]
  {10.1103/PhysRevD.105.023520}, \href
  {https://ui.adsabs.harvard.edu/abs/2022PhRvD.105b3520A} {105, 023520}

\bibitem[\protect\citeauthoryear{{Abbott} et~al.,}{{Abbott}
  et~al.}{2022b}]{DES22}
{Abbott} T.~M.~C.,  et~al., 2022b, \mn@doi [\prd]
  {10.1103/PhysRevD.105.023520}, \href
  {https://ui.adsabs.harvard.edu/abs/2022PhRvD.105b3520A} {105, 023520}

\bibitem[\protect\citeauthoryear{{Almgren}, {Bell}, {Lijewski}, {Luki{\'c}}  \&
  {Van Andel}}{{Almgren} et~al.}{2013}]{Almgren13}
{Almgren} A.~S.,  {Bell} J.~B.,  {Lijewski} M.~J.,  {Luki{\'c}} Z.,   {Van
  Andel} E.,  2013, \mn@doi [\apj] {10.1088/0004-637X/765/1/39}, \href
  {http://adsabs.harvard.edu/abs/2013ApJ...765...39A} {765, 39}

\bibitem[\protect\citeauthoryear{{Astropy Collaboration} et~al.,}{{Astropy
  Collaboration} et~al.}{2013}]{Astropy2013}
{Astropy Collaboration} et~al., 2013, \mn@doi [\aap]
  {10.1051/0004-6361/201322068}, \href
  {https://ui.adsabs.harvard.edu/abs/2013A&A...558A..33A} {558, A33}

\bibitem[\protect\citeauthoryear{{Astropy Collaboration} et~al.,}{{Astropy
  Collaboration} et~al.}{2018}]{Astropy2018}
{Astropy Collaboration} et~al., 2018, \mn@doi [\aj] {10.3847/1538-3881/aabc4f},
  \href {https://ui.adsabs.harvard.edu/abs/2018AJ....156..123A} {156, 123}

\bibitem[\protect\citeauthoryear{{Astropy Collaboration} et~al.,}{{Astropy
  Collaboration} et~al.}{2022}]{Astropy2022}
{Astropy Collaboration} et~al., 2022, \mn@doi [\apj]
  {10.3847/1538-4357/ac7c74}, \href
  {https://ui.adsabs.harvard.edu/abs/2022ApJ...935..167A} {935, 167}

\bibitem[\protect\citeauthoryear{{Ba{\~n}ados} et~al.,}{{Ba{\~n}ados}
  et~al.}{2018}]{Banados18}
{Ba{\~n}ados} E.,  et~al., 2018, \mn@doi [\nat] {10.1038/nature25180}, \href
  {https://ui.adsabs.harvard.edu/abs/2018Natur.553..473B} {553, 473}

\bibitem[\protect\citeauthoryear{Baydin, Pearlmutter, Radul  \& Siskind}{Baydin
  et~al.}{2018}]{autograd}
Baydin A.~G.,  Pearlmutter B.~A.,  Radul A.~A.,   Siskind J.~M.,  2018, Journal
  of Machine Learning Research, 18, 1

\bibitem[\protect\citeauthoryear{{Becker}, {Hewett}, {Worseck}  \&
  {Prochaska}}{{Becker} et~al.}{2013}]{Becker13}
{Becker} G.~D.,  {Hewett} P.~C.,  {Worseck} G.,   {Prochaska} J.~X.,  2013,
  \mn@doi [\mnras] {10.1093/mnras/stt031}, \href
  {https://ui.adsabs.harvard.edu/abs/2013MNRAS.430.2067B} {430, 2067}

\bibitem[\protect\citeauthoryear{{Becker} et~al.,}{{Becker}
  et~al.}{2019}]{Becker19}
{Becker} G.~D.,  et~al., 2019, \mn@doi [\apj] {10.3847/1538-4357/ab3eb5}, \href
  {https://ui.adsabs.harvard.edu/abs/2019ApJ...883..163B} {883, 163}

\bibitem[\protect\citeauthoryear{{Becker}, {D'Aloisio}, {Christenson}, {Zhu},
  {Worseck}  \& {Bolton}}{{Becker} et~al.}{2021}]{Becker21}
{Becker} G.~D.,  {D'Aloisio} A.,  {Christenson} H.~M.,  {Zhu} Y.,  {Worseck}
  G.,   {Bolton} J.~S.,  2021, \mn@doi [\mnras] {10.1093/mnras/stab2696}, \href
  {https://ui.adsabs.harvard.edu/abs/2021MNRAS.508.1853B} {508, 1853}

\bibitem[\protect\citeauthoryear{{Becker}, {Bolton}, {Zhu}  \&
  {Hashemi}}{{Becker} et~al.}{2024}]{Becker24}
{Becker} G.~D.,  {Bolton} J.~S.,  {Zhu} Y.,   {Hashemi} S.,  2024, \mn@doi
  [arXiv e-prints] {10.48550/arXiv.2405.08885}, \href
  {https://ui.adsabs.harvard.edu/abs/2024arXiv240508885B} {p. arXiv:2405.08885}

\bibitem[\protect\citeauthoryear{{Betancourt}}{{Betancourt}}{2017}]{HMC2}
{Betancourt} M.,  2017, \mn@doi [arXiv e-prints] {10.48550/arXiv.1701.02434},
  \href {https://ui.adsabs.harvard.edu/abs/2017arXiv170102434B} {p.
  arXiv:1701.02434}

\bibitem[\protect\citeauthoryear{{Bingham} et~al.,}{{Bingham}
  et~al.}{2018}]{pyro}
{Bingham} E.,  et~al., 2018, \mn@doi [arXiv e-prints]
  {10.48550/arXiv.1810.09538}, \href
  {https://ui.adsabs.harvard.edu/abs/2018arXiv181009538B} {p. arXiv:1810.09538}

\bibitem[\protect\citeauthoryear{Bingham et~al.,}{Bingham
  et~al.}{2019}]{NumPyro2019b}
Bingham E.,  et~al., 2019, J. Mach. Learn. Res., 20, 28:1

\bibitem[\protect\citeauthoryear{{Bolton} \& {Haehnelt}}{{Bolton} \&
  {Haehnelt}}{2007a}]{Bolton07a}
{Bolton} J.~S.,  {Haehnelt} M.~G.,  2007a, \mn@doi [\mnras]
  {10.1111/j.1365-2966.2006.11176.x}, \href
  {http://adsabs.harvard.edu/abs/2007MNRAS.374..493B} {374, 493}

\bibitem[\protect\citeauthoryear{{Bolton} \& {Haehnelt}}{{Bolton} \&
  {Haehnelt}}{2007b}]{Bolton07b}
{Bolton} J.~S.,  {Haehnelt} M.~G.,  2007b, \mn@doi [\mnras]
  {10.1111/j.1745-3933.2007.00361.x}, \href
  {http://adsabs.harvard.edu/abs/2007MNRAS.381L..35B} {381, L35}

\bibitem[\protect\citeauthoryear{{Bolton} \& {Haehnelt}}{{Bolton} \&
  {Haehnelt}}{2013}]{Bolton13}
{Bolton} J.~S.,  {Haehnelt} M.~G.,  2013, \mn@doi [\mnras]
  {10.1093/mnras/sts455}, \href
  {https://ui.adsabs.harvard.edu/abs/2013MNRAS.429.1695B} {429, 1695}

\bibitem[\protect\citeauthoryear{{Bolton}, {Haehnelt}, {Warren}, {Hewett},
  {Mortlock}, {Venemans}, {McMahon}  \& {Simpson}}{{Bolton}
  et~al.}{2011}]{Bolton11}
{Bolton} J.~S.,  {Haehnelt} M.~G.,  {Warren} S.~J.,  {Hewett} P.~C.,
  {Mortlock} D.~J.,  {Venemans} B.~P.,  {McMahon} R.~G.,   {Simpson} C.,  2011,
  \mn@doi [\mnras] {10.1111/j.1745-3933.2011.01100.x}, \href
  {http://adsabs.harvard.edu/abs/2011MNRAS.416L..70B} {416, L70}

\bibitem[\protect\citeauthoryear{{Boroson} \& {Green}}{{Boroson} \&
  {Green}}{1992}]{Boroson92}
{Boroson} T.~A.,  {Green} R.~F.,  1992, \mn@doi [\apjs] {10.1086/191661}, \href
  {https://ui.adsabs.harvard.edu/abs/1992ApJS...80..109B} {80, 109}

\bibitem[\protect\citeauthoryear{{Bosman}, {Kakiichi}, {Meyer}, {Gronke},
  {Laporte}  \& {Ellis}}{{Bosman} et~al.}{2020}]{Bosman20}
{Bosman} S. E.~I.,  {Kakiichi} K.,  {Meyer} R.~A.,  {Gronke} M.,  {Laporte} N.,
    {Ellis} R.~S.,  2020, \mn@doi [\apj] {10.3847/1538-4357/ab85cd}, \href
  {https://ui.adsabs.harvard.edu/abs/2020ApJ...896...49B} {896, 49}

\bibitem[\protect\citeauthoryear{{Bosman}, {{\v{D}}urov{\v{c}}{\'\i}kov{\'a}},
  {Davies}  \& {Eilers}}{{Bosman} et~al.}{2021}]{Bosman21}
{Bosman} S. E.~I.,  {{\v{D}}urov{\v{c}}{\'\i}kov{\'a}} D.,  {Davies} F.~B.,
  {Eilers} A.-C.,  2021, \mn@doi [\mnras] {10.1093/mnras/stab572}, \href
  {https://ui.adsabs.harvard.edu/abs/2021MNRAS.503.2077B} {503, 2077}

\bibitem[\protect\citeauthoryear{Bradbury et~al.,}{Bradbury
  et~al.}{2018a}]{JAX}
Bradbury J.,  et~al., 2018a, {JAX}: composable transformations of
  {P}ython+{N}um{P}y programs, \url {http://github.com/google/jax}

\bibitem[\protect\citeauthoryear{Bradbury et~al.,}{Bradbury
  et~al.}{2018b}]{Jax2018}
Bradbury J.,  et~al., 2018b, {JAX}: composable transformations of
  {P}ython+{N}um{P}y programs, \url {http://github.com/google/jax}

\bibitem[\protect\citeauthoryear{{Byler}, {Dalcanton}, {Conroy}  \&
  {Johnson}}{{Byler} et~al.}{2017}]{Byler17}
{Byler} N.,  {Dalcanton} J.~J.,  {Conroy} C.,   {Johnson} B.~D.,  2017, \mn@doi
  [\apj] {10.3847/1538-4357/aa6c66}, \href
  {https://ui.adsabs.harvard.edu/abs/2017ApJ...840...44B} {840, 44}

\bibitem[\protect\citeauthoryear{{Carswell}, {Whelan}, {Smith}, {Boksenberg}
  \& {Tytler}}{{Carswell} et~al.}{1982}]{Carswell82}
{Carswell} R.~F.,  {Whelan} J.~A.~J.,  {Smith} M.~G.,  {Boksenberg} A.,
  {Tytler} D.,  1982, \mn@doi [\mnras] {10.1093/mnras/198.1.91}, \href
  {https://ui.adsabs.harvard.edu/abs/1982MNRAS.198...91C} {198, 91}

\bibitem[\protect\citeauthoryear{{Cen} \& {Haiman}}{{Cen} \&
  {Haiman}}{2000}]{CenHaiman00}
{Cen} R.,  {Haiman} Z.,  2000, \mn@doi [\apjl] {10.1086/312937}, \href
  {http://adsabs.harvard.edu/abs/2000ApJ...542L..75C} {542, L75}

\bibitem[\protect\citeauthoryear{{Chen}}{{Chen}}{2023}]{Chen23a}
{Chen} H.,  2023, \mn@doi [arXiv e-prints] {10.48550/arXiv.2307.04797}, \href
  {https://ui.adsabs.harvard.edu/abs/2023arXiv230704797C} {p. arXiv:2307.04797}

\bibitem[\protect\citeauthoryear{{Chen} \& {Gnedin}}{{Chen} \&
  {Gnedin}}{2021}]{Chen21a}
{Chen} H.,  {Gnedin} N.~Y.,  2021, \mn@doi [\apj] {10.3847/1538-4357/abe7e7},
  \href {https://ui.adsabs.harvard.edu/abs/2021ApJ...911...60C} {911, 60}

\bibitem[\protect\citeauthoryear{{Chen} et~al.,}{{Chen} et~al.}{2022}]{Chen22}
{Chen} H.,  et~al., 2022, \mn@doi [\apj] {10.3847/1538-4357/ac658d}, \href
  {https://ui.adsabs.harvard.edu/abs/2022ApJ...931...29C} {931, 29}

\bibitem[\protect\citeauthoryear{{Chen}, {Speagle}  \& {Rogers}}{{Chen}
  et~al.}{2023}]{Chen23b}
{Chen} H.,  {Speagle} J.,   {Rogers} K.~K.,  2023, \mn@doi [arXiv e-prints]
  {10.48550/arXiv.2311.16238}, \href
  {https://ui.adsabs.harvard.edu/abs/2023arXiv231116238C} {p. arXiv:2311.16238}

\bibitem[\protect\citeauthoryear{{Chen}, {Stark}, {Mason}, {Topping},
  {Whitler}, {Tang}, {Endsley}  \& {Charlot}}{{Chen} et~al.}{2024}]{Chen24}
{Chen} Z.,  {Stark} D.~P.,  {Mason} C.,  {Topping} M.~W.,  {Whitler} L.,
  {Tang} M.,  {Endsley} R.,   {Charlot} S.,  2024, \mn@doi [\mnras]
  {10.1093/mnras/stae455}, \href
  {https://ui.adsabs.harvard.edu/abs/2024MNRAS.528.7052C} {528, 7052}

\bibitem[\protect\citeauthoryear{{Cheng} et~al.,}{{Cheng}
  et~al.}{2018}]{Cheng18}
{Cheng} C.,  et~al., 2018, \mn@doi [\apj] {10.3847/1538-4357/aae833}, \href
  {https://ui.adsabs.harvard.edu/abs/2018ApJ...868...26C} {868, 26}

\bibitem[\protect\citeauthoryear{{Christensen} et~al.,}{{Christensen}
  et~al.}{2023}]{Christensen23}
{Christensen} L.,  et~al., 2023, \mn@doi [arXiv e-prints]
  {10.48550/arXiv.2309.06470}, \href
  {https://ui.adsabs.harvard.edu/abs/2023arXiv230906470C} {p. arXiv:2309.06470}

\bibitem[\protect\citeauthoryear{Collette}{Collette}{2013}]{h5py2013}
Collette A.,  2013, Python and HDF5.
O'Reilly

\bibitem[\protect\citeauthoryear{{Cranmer}, {Brehmer}  \& {Louppe}}{{Cranmer}
  et~al.}{2020}]{Cranmer20}
{Cranmer} K.,  {Brehmer} J.,   {Louppe} G.,  2020, \mn@doi [Proceedings of the
  National Academy of Science] {10.1073/pnas.1912789117}, \href
  {https://ui.adsabs.harvard.edu/abs/2020PNAS..11730055C} {117, 30055}

\bibitem[\protect\citeauthoryear{{DESI Collaboration} et~al.,}{{DESI
  Collaboration} et~al.}{2024}]{DESI24BAO}
{DESI Collaboration} et~al., 2024, \mn@doi [arXiv e-prints]
  {10.48550/arXiv.2404.03000}, \href
  {https://ui.adsabs.harvard.edu/abs/2024arXiv240403000D} {p. arXiv:2404.03000}

\bibitem[\protect\citeauthoryear{{D'Eugenio} et~al.,}{{D'Eugenio}
  et~al.}{2023}]{Deugenio23}
{D'Eugenio} F.,  et~al., 2023, \mn@doi [arXiv e-prints]
  {10.48550/arXiv.2311.09908}, \href
  {https://ui.adsabs.harvard.edu/abs/2023arXiv231109908D} {p. arXiv:2311.09908}

\bibitem[\protect\citeauthoryear{{Dall'Aglio}, {Wisotzki}  \&
  {Worseck}}{{Dall'Aglio} et~al.}{2008}]{Dallaglio08}
{Dall'Aglio} A.,  {Wisotzki} L.,   {Worseck} G.,  2008, \mn@doi [\aap]
  {10.1051/0004-6361:200810724}, \href
  {https://ui.adsabs.harvard.edu/abs/2008A&A...491..465D} {491, 465}

\bibitem[\protect\citeauthoryear{{Davies} \& {Furlanetto}}{{Davies} \&
  {Furlanetto}}{2022}]{Davies22}
{Davies} F.~B.,  {Furlanetto} S.~R.,  2022, \mn@doi [\mnras]
  {10.1093/mnras/stac1005}, \href
  {https://ui.adsabs.harvard.edu/abs/2022MNRAS.514.1302D} {514, 1302}

\bibitem[\protect\citeauthoryear{{Davies}, {Furlanetto}  \& {McQuinn}}{{Davies}
  et~al.}{2016}]{Davies16a}
{Davies} F.~B.,  {Furlanetto} S.~R.,   {McQuinn} M.,  2016, \mn@doi [\mnras]
  {10.1093/mnras/stw055}, \href
  {https://ui.adsabs.harvard.edu/abs/2016MNRAS.457.3006D} {457, 3006}

\bibitem[\protect\citeauthoryear{{Davies}, {Hennawi}, {Eilers}  \&
  {Luki{\'c}}}{{Davies} et~al.}{2018a}]{Davies18ABC}
{Davies} F.~B.,  {Hennawi} J.~F.,  {Eilers} A.-C.,   {Luki{\'c}} Z.,  2018a,
  \mn@doi [\apj] {10.3847/1538-4357/aaaf70}, \href
  {https://ui.adsabs.harvard.edu/abs/2018ApJ...855..106D} {855, 106}

\bibitem[\protect\citeauthoryear{{Davies} et~al.,}{{Davies}
  et~al.}{2018b}]{Davies18b}
{Davies} F.~B.,  et~al., 2018b, \mn@doi [\apj] {10.3847/1538-4357/aad6dc},
  \href {https://ui.adsabs.harvard.edu/abs/2018ApJ...864..142D} {864, 142}

\bibitem[\protect\citeauthoryear{{Davies} et~al.,}{{Davies}
  et~al.}{2018c}]{Davies18a}
{Davies} F.~B.,  et~al., 2018c, \mn@doi [\apj] {10.3847/1538-4357/aad7f8},
  \href {https://ui.adsabs.harvard.edu/abs/2018ApJ...864..143D} {864, 143}

\bibitem[\protect\citeauthoryear{{Davies}, {Hennawi}  \& {Eilers}}{{Davies}
  et~al.}{2019}]{DaviesBH19}
{Davies} F.~B.,  {Hennawi} J.~F.,   {Eilers} A.-C.,  2019, \mn@doi [\apjl]
  {10.3847/2041-8213/ab42e3}, \href
  {https://ui.adsabs.harvard.edu/abs/2019ApJ...884L..19D} {884, L19}

\bibitem[\protect\citeauthoryear{{Davies}, {Hennawi}  \& {Eilers}}{{Davies}
  et~al.}{2020}]{Davies20}
{Davies} F.~B.,  {Hennawi} J.~F.,   {Eilers} A.-C.,  2020, \mn@doi [\mnras]
  {10.1093/mnras/stz3303}, \href
  {https://ui.adsabs.harvard.edu/abs/2020MNRAS.493.1330D} {493, 1330}

\bibitem[\protect\citeauthoryear{{Davies}, {Ba{\~n}ados}, {Hennawi}  \&
  {Bosman}}{{Davies} et~al.}{2023}]{DaviesPisco23}
{Davies} F.~B.,  {Ba{\~n}ados} E.,  {Hennawi} J.~F.,   {Bosman} S. E.~I.,
  2023, \mn@doi [arXiv e-prints] {10.48550/arXiv.2312.06747}, \href
  {https://ui.adsabs.harvard.edu/abs/2023arXiv231206747D} {p. arXiv:2312.06747}

\bibitem[\protect\citeauthoryear{{Davies} et~al.,}{{Davies}
  et~al.}{2024}]{Davies24}
{Davies} F.~B.,  et~al., 2024, \mn@doi [\apj] {10.3847/1538-4357/ad1d5d}, \href
  {https://ui.adsabs.harvard.edu/abs/2024ApJ...965..134D} {965, 134}

\bibitem[\protect\citeauthoryear{{Dawson} et~al.,}{{Dawson}
  et~al.}{2013}]{Dawson13}
{Dawson} K.~S.,  et~al., 2013, \mn@doi [\aj] {10.1088/0004-6256/145/1/10},
  \href {https://ui.adsabs.harvard.edu/abs/2013AJ....145...10D} {145, 10}

\bibitem[\protect\citeauthoryear{{Dawson} et~al.,}{{Dawson}
  et~al.}{2016}]{Dawson16}
{Dawson} K.~S.,  et~al., 2016, \mn@doi [\aj] {10.3847/0004-6256/151/2/44},
  \href {https://ui.adsabs.harvard.edu/abs/2016AJ....151...44D} {151, 44}

\bibitem[\protect\citeauthoryear{{Delchambre}}{{Delchambre}}{2015}]{WPCA}
{Delchambre} L.,  2015, \mn@doi [\mnras] {10.1093/mnras/stu2219}, \href
  {https://ui.adsabs.harvard.edu/abs/2015MNRAS.446.3545D} {446, 3545}

\bibitem[\protect\citeauthoryear{{Di Valentino} \& {Melchiorri}}{{Di Valentino}
  \& {Melchiorri}}{2022}]{DiValentino22}
{Di Valentino} E.,  {Melchiorri} A.,  2022, \mn@doi [\apjl]
  {10.3847/2041-8213/ac6ef5}, \href
  {https://ui.adsabs.harvard.edu/abs/2022ApJ...931L..18D} {931, L18}

\bibitem[\protect\citeauthoryear{{Dijkstra}, {Mesinger}  \&
  {Wyithe}}{{Dijkstra} et~al.}{2011}]{Dijkstra11}
{Dijkstra} M.,  {Mesinger} A.,   {Wyithe} J. S.~B.,  2011, \mn@doi [\mnras]
  {10.1111/j.1365-2966.2011.18530.x}, \href
  {https://ui.adsabs.harvard.edu/abs/2011MNRAS.414.2139D} {414, 2139}

\bibitem[\protect\citeauthoryear{{Dor{\'e}}, {Hennawi}  \&
  {Spergel}}{{Dor{\'e}} et~al.}{2004}]{Dore04}
{Dor{\'e}} O.,  {Hennawi} J.~F.,   {Spergel} D.~N.,  2004, \mn@doi [\apj]
  {10.1086/382946}, \href
  {https://ui.adsabs.harvard.edu/abs/2004ApJ...606...46D} {606, 46}

\bibitem[\protect\citeauthoryear{{Duane}, {Kennedy}, {Pendleton}  \&
  {Roweth}}{{Duane} et~al.}{1987}]{HMC1}
{Duane} S.,  {Kennedy} A.~D.,  {Pendleton} B.~J.,   {Roweth} D.,  1987, \mn@doi
  [Physics Letters B] {10.1016/0370-2693(87)91197-X}, \href
  {https://ui.adsabs.harvard.edu/abs/1987PhLB..195..216D} {195, 216}

\bibitem[\protect\citeauthoryear{{Eilers}, {Davies}, {Hennawi}, {Prochaska},
  {Luki{\'c}}  \& {Mazzucchelli}}{{Eilers} et~al.}{2017a}]{Eilers17a}
{Eilers} A.-C.,  {Davies} F.~B.,  {Hennawi} J.~F.,  {Prochaska} J.~X.,
  {Luki{\'c}} Z.,   {Mazzucchelli} C.,  2017a, \mn@doi [\apj]
  {10.3847/1538-4357/aa6c60}, \href
  {http://adsabs.harvard.edu/abs/2017ApJ...840...24E} {840, 24}

\bibitem[\protect\citeauthoryear{{Eilers}, {Hennawi}  \& {Lee}}{{Eilers}
  et~al.}{2017b}]{Eilers17b}
{Eilers} A.-C.,  {Hennawi} J.~F.,   {Lee} K.-G.,  2017b, \mn@doi [\apj]
  {10.3847/1538-4357/aa7e31}, \href
  {http://adsabs.harvard.edu/abs/2017ApJ...844..136E} {844, 136}

\bibitem[\protect\citeauthoryear{{Eilers}, {Hennawi}  \& {Davies}}{{Eilers}
  et~al.}{2018}]{Eilers18}
{Eilers} A.-C.,  {Hennawi} J.~F.,   {Davies} F.~B.,  2018, \mn@doi [The
  Astrophysical Journal] {10.3847/1538-4357/aae081}, \href
  {https://ui.adsabs.harvard.edu/abs/2018ApJ...867...30E} {867, 30}

\bibitem[\protect\citeauthoryear{{Eilers} et~al.,}{{Eilers}
  et~al.}{2020}]{Eilers20a}
{Eilers} A.-C.,  et~al., 2020, \mn@doi [\apj] {10.3847/1538-4357/aba52e}, \href
  {https://ui.adsabs.harvard.edu/abs/2020ApJ...900...37E} {900, 37}

\bibitem[\protect\citeauthoryear{{Eilers}, {Hennawi}, {Davies}  \&
  {Simcoe}}{{Eilers} et~al.}{2021}]{Eilers21}
{Eilers} A.-C.,  {Hennawi} J.~F.,  {Davies} F.~B.,   {Simcoe} R.~A.,  2021,
  \mn@doi [\apj] {10.3847/1538-4357/ac0a76}, \href
  {https://ui.adsabs.harvard.edu/abs/2021ApJ...917...38E} {917, 38}

\bibitem[\protect\citeauthoryear{{Eilers}, {Hogg}, {Sch{\"o}lkopf},
  {Foreman-Mackey}, {Davies}  \& {Schindler}}{{Eilers}
  et~al.}{2022}]{EilersGPLVM22}
{Eilers} A.-C.,  {Hogg} D.~W.,  {Sch{\"o}lkopf} B.,  {Foreman-Mackey} D.,
  {Davies} F.~B.,   {Schindler} J.-T.,  2022, \mn@doi [\apj]
  {10.3847/1538-4357/ac8ead}, \href
  {https://ui.adsabs.harvard.edu/abs/2022ApJ...938...17E} {938, 17}

\bibitem[\protect\citeauthoryear{{Euclid Collaboration} et~al.,}{{Euclid
  Collaboration} et~al.}{2019}]{Warren19}
{Euclid Collaboration} et~al., 2019, \mn@doi [\aap]
  {10.1051/0004-6361/201936427}, \href
  {https://ui.adsabs.harvard.edu/abs/2019A&A...631A..85E} {631, A85}

\bibitem[\protect\citeauthoryear{{Fan}, {Strauss}, {Becker}  \& {et al.}}{{Fan}
  et~al.}{2006}]{Fan06}
{Fan} X.,  {Strauss} M.~A.,  {Becker} R.~H.,   {et al.} 2006, \mn@doi [\aj]
  {10.1086/504836}, \href {http://adsabs.harvard.edu/abs/2006AJ....132..117F}
  {132, 117}

\bibitem[\protect\citeauthoryear{{Fathivavsari}}{{Fathivavsari}}{2020}]{Fathi20}
{Fathivavsari} H.,  2020, \mn@doi [\apj] {10.3847/1538-4357/ab9b7d}, \href
  {https://ui.adsabs.harvard.edu/abs/2020ApJ...898..114F} {898, 114}

\bibitem[\protect\citeauthoryear{{Ferraro} \& {Smith}}{{Ferraro} \&
  {Smith}}{2018}]{Ferraro18}
{Ferraro} S.,  {Smith} K.~M.,  2018, \mn@doi [\prd]
  {10.1103/PhysRevD.98.123519}, \href
  {https://ui.adsabs.harvard.edu/abs/2018PhRvD..98l3519F} {98, 123519}

\bibitem[\protect\citeauthoryear{{Foreman-Mackey}}{{Foreman-Mackey}}{2016}]{Corner2016}
{Foreman-Mackey} D.,  2016, \mn@doi [The Journal of Open Source Software]
  {10.21105/joss.00024}, \href
  {https://ui.adsabs.harvard.edu/abs/2016JOSS....1...24F} {1, 24}

\bibitem[\protect\citeauthoryear{{Francis}, {Hewett}, {Foltz}  \&
  {Chaffee}}{{Francis} et~al.}{1992}]{Francis92}
{Francis} P.~J.,  {Hewett} P.~C.,  {Foltz} C.~B.,   {Chaffee} F.~H.,  1992,
  \mn@doi [\apj] {10.1086/171870}, \href
  {https://ui.adsabs.harvard.edu/abs/1992ApJ...398..476F} {398, 476}

\bibitem[\protect\citeauthoryear{{Gaikwad} et~al.,}{{Gaikwad}
  et~al.}{2023}]{Gaikwad23}
{Gaikwad} P.,  et~al., 2023, \mn@doi [\mnras] {10.1093/mnras/stad2566}, \href
  {https://ui.adsabs.harvard.edu/abs/2023MNRAS.525.4093G} {525, 4093}

\bibitem[\protect\citeauthoryear{{Gardner}, {Pleiss}, {Bindel}, {Weinberger}
  \& {Wilson}}{{Gardner} et~al.}{2018}]{gpytorch}
{Gardner} J.~R.,  {Pleiss} G.,  {Bindel} D.,  {Weinberger} K.~Q.,   {Wilson}
  A.~G.,  2018, arXiv e-prints, \href
  {https://ui.adsabs.harvard.edu/abs/2018arXiv180911165G} {p. arXiv:1809.11165}

\bibitem[\protect\citeauthoryear{{Garnett}, {Ho}, {Bird}  \&
  {Schneider}}{{Garnett} et~al.}{2017}]{Garnett17}
{Garnett} R.,  {Ho} S.,  {Bird} S.,   {Schneider} J.,  2017, \mn@doi [\mnras]
  {10.1093/mnras/stx1958}, \href
  {https://ui.adsabs.harvard.edu/abs/2017MNRAS.472.1850G} {472, 1850}

\bibitem[\protect\citeauthoryear{{George} et~al.,}{{George}
  et~al.}{2015}]{George15}
{George} E.~M.,  et~al., 2015, \mn@doi [\apj] {10.1088/0004-637X/799/2/177},
  \href {http://adsabs.harvard.edu/abs/2015ApJ...799..177G} {799, 177}

\bibitem[\protect\citeauthoryear{{Greig}, {Mesinger}, {McGreer}, {Gallerani}
  \& {Haiman}}{{Greig} et~al.}{2017a}]{Greig17a}
{Greig} B.,  {Mesinger} A.,  {McGreer} I.~D.,  {Gallerani} S.,   {Haiman} Z.,
  2017a, \mn@doi [\mnras] {10.1093/mnras/stw3210}, \href
  {http://adsabs.harvard.edu/abs/2017MNRAS.466.1814G} {466, 1814}

\bibitem[\protect\citeauthoryear{{Greig}, {Mesinger}, {Haiman}  \&
  {Simcoe}}{{Greig} et~al.}{2017b}]{Greig17b}
{Greig} B.,  {Mesinger} A.,  {Haiman} Z.,   {Simcoe} R.~A.,  2017b, \mn@doi
  [\mnras] {10.1093/mnras/stw3351}, \href
  {http://adsabs.harvard.edu/abs/2017MNRAS.466.4239G} {466, 4239}

\bibitem[\protect\citeauthoryear{{Greig}, {Mesinger}, {Davies}, {Wang}, {Yang}
  \& {Hennawi}}{{Greig} et~al.}{2022}]{Greig22}
{Greig} B.,  {Mesinger} A.,  {Davies} F.~B.,  {Wang} F.,  {Yang} J.,
  {Hennawi} J.~F.,  2022, \mn@doi [\mnras] {10.1093/mnras/stac825}, \href
  {https://ui.adsabs.harvard.edu/abs/2022MNRAS.512.5390G} {512, 5390}

\bibitem[\protect\citeauthoryear{{Greig} et~al.,}{{Greig}
  et~al.}{2024a}]{Greig24a}
{Greig} B.,  et~al., 2024a, \mn@doi [\mnras] {10.1093/mnras/stae1080}, \href
  {https://ui.adsabs.harvard.edu/abs/2024MNRAS.tmp.1103G} {}

\bibitem[\protect\citeauthoryear{{Greig} et~al.,}{{Greig}
  et~al.}{2024b}]{Greig24b}
{Greig} B.,  et~al., 2024b, \mn@doi [arXiv e-prints]
  {10.48550/arXiv.2404.01556}, \href
  {https://ui.adsabs.harvard.edu/abs/2024arXiv240401556G} {p. arXiv:2404.01556}

\bibitem[\protect\citeauthoryear{{HERA} et~al.,}{{HERA} et~al.}{2022}]{HERA22}
{HERA} et~al., 2022, \mn@doi [\apj] {10.3847/1538-4357/ac1c78}, \href
  {https://ui.adsabs.harvard.edu/abs/2022ApJ...925..221A} {925, 221}

\bibitem[\protect\citeauthoryear{{Harris} et~al.,}{{Harris}
  et~al.}{2020}]{Numpy2020}
{Harris} C.~R.,  et~al., 2020, \mn@doi [\nat] {10.1038/s41586-020-2649-2},
  \href {https://ui.adsabs.harvard.edu/abs/2020Natur.585..357H} {585, 357}

\bibitem[\protect\citeauthoryear{{Heintz} et~al.,}{{Heintz}
  et~al.}{2023}]{Heintz23}
{Heintz} K.~E.,  et~al., 2023, \mn@doi [arXiv e-prints]
  {10.48550/arXiv.2306.00647}, \href
  {https://ui.adsabs.harvard.edu/abs/2023arXiv230600647H} {p. arXiv:2306.00647}

\bibitem[\protect\citeauthoryear{{Heintz} et~al.,}{{Heintz}
  et~al.}{2024}]{Heintz24b}
{Heintz} K.~E.,  et~al., 2024, \mn@doi [arXiv e-prints]
  {10.48550/arXiv.2404.02211}, \href
  {https://ui.adsabs.harvard.edu/abs/2024arXiv240402211H} {p. arXiv:2404.02211}

\bibitem[\protect\citeauthoryear{{Hensman}, {Fusi}  \& {Lawrence}}{{Hensman}
  et~al.}{2013}]{Hensman13}
{Hensman} J.,  {Fusi} N.,   {Lawrence} N.~D.,  2013, arXiv e-prints, \href
  {https://ui.adsabs.harvard.edu/abs/2013arXiv1309.6835H} {p. arXiv:1309.6835}

\bibitem[\protect\citeauthoryear{{Heymans} et~al.,}{{Heymans}
  et~al.}{2021}]{KIDS21}
{Heymans} C.,  et~al., 2021, \mn@doi [\aap] {10.1051/0004-6361/202039063},
  \href {https://ui.adsabs.harvard.edu/abs/2021A&A...646A.140H} {646, A140}

\bibitem[\protect\citeauthoryear{Higgins, Matthey, Pal, Burgess, Glorot,
  Botvinick, Mohamed  \& Lerchner}{Higgins et~al.}{2017}]{bVAE}
Higgins I.,  Matthey L.,  Pal A.,  Burgess C.,  Glorot X.,  Botvinick M.,
  Mohamed S.,   Lerchner A.,  2017, in International Conference on Learning
  Representations. \url {https://openreview.net/forum?id=Sy2fzU9gl}

\bibitem[\protect\citeauthoryear{{Hikage} et~al.,}{{Hikage}
  et~al.}{2019}]{HSCWL19}
{Hikage} C.,  et~al., 2019, \mn@doi [\pasj] {10.1093/pasj/psz010}, \href
  {https://ui.adsabs.harvard.edu/abs/2019PASJ...71...43H} {71, 43}

\bibitem[\protect\citeauthoryear{{Hoag} et~al.,}{{Hoag} et~al.}{2019}]{Hoag19}
{Hoag} A.,  et~al., 2019, \mn@doi [\apj] {10.3847/1538-4357/ab1de7}, \href
  {https://ui.adsabs.harvard.edu/abs/2019ApJ...878...12H} {878, 12}

\bibitem[\protect\citeauthoryear{Hoffman, Blei, Wang  \& Paisley}{Hoffman
  et~al.}{2013}]{SVI}
Hoffman M.~D.,  Blei D.~M.,  Wang C.,   Paisley J.,  2013, Journal of Machine
  Learning Research, 14, 1303

\bibitem[\protect\citeauthoryear{Hoffman, Gelman  et~al.}{Hoffman
  et~al.}{2014}]{NUTS}
Hoffman M.~D.,  Gelman A.,   et~al., 2014, J. Mach. Learn. Res., 15, 1593

\bibitem[\protect\citeauthoryear{{Horowitz}, {Lee}, {White}, {Krolewski}  \&
  {Ata}}{{Horowitz} et~al.}{2019}]{Horowitz19}
{Horowitz} B.,  {Lee} K.-G.,  {White} M.,  {Krolewski} A.,   {Ata} M.,  2019,
  \mn@doi [\apj] {10.3847/1538-4357/ab4d4c}, \href
  {https://ui.adsabs.harvard.edu/abs/2019ApJ...887...61H} {887, 61}

\bibitem[\protect\citeauthoryear{{Hunter}}{{Hunter}}{2007}]{Matplotlib2007}
{Hunter} J.~D.,  2007, \mn@doi [Computing in Science and Engineering]
  {10.1109/MCSE.2007.55}, \href
  {https://ui.adsabs.harvard.edu/abs/2007CSE.....9...90H} {9, 90}

\bibitem[\protect\citeauthoryear{{Jin} et~al.,}{{Jin} et~al.}{2023}]{Jin23}
{Jin} X.,  et~al., 2023, \mn@doi [\apj] {10.3847/1538-4357/aca678}, \href
  {https://ui.adsabs.harvard.edu/abs/2023ApJ...942...59J} {942, 59}

\bibitem[\protect\citeauthoryear{{Jung} et~al.,}{{Jung} et~al.}{2020}]{Jung20}
{Jung} I.,  et~al., 2020, \mn@doi [\apj] {10.3847/1538-4357/abbd44}, \href
  {https://ui.adsabs.harvard.edu/abs/2020ApJ...904..144J} {904, 144}

\bibitem[\protect\citeauthoryear{{Keating}, {Haehnelt}, {Cantalupo}  \&
  {Puchwein}}{{Keating} et~al.}{2015}]{Keating15}
{Keating} L.~C.,  {Haehnelt} M.~G.,  {Cantalupo} S.,   {Puchwein} E.,  2015,
  \mn@doi [\mnras] {10.1093/mnras/stv2020}, \href
  {https://ui.adsabs.harvard.edu/abs/2015MNRAS.454..681K} {454, 681}

\bibitem[\protect\citeauthoryear{{Keating}, {Bolton}, {Cullen}, {Haehnelt},
  {Puchwein}  \& {Kulkarni}}{{Keating} et~al.}{2023}]{Keating23a}
{Keating} L.~C.,  {Bolton} J.~S.,  {Cullen} F.,  {Haehnelt} M.~G.,  {Puchwein}
  E.,   {Kulkarni} G.,  2023, \mn@doi [arXiv e-prints]
  {10.48550/arXiv.2308.05800}, \href
  {https://ui.adsabs.harvard.edu/abs/2023arXiv230805800K} {p. arXiv:2308.05800}

\bibitem[\protect\citeauthoryear{{Keating}, {Puchwein}, {Bolton}, {Haehnelt}
  \& {Kulkarni}}{{Keating} et~al.}{2024}]{Keating24}
{Keating} L.~C.,  {Puchwein} E.,  {Bolton} J.~S.,  {Haehnelt} M.~G.,
  {Kulkarni} G.,  2024, \mn@doi [\mnras] {10.1093/mnrasl/slae022}, \href
  {https://ui.adsabs.harvard.edu/abs/2024MNRAS.531L..34K} {531, L34}

\bibitem[\protect\citeauthoryear{{Khrykin}, {Hennawi}, {McQuinn}  \&
  {Worseck}}{{Khrykin} et~al.}{2015}]{Khrykin15}
{Khrykin} I.~S.,  {Hennawi} J.~F.,  {McQuinn} M.,   {Worseck} G.,  2015,
  preprint, \href {http://adsabs.harvard.edu/abs/2015arXiv151103659K} {}
  (\mn@eprint {arXiv} {1511.03659})

\bibitem[\protect\citeauthoryear{{Khrykin}, {Hennawi}, {Worseck}  \&
  {Davies}}{{Khrykin} et~al.}{2021}]{Khrykin21}
{Khrykin} I.~S.,  {Hennawi} J.~F.,  {Worseck} G.,   {Davies} F.~B.,  2021,
  \mn@doi [\mnras] {10.1093/mnras/stab1288}, \href
  {https://ui.adsabs.harvard.edu/abs/2021MNRAS.505..649K} {505, 649}

\bibitem[\protect\citeauthoryear{Kingma \& Welling}{Kingma \&
  Welling}{2013}]{VAE}
Kingma D.~P.,  Welling M.,  2013, arXiv preprint arXiv:1312.6114

\bibitem[\protect\citeauthoryear{{Kirkman} et~al.,}{{Kirkman}
  et~al.}{2005}]{Kirkman05}
{Kirkman} D.,  et~al., 2005, \mn@doi [\mnras]
  {10.1111/j.1365-2966.2005.09126.x}, \href
  {https://ui.adsabs.harvard.edu/abs/2005MNRAS.360.1373K} {360, 1373}

\bibitem[\protect\citeauthoryear{{Kist}, {Hennawi}  \& {Davies}}{{Kist}
  et~al.}{2025}]{Kist25a}
{Kist} T.,  {Hennawi} J.~F.,   {Davies} F.~B.,  2025, \mn@doi [\mnras]
  {10.1093/mnras/staf460}, \href
  {https://ui.adsabs.harvard.edu/abs/2025MNRAS.538.2704K} {538, 2704}

\bibitem[\protect\citeauthoryear{{Kreisch}, {Cyr-Racine}  \&
  {Dor{\'e}}}{{Kreisch} et~al.}{2020}]{Kreish20}
{Kreisch} C.~D.,  {Cyr-Racine} F.-Y.,   {Dor{\'e}} O.,  2020, \mn@doi [\prd]
  {10.1103/PhysRevD.101.123505}, \href
  {https://ui.adsabs.harvard.edu/abs/2020PhRvD.101l3505K} {101, 123505}

\bibitem[\protect\citeauthoryear{{Lalchand}, {Ravuri}  \&
  {Lawrence}}{{Lalchand} et~al.}{2022}]{gplvm_gpytorch}
{Lalchand} V.,  {Ravuri} A.,   {Lawrence} N.~D.,  2022, arXiv e-prints, \href
  {https://ui.adsabs.harvard.edu/abs/2022arXiv220212979L} {p. arXiv:2202.12979}

\bibitem[\protect\citeauthoryear{Lawrence}{Lawrence}{2005}]{GPLVM}
Lawrence N.,  2005, J. Mach. Learn. Res., 6, 1783

\bibitem[\protect\citeauthoryear{{Lee}, {Suzuki}  \& {Spergel}}{{Lee}
  et~al.}{2012}]{Lee12}
{Lee} K.-G.,  {Suzuki} N.,   {Spergel} D.~N.,  2012, \mn@doi [\aj]
  {10.1088/0004-6256/143/2/51}, \href
  {https://ui.adsabs.harvard.edu/abs/2012AJ....143...51L} {143, 51}

\bibitem[\protect\citeauthoryear{{Lee} et~al.,}{{Lee} et~al.}{2013}]{Lee13}
{Lee} K.-G.,  et~al., 2013, \mn@doi [\aj] {10.1088/0004-6256/145/3/69}, \href
  {https://ui.adsabs.harvard.edu/abs/2013AJ....145...69L} {145, 69}

\bibitem[\protect\citeauthoryear{{Lee} et~al.,}{{Lee} et~al.}{2015}]{Lee15}
{Lee} K.-G.,  et~al., 2015, \mn@doi [The Astrophysical Journal]
  {10.1088/0004-637X/799/2/196}, \href
  {https://ui.adsabs.harvard.edu/abs/2015ApJ...799..196L} {799, 196}

\bibitem[\protect\citeauthoryear{{Leschinski}}{{Leschinski}}{2021}]{SkyCalc_ipy2021}
{Leschinski} K.,  2021, {SkyCalc\_ipy: SkyCalc wrapper for interactive Python},
  Astrophysics Source Code Library, record ascl:2109.007 (\mn@eprint {ascl}
  {2109.007})

\bibitem[\protect\citeauthoryear{{Liu} \& {Bordoloi}}{{Liu} \&
  {Bordoloi}}{2021}]{Liu21}
{Liu} B.,  {Bordoloi} R.,  2021, \mn@doi [\mnras] {10.1093/mnras/stab177},
  \href {https://ui.adsabs.harvard.edu/abs/2021MNRAS.502.3510L} {502, 3510}

\bibitem[\protect\citeauthoryear{{L{\'o}pez} et~al.,}{{L{\'o}pez}
  et~al.}{2016}]{XQ100}
{L{\'o}pez} S.,  et~al., 2016, \mn@doi [\aap] {10.1051/0004-6361/201628161},
  \href {https://ui.adsabs.harvard.edu/abs/2016A&A...594A..91L} {594, A91}

\bibitem[\protect\citeauthoryear{{Luki{\'c}}, {Stark}, {Nugent}, {White},
  {Meiksin}  \& {Almgren}}{{Luki{\'c}} et~al.}{2015}]{Lukic15}
{Luki{\'c}} Z.,  {Stark} C.~W.,  {Nugent} P.,  {White} M.,  {Meiksin} A.~A.,
  {Almgren} A.,  2015, \mn@doi [\mnras] {10.1093/mnras/stu2377}, \href
  {http://adsabs.harvard.edu/abs/2015MNRAS.446.3697L} {446, 3697}

\bibitem[\protect\citeauthoryear{{Lusso}, {Worseck}, {Hennawi}, {Prochaska},
  {Vignali}, {Stern}  \& {O'Meara}}{{Lusso} et~al.}{2015}]{Lusso15}
{Lusso} E.,  {Worseck} G.,  {Hennawi} J.~F.,  {Prochaska} J.~X.,  {Vignali} C.,
   {Stern} J.,   {O'Meara} J.~M.,  2015, \mn@doi [\mnras]
  {10.1093/mnras/stv516}, \href
  {http://adsabs.harvard.edu/abs/2015MNRAS.449.4204L} {449, 4204}

\bibitem[\protect\citeauthoryear{{Malloy} \& {Lidz}}{{Malloy} \&
  {Lidz}}{2015}]{Malloy15}
{Malloy} M.,  {Lidz} A.,  2015, \mn@doi [\apj] {10.1088/0004-637X/799/2/179},
  \href {https://ui.adsabs.harvard.edu/abs/2015ApJ...799..179M} {799, 179}

\bibitem[\protect\citeauthoryear{{Mason}, {Treu}, {Dijkstra}, {Mesinger},
  {Trenti}, {Pentericci}, {de Barros}  \& {Vanzella}}{{Mason}
  et~al.}{2018}]{Mason18}
{Mason} C.~A.,  {Treu} T.,  {Dijkstra} M.,  {Mesinger} A.,  {Trenti} M.,
  {Pentericci} L.,  {de Barros} S.,   {Vanzella} E.,  2018, \mn@doi [\apj]
  {10.3847/1538-4357/aab0a7}, \href
  {https://ui.adsabs.harvard.edu/abs/2018ApJ...856....2M} {856, 2}

\bibitem[\protect\citeauthoryear{{Mason} et~al.,}{{Mason}
  et~al.}{2019}]{Mason19}
{Mason} C.~A.,  et~al., 2019, \mn@doi [\mnras] {10.1093/mnras/stz632}, \href
  {https://ui.adsabs.harvard.edu/abs/2019MNRAS.485.3947M} {485, 3947}

\bibitem[\protect\citeauthoryear{{McGreer}, {Mesinger}  \&
  {D'Odorico}}{{McGreer} et~al.}{2015}]{McGreer15}
{McGreer} I.~D.,  {Mesinger} A.,   {D'Odorico} V.,  2015, \mn@doi [\mnras]
  {10.1093/mnras/stu2449}, \href
  {http://adsabs.harvard.edu/abs/2015MNRAS.447..499M} {447, 499}

\bibitem[\protect\citeauthoryear{{Mesinger} \& {Furlanetto}}{{Mesinger} \&
  {Furlanetto}}{2007}]{Mesinger07}
{Mesinger} A.,  {Furlanetto} S.,  2007, \mn@doi [\apj] {10.1086/521806}, \href
  {https://ui.adsabs.harvard.edu/abs/2007ApJ...669..663M} {669, 663}

\bibitem[\protect\citeauthoryear{{Mesinger}, {Furlanetto}  \& {Cen}}{{Mesinger}
  et~al.}{2011}]{Mesinger11}
{Mesinger} A.,  {Furlanetto} S.,   {Cen} R.,  2011, \mn@doi [\mnras]
  {10.1111/j.1365-2966.2010.17731.x}, \href
  {http://adsabs.harvard.edu/abs/2011MNRAS.411..955M} {411, 955}

\bibitem[\protect\citeauthoryear{{Mesinger}, {Aykutalp}, {Vanzella},
  {Pentericci}, {Ferrara}  \& {Dijkstra}}{{Mesinger} et~al.}{2015}]{Mesinger15}
{Mesinger} A.,  {Aykutalp} A.,  {Vanzella} E.,  {Pentericci} L.,  {Ferrara} A.,
    {Dijkstra} M.,  2015, \mn@doi [\mnras] {10.1093/mnras/stu2089}, \href
  {https://ui.adsabs.harvard.edu/abs/2015MNRAS.446..566M} {446, 566}

\bibitem[\protect\citeauthoryear{{Mesinger}, {Greig}  \& {Sobacchi}}{{Mesinger}
  et~al.}{2016}]{Mesinger16}
{Mesinger} A.,  {Greig} B.,   {Sobacchi} E.,  2016, \mn@doi [\mnras]
  {10.1093/mnras/stw831}, \href
  {https://ui.adsabs.harvard.edu/abs/2016MNRAS.459.2342M} {459, 2342}

\bibitem[\protect\citeauthoryear{{Miralda-Escud{\'e}}}{{Miralda-Escud{\'e}}}{1998}]{Miralda98}
{Miralda-Escud{\'e}} J.,  1998, \mn@doi [\apj] {10.1086/305799}, \href
  {http://adsabs.harvard.edu/abs/1998ApJ...501...15M} {501, 15}

\bibitem[\protect\citeauthoryear{{Morey}, {Eilers}, {Davies}, {Hennawi}  \&
  {Simcoe}}{{Morey} et~al.}{2021}]{Morey21}
{Morey} K.~A.,  {Eilers} A.-C.,  {Davies} F.~B.,  {Hennawi} J.~F.,   {Simcoe}
  R.~A.,  2021, \mn@doi [\apj] {10.3847/1538-4357/ac1c70}, \href
  {https://ui.adsabs.harvard.edu/abs/2021ApJ...921...88M} {921, 88}

\bibitem[\protect\citeauthoryear{{Mortlock}, {Warren}  \& {et al.}}{{Mortlock}
  et~al.}{2011}]{Mortlock11}
{Mortlock} D.~J.,  {Warren} S.~J.,   {et al.} 2011, \mn@doi [\nat]
  {10.1038/nature10159}, \href
  {http://adsabs.harvard.edu/abs/2011Natur.474..616M} {474, 616}

\bibitem[\protect\citeauthoryear{{Murakami} et~al.,}{{Murakami}
  et~al.}{2023}]{SHOES23}
{Murakami} Y.~S.,  et~al., 2023, \mn@doi [\jcap]
  {10.1088/1475-7516/2023/11/046}, \href
  {https://ui.adsabs.harvard.edu/abs/2023JCAP...11..046M} {2023, 046}

\bibitem[\protect\citeauthoryear{{Nikoli{\'c}}, {Mesinger}, {Qin}  \&
  {Gorce}}{{Nikoli{\'c}} et~al.}{2023}]{Nikolic23}
{Nikoli{\'c}} I.,  {Mesinger} A.,  {Qin} Y.,   {Gorce} A.,  2023, \mn@doi
  [\mnras] {10.1093/mnras/stad2961}, \href
  {https://ui.adsabs.harvard.edu/abs/2023MNRAS.526.3170N} {526, 3170}

\bibitem[\protect\citeauthoryear{{O{\~n}orbe}, {Hennawi}  \&
  {Luki{\'c}}}{{O{\~n}orbe} et~al.}{2017}]{Onorbe17a}
{O{\~n}orbe} J.,  {Hennawi} J.~F.,   {Luki{\'c}} Z.,  2017, \mn@doi [\apj]
  {10.3847/1538-4357/aa6031}, \href
  {http://adsabs.harvard.edu/abs/2017ApJ...837..106O} {837, 106}

\bibitem[\protect\citeauthoryear{{P{\^a}ris} et~al.,}{{P{\^a}ris}
  et~al.}{2011}]{Paris11}
{P{\^a}ris} I.,  et~al., 2011, \mn@doi [\aap] {10.1051/0004-6361/201016233},
  \href {https://ui.adsabs.harvard.edu/abs/2011A&A...530A..50P} {530, A50}

\bibitem[\protect\citeauthoryear{{P{\^a}ris} et~al.,}{{P{\^a}ris}
  et~al.}{2017}]{Paris17}
{P{\^a}ris} I.,  et~al., 2017, \mn@doi [\aap] {10.1051/0004-6361/201527999},
  \href {https://ui.adsabs.harvard.edu/abs/2017A&A...597A..79P} {597, A79}

\bibitem[\protect\citeauthoryear{{P{\^a}ris} et~al.,}{{P{\^a}ris}
  et~al.}{2018}]{Paris18}
{P{\^a}ris} I.,  et~al., 2018, \mn@doi [\aap] {10.1051/0004-6361/201732445},
  \href {https://ui.adsabs.harvard.edu/abs/2018A&A...613A..51P} {613, A51}

\bibitem[\protect\citeauthoryear{{Pedregosa} et~al.,}{{Pedregosa}
  et~al.}{2011}]{sklearn2011}
{Pedregosa} F.,  et~al., 2011, \mn@doi [Journal of Machine Learning Research]
  {10.48550/arXiv.1201.0490}, \href
  {https://ui.adsabs.harvard.edu/abs/2011JMLR...12.2825P} {12, 2825}

\bibitem[\protect\citeauthoryear{P\'erez \& Granger}{P\'erez \&
  Granger}{2007}]{IPython2007}
P\'erez F.,  Granger B.~E.,  2007, \mn@doi [Computing in Science and
  Engineering] {10.1109/MCSE.2007.53}, 9, 21

\bibitem[\protect\citeauthoryear{Phan, Pradhan  \& Jankowiak}{Phan
  et~al.}{2019a}]{NumPyro2019a}
Phan D.,  Pradhan N.,   Jankowiak M.,  2019a, arXiv preprint arXiv:1912.11554

\bibitem[\protect\citeauthoryear{{Phan}, {Pradhan}  \& {Jankowiak}}{{Phan}
  et~al.}{2019b}]{numpyro}
{Phan} D.,  {Pradhan} N.,   {Jankowiak} M.,  2019b, \mn@doi [arXiv e-prints]
  {10.48550/arXiv.1912.11554}, \href
  {https://ui.adsabs.harvard.edu/abs/2019arXiv191211554P} {p. arXiv:1912.11554}

\bibitem[\protect\citeauthoryear{{Planck Collaboration} et~al.,}{{Planck
  Collaboration} et~al.}{2020}]{Planck20}
{Planck Collaboration} et~al., 2020, \mn@doi [\aap]
  {10.1051/0004-6361/201833910}, \href
  {https://ui.adsabs.harvard.edu/abs/2020A&A...641A...6P} {641, A6}

\bibitem[\protect\citeauthoryear{{Prochaska}}{{Prochaska}}{2017}]{Prochaska17}
{Prochaska} J.~X.,  2017, \mn@doi [Astronomy and Computing]
  {10.1016/j.ascom.2017.03.003}, \href
  {https://ui.adsabs.harvard.edu/abs/2017A&C....19...27P} {19, 27}

\bibitem[\protect\citeauthoryear{{Prochaska}, {Hennawi}, {Westfall}, {Cooke},
  {Wang}, {Hsyu}, {Davies}  \& {Farina}}{{Prochaska} et~al.}{2020}]{PypeIt}
{Prochaska} J.~X.,  {Hennawi} J.~F.,  {Westfall} K.~B.,  {Cooke} R.~J.,  {Wang}
  F.,  {Hsyu} T.,  {Davies} F.~B.,   {Farina} E.~P.,  2020, arXiv e-prints,
  \href {https://ui.adsabs.harvard.edu/abs/2020arXiv200506505P} {p.
  arXiv:2005.06505}

\bibitem[\protect\citeauthoryear{{Raiter}, {Schaerer}  \& {Fosbury}}{{Raiter}
  et~al.}{2010}]{Raiter10}
{Raiter} A.,  {Schaerer} D.,   {Fosbury} R.~A.~E.,  2010, \mn@doi [\aap]
  {10.1051/0004-6361/201015236}, \href
  {https://ui.adsabs.harvard.edu/abs/2010A&A...523A..64R} {523, A64}

\bibitem[\protect\citeauthoryear{{Reiman}}{{Reiman}}{2020}]{Reiman20}
{Reiman} D. e.~a.,  2020, submitted to \apj, \href
  {https://ui.adsabs.harvard.edu/abs/2019AJ....157..236Y} {}

\bibitem[\protect\citeauthoryear{{Riess} et~al.,}{{Riess}
  et~al.}{2022}]{Riess22}
{Riess} A.~G.,  et~al., 2022, \mn@doi [\apjl] {10.3847/2041-8213/ac5c5b}, \href
  {https://ui.adsabs.harvard.edu/abs/2022ApJ...934L...7R} {934, L7}

\bibitem[\protect\citeauthoryear{{Sadoun}, {Zheng}  \&
  {Miralda-Escud{\'e}}}{{Sadoun} et~al.}{2017}]{Sadoun17}
{Sadoun} R.,  {Zheng} Z.,   {Miralda-Escud{\'e}} J.,  2017, \mn@doi [\apj]
  {10.3847/1538-4357/aa683b}, \href
  {https://ui.adsabs.harvard.edu/abs/2017ApJ...839...44S} {839, 44}

\bibitem[\protect\citeauthoryear{{Satyavolu}, {Kulkarni}, {Keating}  \&
  {Haehnelt}}{{Satyavolu} et~al.}{2023}]{Satyavolu23a}
{Satyavolu} S.,  {Kulkarni} G.,  {Keating} L.~C.,   {Haehnelt} M.~G.,  2023,
  \mn@doi [\mnras] {10.1093/mnras/stad729}, \href
  {https://ui.adsabs.harvard.edu/abs/2023MNRAS.521.3108S} {521, 3108}

\bibitem[\protect\citeauthoryear{{Schaller}, {Schaye}, {Kugel}, {Broxterman}
  \& {van Daalen}}{{Schaller} et~al.}{2024}]{Schaller24}
{Schaller} M.,  {Schaye} J.,  {Kugel} R.,  {Broxterman} J.~C.,   {van Daalen}
  M.~P.,  2024, \mn@doi [arXiv e-prints] {10.48550/arXiv.2410.17109}, \href
  {https://ui.adsabs.harvard.edu/abs/2024arXiv241017109S} {p. arXiv:2410.17109}

\bibitem[\protect\citeauthoryear{{Sellentin} \& {Starck}}{{Sellentin} \&
  {Starck}}{2019}]{Sellentin19}
{Sellentin} E.,  {Starck} J.-L.,  2019, \mn@doi [\jcap]
  {10.1088/1475-7516/2019/08/021}, \href
  {https://ui.adsabs.harvard.edu/abs/2019JCAP...08..021S} {2019, 021}

\bibitem[\protect\citeauthoryear{{Shen} et~al.,}{{Shen} et~al.}{2019}]{Shen19}
{Shen} Y.,  et~al., 2019, \mn@doi [\apj] {10.3847/1538-4357/ab03d9}, \href
  {https://ui.adsabs.harvard.edu/abs/2019ApJ...873...35S} {873, 35}

\bibitem[\protect\citeauthoryear{{Sorini}, {O{\~n}orbe}, {Hennawi}  \&
  {Luki{\'c}}}{{Sorini} et~al.}{2018}]{Sorini18}
{Sorini} D.,  {O{\~n}orbe} J.,  {Hennawi} J.~F.,   {Luki{\'c}} Z.,  2018,
  \mn@doi [\apj] {10.3847/1538-4357/aabb52}, \href
  {https://ui.adsabs.harvard.edu/abs/2018ApJ...859..125S} {859, 125}

\bibitem[\protect\citeauthoryear{{Spergel} et~al.,}{{Spergel}
  et~al.}{2003}]{Spergel03}
{Spergel} D.~N.,  et~al., 2003, \mn@doi [\apjs] {10.1086/377226}, \href
  {http://adsabs.harvard.edu/abs/2003ApJS..148..175S} {148, 175}

\bibitem[\protect\citeauthoryear{{Spina}, {Bosman}, {Davies}, {Gaikwad}  \&
  {Zhu}}{{Spina} et~al.}{2024}]{Spina24}
{Spina} B.,  {Bosman} S. E.~I.,  {Davies} F.~B.,  {Gaikwad} P.,   {Zhu} Y.,
  2024, \mn@doi [arXiv e-prints] {10.48550/arXiv.2405.12273}, \href
  {https://ui.adsabs.harvard.edu/abs/2024arXiv240512273S} {p. arXiv:2405.12273}

\bibitem[\protect\citeauthoryear{{Sun}, {Ting}  \& {Cai}}{{Sun}
  et~al.}{2023}]{Sun23}
{Sun} Z.,  {Ting} Y.-S.,   {Cai} Z.,  2023, \mn@doi [\apjs]
  {10.3847/1538-4365/acf2f1}, \href
  {https://ui.adsabs.harvard.edu/abs/2023ApJS..269....4S} {269, 4}

\bibitem[\protect\citeauthoryear{{Suzuki}}{{Suzuki}}{2006}]{Suzuki06}
{Suzuki} N.,  2006, \mn@doi [\apjs] {10.1086/499272}, \href
  {http://adsabs.harvard.edu/cgi-bin/nph-bib\_query?bibcode=2006ApJS..163..110S\&db\_key=AST}
  {163, 110}

\bibitem[\protect\citeauthoryear{{Suzuki}, {Tytler}, {Kirkman}, {O'Meara}  \&
  {Lubin}}{{Suzuki} et~al.}{2005}]{Suzuki05}
{Suzuki} N.,  {Tytler} D.,  {Kirkman} D.,  {O'Meara} J.~M.,   {Lubin} D.,
  2005, \mn@doi [\apj] {10.1086/426062}, \href
  {https://ui.adsabs.harvard.edu/abs/2005ApJ...618..592S} {618, 592}

\bibitem[\protect\citeauthoryear{Titsias \& Lawrence}{Titsias \&
  Lawrence}{2010}]{Titsias10}
Titsias M.,  Lawrence N.~D.,  2010, in Teh Y.~W.,  Titterington M.,  eds,
  Proceedings of Machine Learning Research Vol. 9, Proceedings of the
  Thirteenth International Conference on Artificial Intelligence and
  Statistics. PMLR, Chia Laguna Resort, Sardinia, Italy, pp 844--851, \url
  {https://proceedings.mlr.press/v9/titsias10a.html}

\bibitem[\protect\citeauthoryear{Titsias \& L{\'a}zaro-Gredilla}{Titsias \&
  L{\'a}zaro-Gredilla}{2014}]{Titsias14}
Titsias M.,  L{\'a}zaro-Gredilla M.,  2014, in International conference on
  machine learning. pp 1971--1979

\bibitem[\protect\citeauthoryear{{Umeda}, {Ouchi}, {Nakajima}, {Harikane},
  {Ono}, {Xu}, {Isobe}  \& {Zhang}}{{Umeda} et~al.}{2023}]{Umeda23}
{Umeda} H.,  {Ouchi} M.,  {Nakajima} K.,  {Harikane} Y.,  {Ono} Y.,  {Xu} Y.,
  {Isobe} Y.,   {Zhang} Y.,  2023, \mn@doi [arXiv e-prints]
  {10.48550/arXiv.2306.00487}, \href
  {https://ui.adsabs.harvard.edu/abs/2023arXiv230600487U} {p. arXiv:2306.00487}

\bibitem[\protect\citeauthoryear{{Virtanen} et~al.,}{{Virtanen}
  et~al.}{2020}]{Scipy2020}
{Virtanen} P.,  et~al., 2020, \mn@doi [Nature Methods]
  {10.1038/s41592-019-0686-2}, \href
  {https://ui.adsabs.harvard.edu/abs/2020NatMe..17..261V} {17, 261}

\bibitem[\protect\citeauthoryear{{Wang} et~al.,}{{Wang} et~al.}{2020}]{Wang20a}
{Wang} F.,  et~al., 2020, \mn@doi [\apj] {10.3847/1538-4357/ab8c45}, \href
  {https://ui.adsabs.harvard.edu/abs/2020ApJ...896...23W} {896, 23}

\bibitem[\protect\citeauthoryear{{Wolfson}, {Hennawi}, {Davies}  \&
  {O{\~n}orbe}}{{Wolfson} et~al.}{2023}]{Wolfson23}
{Wolfson} M.,  {Hennawi} J.~F.,  {Davies} F.~B.,   {O{\~n}orbe} J.,  2023,
  \mn@doi [\mnras] {10.1093/mnras/stad701}, \href
  {https://ui.adsabs.harvard.edu/abs/2023MNRAS.tmp..677W} {}

\bibitem[\protect\citeauthoryear{{Yang} et~al.,}{{Yang} et~al.}{2020}]{Yang20b}
{Yang} J.,  et~al., 2020, \mn@doi [\apjl] {10.3847/2041-8213/ab9c26}, \href
  {https://ui.adsabs.harvard.edu/abs/2020ApJ...897L..14Y} {897, L14}

\bibitem[\protect\citeauthoryear{{Yip} et~al.,}{{Yip} et~al.}{2004}]{Yip04}
{Yip} C.~W.,  et~al., 2004, \mn@doi [\aj] {10.1086/425626}, \href
  {https://ui.adsabs.harvard.edu/abs/2004AJ....128.2603Y} {128, 2603}

\bibitem[\protect\citeauthoryear{{Young}, {Sargent}, {Boksenberg}, {Carswell}
  \& {Whelan}}{{Young} et~al.}{1979}]{Young79}
{Young} P.~J.,  {Sargent} W.~L.~W.,  {Boksenberg} A.,  {Carswell} R.~F.,
  {Whelan} J.~A.~J.,  1979, \mn@doi [\apj] {10.1086/157024}, \href
  {https://ui.adsabs.harvard.edu/abs/1979ApJ...229..891Y} {229, 891}

\bibitem[\protect\citeauthoryear{{Zahn} et~al.,}{{Zahn} et~al.}{2012}]{Zahn12}
{Zahn} O.,  et~al., 2012, \mn@doi [\apj] {10.1088/0004-637X/756/1/65}, \href
  {http://adsabs.harvard.edu/abs/2012ApJ...756...65Z} {756, 65}

\bibitem[\protect\citeauthoryear{{Zel'Dovich}}{{Zel'Dovich}}{1970}]{Zeldovich70}
{Zel'Dovich} Y.~B.,  1970, \aap, \href
  {https://ui.adsabs.harvard.edu/abs/1970A&A.....5...84Z} {500, 13}

\bibitem[\protect\citeauthoryear{{Zhou}, {Chen}, {Di Matteo}, {Ni}, {Croft}  \&
  {Bird}}{{Zhou} et~al.}{2023}]{Zhou23}
{Zhou} Y.,  {Chen} H.,  {Di Matteo} T.,  {Ni} Y.,  {Croft} R. A.~C.,   {Bird}
  S.,  2023, \mn@doi [arXiv e-prints] {10.48550/arXiv.2309.11571}, \href
  {https://ui.adsabs.harvard.edu/abs/2023arXiv230911571Z} {p. arXiv:2309.11571}

\bibitem[\protect\citeauthoryear{{Zhu} et~al.,}{{Zhu} et~al.}{2023}]{Zhu23}
{Zhu} Y.,  et~al., 2023, \mn@doi [\apj] {10.3847/1538-4357/aceef4}, \href
  {https://ui.adsabs.harvard.edu/abs/2023ApJ...955..115Z} {955, 115}

\bibitem[\protect\citeauthoryear{{Zhu} et~al.,}{{Zhu} et~al.}{2024}]{Zhu24}
{Zhu} Y.,  et~al., 2024, \mn@doi [arXiv e-prints] {10.48550/arXiv.2405.12275},
  \href {https://ui.adsabs.harvard.edu/abs/2024arXiv240512275Z} {p.
  arXiv:2405.12275}

\bibitem[\protect\citeauthoryear{{{\v{D}}urov{\v{c}}{\'\i}kov{\'a}}, {Katz},
  {Bosman}, {Davies}, {Devriendt}  \&
  {Slyz}}{{{\v{D}}urov{\v{c}}{\'\i}kov{\'a}} et~al.}{2020}]{Dominika20}
{{\v{D}}urov{\v{c}}{\'\i}kov{\'a}} D.,  {Katz} H.,  {Bosman} S. E.~I.,
  {Davies} F.~B.,  {Devriendt} J.,   {Slyz} A.,  2020, \mn@doi [\mnras]
  {10.1093/mnras/staa505}, \href
  {https://ui.adsabs.harvard.edu/abs/2020MNRAS.493.4256D} {493, 4256}

\bibitem[\protect\citeauthoryear{{{\v{D}}urov{\v{c}}{\'\i}kov{\'a}}
  et~al.,}{{{\v{D}}urov{\v{c}}{\'\i}kov{\'a}} et~al.}{2024}]{Dominika24}
{{\v{D}}urov{\v{c}}{\'\i}kov{\'a}} D.,  et~al., 2024, \mn@doi [arXiv e-prints]
  {10.48550/arXiv.2401.10328}, \href
  {https://ui.adsabs.harvard.edu/abs/2024arXiv240110328D} {p. arXiv:2401.10328}

\bibitem[\protect\citeauthoryear{{van den Busch} et~al.,}{{van den Busch}
  et~al.}{2022}]{KIDSWL22}
{van den Busch} J.~L.,  et~al., 2022, \mn@doi [\aap]
  {10.1051/0004-6361/202142083}, \href
  {https://ui.adsabs.harvard.edu/abs/2022A&A...664A.170V} {664, A170}

\makeatother
\end{thebibliography}

%%%%%%%%%%%%%%%%%%%%%%%%%%%%%%%%%%%%%%%%%%%%%%%%%%

%%%%%%%%%%%%%%%%% APPENDICES %%%%%%%%%%%%%%%%%%%%%

\appendix

\section{Coverage Tests}

In this Appendix we provide details on the coverage test presented in \S~\ref{sec:cov_test}, the
reweighting scheme described in \S~\ref{sec:reweighting}, and the application of both to marginal posterior distributions. 
\label{appendix:coverage}

\subsection{An Algorithm for Performing a Coverage Test}
\label{sec:algorithm}
Below we provide a  description of an algorithm for carrying out a coverage test.

\begin{enumerate}
\item Draw $N$ parameter vectors $\bTheta_{{\rm true},j}$ from the prior distribution  $P(\bTheta)$. These are the `true' parameters that generate the mock datasets used to perform the coverage test. 

\item Using a forward simulator, generate a set of $N$ mock datsets, $\bx_j$, for
  each of the parameter vectors  $\bTheta_{{\rm true},j}$.

\item Inference is carried out on each dataset resulting in a set of $N$ posterior distributions
  $P(\bTheta | \bx_j)$.
\item Consider a set of $M$ credibility contour levels $\alpha \in [0,1]$.  For each value $\alpha$ and each mock
  dataset $(\bTheta_{{\rm true},j}, \bx_j)$, one tests whether the true parameter value $\bTheta_{{\rm true},j}$ resides within the volume $V_{\alpha}$
  enclosed by the $\alpha$-th contour,
  defined by  
  \be
  \int_{V_{\alpha}} P(\btheta| \bx_j) \mathrm{d}\btheta = \alpha\label{eqn:post_volume}. 
  \ee
  For each $\alpha$, the  coverage probability $C(\alpha)$ is the fraction of the $N$ mock datasets for which the
  true value $\btheta_{{\rm true},j}$ lies within the volume $V_{\alpha}$.
\end{enumerate}
The result of the coverage test is the relation $C(\alpha)$ versus
$\alpha$. A perfect inference procedure would yield $C(\alpha) =
\alpha$ in the limit $N\rightarrow \infty$. An overconfident inference
pipeline will yield a curve $C(\alpha)$ versus $\alpha$ that lies
systematically below the line $y=x$, whereas for an underconfident
inference procedure the $C(\alpha$) will lie
above $y=x$.  This algorithm yields an unbiased estimate $C(\alpha)$
of the underlying coverage probability from a finite set of mock
datasets $N$. Since $C(\alpha)$ counts how often the true
parameters fall inside the $\alpha$-th contour, it is the number of
successes in a sequence of $N$ independent experiments each asking a
yes-or-no question --- success occuring with a probability $p=C(\alpha)$
and failure occuring with probability $q=1-C(\alpha)$. Thus by
definition our estimate $C(\alpha)$ must follow the Binomial
distribution $B(N, C(\alpha))$, which can be used to assign error bars
to $C(\alpha)$ resulting from the finite number of mock datasets $N$.

Underlying this coverage test algorithm is a procedure for testing whether a true parameter vector resides within the volume $V_{\alpha}$ enclosed by a contour corresponding to credibility level $\alpha$. A contour of the posterior $P(\bTheta| \bx_j)$ containing
a fraction $\alpha$ of the total probability slices the posterior along an isodensity level $P_\alpha$, such
that the volume $V_\alpha$ in eqn.~(\ref{eqn:post_volume}) is defined by
\be
V_\alpha = \{ \bTheta ~|~P(\bTheta| \bx_j) \ge P_\alpha \}\label{eqn:domain}. 
\ee
A parameter value $\bTheta^\prime$ will lie within the volume enclosed by the $\alpha$-th contour provided that $P(\bTheta^\prime| \bx_j) \ge P_\alpha$,
so the procedure boils down to estimating the set of isodensity levels $P_\alpha$ corresponding to the set of credibility levels 
$\alpha$.

In practice, one typically has a number of samples, $n$, from
the  posterior from a run of an MCMC or HMC sampling algorithm.
If we
rewrite eqn.~(\ref{eqn:post_volume}) as an integral over the entire
parameter space $\bTheta$:
\be
\int H[P(\bTheta| \bx_j) - P_\alpha]
P(\bTheta| \bx_j) \mathrm{d}\bTheta = \alpha\label{eqn:heaviside},
\ee
where the Heaviside function, $H(x)$, enforces the condition on the volume from 
eqn.~(\ref{eqn:domain}), then it can be evaluated via
Monte Carlo integration according to
\be
\int H[P(\bTheta| \bx_j) -
P_\alpha] P(\bTheta| \bx_j) \mathrm{d}\bTheta = \frac{1}{n} \sum_i^{n} H[P(\bTheta_i| \bx_j) - P_\alpha]\label{eqn:mcmc_integ}, 
\ee
where the sum is over the $n$ MCMC or HMC posterior samples. 
Hence we can determine the isodensity level, $P_\alpha$, corresponding to each
credibility level, $\alpha$, by solving the equation:
\begin{align}
\frac{1}{n} \sum_i^n H[P(\bTheta_i| \bx_j) -
P_\alpha] & = \frac{\#~{\rm of~samples~with~} P(\bTheta| \bx_j) \ge
  P_\alpha}{n} = \alpha\nonumber\\
 & = {\rm CDF}(\ge P_\alpha) = \alpha. \label{eqn:Palpha}
\end{align}
The second equality in eqn.~(\ref{eqn:Palpha}) indicates that credibility contour definition
amounts to computing the cumulative probability distribution, ${\rm CDF}(\ge P_\alpha)$, of the posterior
distribution at the MCMC or HMC samples, $\bTheta_i$.
One can
then invert the CDF 
\be
P_\alpha = {\rm CDF}^{-1}(\alpha), 
\ee
to determine the corresponding isodensity levels.
This procedure can be employed to test
each of the $N$ posteriors in the inference test for all credibility
levels.  If the true model parameters, $\bTheta_{{\rm true},j}$, lie inside the
volume of credibility contour $\alpha$ then the condition
$P(\bTheta_{{\rm true, j}}| \bx_j) \ge P_\alpha$ will be satisfied.
MCMC/HMC samplers typically can return the value of $P(\bTheta_i| \bx)$ at every
sample in the chain, allowing one to easily estimate ${\rm CDF}(\ge P_\alpha)$.
Then one only needs to evaluate
the $P(\bTheta_{{\rm true, j}}| \bx_j)$,
which is straightforward since the function $P(\bTheta | \bx_j)$ is a prerequisite
for performing inference.

\subsection{Reweighting Posterior Samples to Pass a Coverage Test}

Consider a scenario where our current inference pipeline fails a coverage test because of
some imperfection in our probabilistic model of the measurement process which it is not straightforward to correct.
For concreteness, imagine we adopted an approximate form for the posterior distribution because the true form is not analytically
tractable, and that our coverage test indicates these approximate posteriors are overconfident, which is to say that
$C(\alpha)$ versus $\alpha$ lies systematically below the line $y=x$. How can we nevertheless perform statisically reliable inference?
\citet{Sellentin19} advocate that one simply relabel the contours to reflect this unfortunate reality. For example, if the
68th percentile credibility contour, $\alpha = 0.68$,  actually only contains the true model 48\% of the time, i.e. $C(\alpha) = 0.48 < \alpha$, then label this contour as the 48th percentile rather than the 68th. The real 68th percentile contour 
containing the true parameters 68\% of the time under the inference test might say correspond to the value $\alpha=0.85 = C^{-1}(0.68)$
contour for the original approximate inference, which in turn maps to a lower isodensity level of the approximate posterior  $P(\btheta| \bx)$, i.e. $P_{0.85} < P_{0.68}$. In other words,
by choosing a lower isodensity threshold $P_\alpha$ we expand the contours to contain the true model the empirically correct fraction
of the time. 

In general, this remapping of the credibility levels $\alpha$ into
true coverage probabilities $C(\alpha)$ can be achieved by assigning a
set of weights to the samples from the approximate posterior. The
purpose of MCMC or HMC samples from a posterior is
to estimate credibility intervals on parameters,
perform marginalization integrals, and compute `moments' of the posterior via
Monte Carlo integration. If we can determine the set of weights that corrects the imperfect inference
such that it passes an inference test, these weights can then be used in all of the downstream computations that one performs
with the samples. To achieve this
we generalize eqn.~(\ref{eqn:Palpha}) to the case of defining contour levels
from reweighted samples
\be
\sum_i^n w_i H[P(\bTheta_i| \bx_j)
- P^\prime_\alpha]  = {\rm wCDF}(\ge P^\prime_\alpha)  = \alpha\label{eqn:reweight}, 
\ee
where we have simply absorbed the constant $1\slash n$ normalization factor into the definition of the weights. Here
${\rm wCDF}(\ge P^\prime_\alpha)$ is the weighted cumulative distribution function of the original posterior evaluated
at the samples, $\bTheta_i$.  Similar to before, the isodensity levels of the reweighted posterior can be determined according
\be
P^\prime_\alpha = {\rm wCDF}^{-1}(\alpha)\label{eqn:Pprime}.  
\ee
In order for the weighting to correct the inference for e.g. credibility level
$\alpha=0.68$,  we need the solution to eqn.~(\ref{eqn:Pprime}) to yield
$P^\prime_{0.68} = P_{0.85}$ where $0.85 = C^{-1}(0.68)$ and $P_{0.85} = {\rm CDF}^{-1}(0.85)$ is the solution for the
isodensity threshold under the original approximate inference from eqn.~(\ref{eqn:Palpha}). Furthermore, this remapping 
must hold for all of the isodensity levels $P^\prime_\alpha =P_{C^{-1}(\alpha)} = {\rm CDF}^{-1}(C^{-1}(\alpha))$.

To build intuition, first consider the situtation where
$C(\alpha)=\alpha$, i.e.  where we pass the inference test
perfectly. In this limiting case, it is clear that the weighting should
be uniform and hence $w_i = 1\slash n$. Without loss of generality we
can sort the samples in the sum in eqn.~(\ref{eqn:reweight}) in order
of increasing $P(\bTheta_i| \bx_j)$, adopting the convention that
$P(\bTheta_1| \bx_j)) \le P(\bTheta_2| \bx_j)) \le \dots \le P(\bTheta_n | \bx_j)$ and the corresponding weights are similarly ordered
such that $w_1$ is the weight assigned to the sample with smallest
value $P(\bTheta_1| \bx_j)$, etc.  Since we need to solve for $n$
weights, $w_i$, we need the same number of values of $\alpha$ as
constraints. If we choose $n$ linearly decreasing values of $\alpha$
spanning the range $\alpha =[1, 1\slash n]$, then
eqn.~(\ref{eqn:reweight}) implies the following set of linear
equations:
\begin{equation}
\begin{array}{ccccc}
  w_1  + w_2~~ &\!\! +~~\dots~~+ &  w_{n-1} + w_n & = &1\\
  \phantom{w_1 +\,\,\,} w_2~~ & \!\! +~~\dots~~+ &  w_{n-1} + w_n & = &1-1\slash n\\
  &\vdots&\vdots&\vdots&\vdots\\
  &&   w_{n -1} + w_n & = & 2\slash n\\
  &&   \phantom{w_{n-1}} \phantom{+\,\,\,} w_n & = & 1\slash n,\label{eqn:linear}\\ 
\end{array} 
\end{equation}
or equivalently in matrix form
\be
\mathbf{U}\mathbf{w} = \boldsymbol{\alpha}, 
\ee
where $\mathbf{U}$ is an upper triangular matrix with all non-zero elements equal to unity, $\mathbf{w}$ is a vector of weights to be
assigned to the samples, and $\balpha$ is the vector of uniformly spaced credibility levels. Since the
$\balpha$ constraints that we chose can also be interpreted as the cumulative distribution of the posterior evaluated at the samples, i.e.  $\alpha_i = {\rm CDF}(\ge P_i)$, eqn.~(\ref{eqn:linear}) is equivalent to 
\be
{\rm wCDF}(\ge P_i) = {\rm CDF}(\ge P_i). 
\ee
It is thus no surprise that the solution to this linear system  is obviously $w_i = 1\slash n$ yielding the uniform weighting we expect.

To generalize to the case $C(\alpha) \ne \alpha$, we solve for the weights that satisfy
\be
\mathbf{U}\mathbf{w} = \mathbf{C}(\balpha)\label{eqn:matrix}, 
\ee
where $\mathbf{C}(\balpha)$ is now a vector of coverage values evaluated at the vector of uniformly spaced credibility
levels $\balpha$. To see why this works consider for example the $k$th equation in the linear system in eqn.~(\ref{eqn:matrix}) where
$k$ is chosen to be the sum over the 85th percentile highest rank weights and  $P(\bTheta_i| \bx_j)$ values (recall the weights are sorted in order of increasing $P(\bTheta_i| \bx_j)$)
\be
w_k + w_{k+1} + \dots + w_n = C(0.85) = 0.68. 
\ee
or equivalently 
\be
{\rm wCDF}(\ge P^\prime_{0.68}) = 0.68. 
\ee
Because we chose $\balpha$ to be linearly decreasing, the $k$th ranked sample will correspond to the 85th percentile of the CDF of the
original approximate posterior values evaluated at the MCMC/HMC samples, and hence it is guaranteed that $P^\prime_{0.68} = P_{0.85} =  {\rm CDF}^{-1}(0.85)$. In other words, the 68th percentile contour of the new reweighted posterior distribution will correspond to the 85th percentile contour of the original approximate posterior, as desired.  The rest of the equations in the linear system similarly enforce the constraints that $P^\prime_\alpha =P_{C^{-1}(\alpha)} = {\rm CDF}^{-1}(C^{-1}(\alpha))$ for the other credibility levels $\alpha$.  Since the determinant of an upper triangular matrix is simply the product of the diagonal elements, $\det{\mathbf{U}} = 1$, and the linear system in eqn.~(\ref{eqn:matrix}) will always yield a unique solution for the vector of weights $\mathbf{w}$ which imposes these constraints.

\subsection{Coverage Tests for Marginal Distributions}

Performing inference requires that one can evaluate $P(\bTheta| \bx_j)$
at any location within the parameter space. Furthermore, MCMC and HMC
samplers typically return the value of $P(\bTheta| \bx_j)$ at every
sample in the chain, allowing one to easily determine the set of
isodensity levels for each posterior and test $P(\bTheta_{{\rm true},j}| \bx_j) \ge
P_\alpha$. But what about performing inference tests on \emph{marginal
distributions}?  To make the discussion more concrete let us imagine
that our parameter vector $\bTheta$ can be split into a set of
$\btheta$ physical parameters of interest and a set of
$\boeta$ nuisance parameters. It is easy to see that one
can pass a coverage test for the full posterior $P(\bTheta| \bx_j)$ but
nevertheless fail a coverage test for the marginal posterior
$P(\btheta | \bx_j)$\footnote{Consider a thought experiment
where there is a single physical parameter and a million nuisance
parameters. Imagine a trivial true posterior independently described by
$\mathcal{N}(\theta_i | 0, 1)$ for each element of $\bTheta$. 
In conducting inference, suppose we use the correct
unit variance for each of the million elements of $\boeta$,
but erroneously adopt $\sigma_\theta = 1\slash 2$ for the Gaussian
describing $\theta$. It is obvious that the incorrect form
for the $\theta$ component will be diluted by contributions
from the million nuisance parameters, and thus have a negligible
impact on the total posterior. Within reasonable numerical precision
the incorrect form of $P(\bTheta| \bx_j)$ with $\sigma_\theta =
1\slash 2$ will pass a coverage test. But since the marginalization
integral over the million nuisance parameters here trivially evaluates
to unity, the marginal posterior is $P(\theta | \bx_j) =
\mathcal{N}(\theta | 0, 1\slash 2)$, and it is manifestly clear that it will fail
the marginal coverage test.}.  However, the marginal physical parameter posterior, $P(\btheta | \bx_j)$, is what
we actually care about, whereas we would be more willing
to tolerate overconfident (or underconfident) total posteriors and
marginal nuisance parameter posteriors.

The same coverage test algorithm described in \S~\ref{sec:algorithm}
applies to the marginal case. Specifically, all of the steps of the
procedure are the same with the exception of step iv), where one must
now test whether the true physical parameter values $\btheta_{{\rm
    true}, j}$ lie within contours of the marginal posterior
$P(\btheta | \bx_j)$ for each dataset. There is however an
important technical difference. The procedure outlined in
eqn.~(\ref{eqn:Palpha}) for determining isodensity levels and testing
whether the true physical parameters lie within volumes enclosed by
the credibility contours specified by $\alpha$ presumes that one can
evaluate $P(\btheta | \bx_j)$ at every sample in the MCMC/HMC chain 
as well as at the true parameter location. However, in practice
evaluating $P(\btheta | \bx_j)$ would require performing a
typically intractable marginalization integral over the nuisance
parameters. Instead, this intractable marginalization can be performed
via Monte Carlo integration. By applying a density estimation
algorithm (i.e. histogram, kernel density estimation, or Gaussian
mixture model) to the MCMC or HMC samples marginalized over the nuisance parameters, 
an expression for $P(\btheta| \bx_j)$ can then be determined allowing one to test whether
the true physical parameter values lie within the contours of the
marginal posteriors. While this Monte Carlo integration plus density
estimation procedure may sound complex, it is exactly the procedure
adopted when making a corner plot of MCMC/HMC samples.

%%%%%%%%%%%%%%%%%%%%%%%%%%%%%%%%%%%%%%%%%%%%%%%%%%

% Don't change these lines
\bsp	% typesetting comment
\label{lastpage}
\end{document}